\documentclass[12pt]{article}
\usepackage{epsf}

\newcommand{\be}{\begin{equation}}
\newcommand{\ee}{\end{equation}}
\newcommand{\bea}{\begin{eqnarray}}
\newcommand{\eea}{\end{eqnarray}}
\newcommand{\vap}{\varepsilon}
\newcommand{\pary}{\partial_{y}}
\newcommand{\parp}{\partial}

\begin{document}

\title{
\begin{flushright}
{\small SU{-}4240{-}714}
\end{flushright} 
The Statistical Mechanics of Membranes}

\author{\small \\ Mark~J. Bowick\thanks{\tt 
bowick@physics.syr.edu} \, 
and Alex Travesset\thanks{\tt alex@suhep.phy.syr.edu}
\\ $^1$Physics Department, Syracuse University,\\
Syracuse, NY 13244-1130, USA  \\ }

\date{}

\maketitle

\begin{abstract}

The fluctuations of two-dimensional extended objects ({\em membranes}) is a
rich and exciting field with many solid results and a wide range of 
open issues. 
We review the distinct universality classes of membranes, determined by
the local order, and the associated phase diagrams.  
After a discussion of several physical examples of membranes 
we turn to the physics of {\em crystalline} (or {\em polymerized})
membranes in which the individual monomers are rigidly bound. 
We discuss the phase diagram with particular attention to the
dependence on the degree of {\em self-avoidance} and  {\em
anisotropy}. In each case we review and discuss
analytic, numerical and experimental predictions of critical
exponents and other key observables. 
Particular emphasis is given to the results obtained from the  
renormalization group $\vap$-expansion.
The resulting renormalization group flows and fixed points are illustrated
graphically. The full technical details necessary to perform actual
calculations are presented in the Appendices.   
We then turn to a discussion of the role of topological defects 
whose liberation leads to the {\em hexatic} and {\em fluid} universality
classes. We finish with conclusions and a discussion of promising open
directions for the future.

\end{abstract}

\vfill\newpage

\pagebreak

\tableofcontents

\newpage

\section{Introduction}
\label{SECT__Intro}

\bibliographystyle{unsrt}

The statistical mechanics of one-dimensional structures (polymers) is
fascinating and has proved to be fruitful from the fundamental and
applied points of view \cite{deGennes,desCloiseaux}. 
The key reasons for this success lie in the
notion of {\em universality} and the relative simplicity of
one-dimensional geometry. Many features of the long-wavelength behavior
of polymers are independent of the detailed physical and chemical
nature of the monomers that constitute the polymer building blocks and
their bonding into macromolecules. These microscopic details simply
wash out in the thermodynamic limit of large systems and allow
predictions of critical exponents that should apply to a wide class of
microscopically distinct polymeric systems. Polymers are also
sufficiently simple that considerable analytic and numerical progress
has been possible. Their statistical mechanics is essentially that of
ensembles of various classes of {\em random walks} in some
$d$-dimensional bulk or embedding space. 

A natural extension of these systems is to intrinsic two-dimensional
structures which we may call generically call {\em membranes}. The
statistical mechanics of these {\em random surfaces} is far more
complex than that of polymers because two-dimensional geometry is far
richer than the very restricted geometry of lines. Even planar
two-dimensional {---} monolayers {---} are complex, as evidenced by
the KTNHY \cite{KT,NH:79,Young} theory of defect-mediated melting of 
monolayers with two distinct {\em continuous} phase transitions 
separating an intermediate
hexatic phase, characterized by quasi-long-range bond orientational
order, from both a low-temperature {\em crystalline} phase and a
high-temperature {\em fluid} phase.  But full-fledged membranes are
subject also to shape fluctuations and their macroscopic behavior is
determined by a subtle interplay between their particular microscopic
order and the entropy of shape and elastic deformations. For
membranes, unlike polymers, distinct types of microscopic order
(crystalline, hexatic, fluid) will lead to distinct long-wavelength
behavior and consequently a rich set of universality classes. 

Flexible membranes are an important member of the enormous class of 
{\em soft} condensed matter systems \cite{deGennes2,Lub:97,GuPa:92,CL:95},
those which respond easily to external forces. Their physical
properties are to a considerable extent dominated by the entropy of thermal
fluctuations. 
 
In this review we will describe some of the presently understood behavior of
crystalline (fixed-connectivity), hexatic and fluid membranes,
including the relevance of self-avoidance, intrinsic anisotropy and 
topological defects. Emphasis will be given to the role
of the renormalization group in elucidating the critical behavior of
membranes. The polymer pastures may be lovely but a dazzling world
awaits those who wander into the membrane meadows. 

The outline of the review is the following.
In sec.~\ref{SECT__Examples} we describe a variety of important
physical examples of membranes, with representatives from the key
universality classes. In sec.~\ref{SECT__RG} we introduce basic
notions from the renormalization group and some formalism that we will
use in the rest of the review. In sec.~\ref{SECT__POLYMEM} we review
the phase structure of crystalline membranes for both phantom and
self-avoiding membranes, including a thorough discussion of the
fixed-point structure, RG flows and critical exponents of each global
phase. In sec.~\ref{SECT__POLYMEM_ANI} we turn to the same issues
for intrinsically anisotropic membranes, with the new feature of the
{\em tubular} phase. In sec.~\ref{SECT__Defects} we address the
consequences of allowing for membrane defects, leading to a discussion
of the hexatic membrane universality class. We end with a
brief discussion of fluid membranes in sec.~\ref{SECT__Fluid}
and conclusions. 

\section{Physical examples of membranes}
\label{SECT__Examples}

\begin{figure}[htb]
\epsfxsize=3in
\centerline{\epsfbox{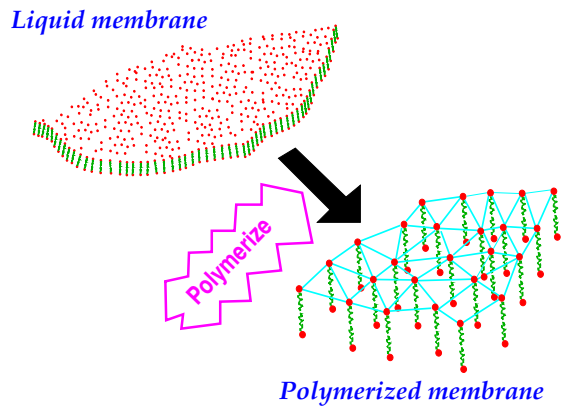}}
\caption{The polymerization of fluid-like membrane to a crystalline membrane.}
\label{fig__polymerize}
\end{figure}

There are many concrete realizations of membranes in
nature, which greatly enhances the significance of their study.
Crystalline membranes, sometimes termed {\em tethered} or {\em polymerized}
membranes, are the natural generalization of linear polymer chains to
intrinsically two-dimensional structures. They possess in-plane
elastic moduli as well as bending rigidity and are characterized by
broken translational invariance in the plane and fixed connectivity
resulting from relatively strong bonding. Geometrically speaking they
have a preferred two-dimensional metric.  
Let's look at some of the examples. 
One can polymerize suitable chiral oligomeric precursors to form
molecular sheets \cite{Stupp}. This approach is based directly on the
idea of creating an intrinsically two-dimensional polymer. 
Alternatively one can permanently cross-link
fluid-like Langmuir-Blodgett films or amphiphilic bilayers by adding
certain functional groups to the hydrocarbon tails and/or the 
polar heads \cite{Fendler1,Fendler2} as shown schematically in 
Fig.~\ref{fig__polymerize}.   

\begin{figure}[hp]
\epsfxsize=3in
\centerline{\epsfbox{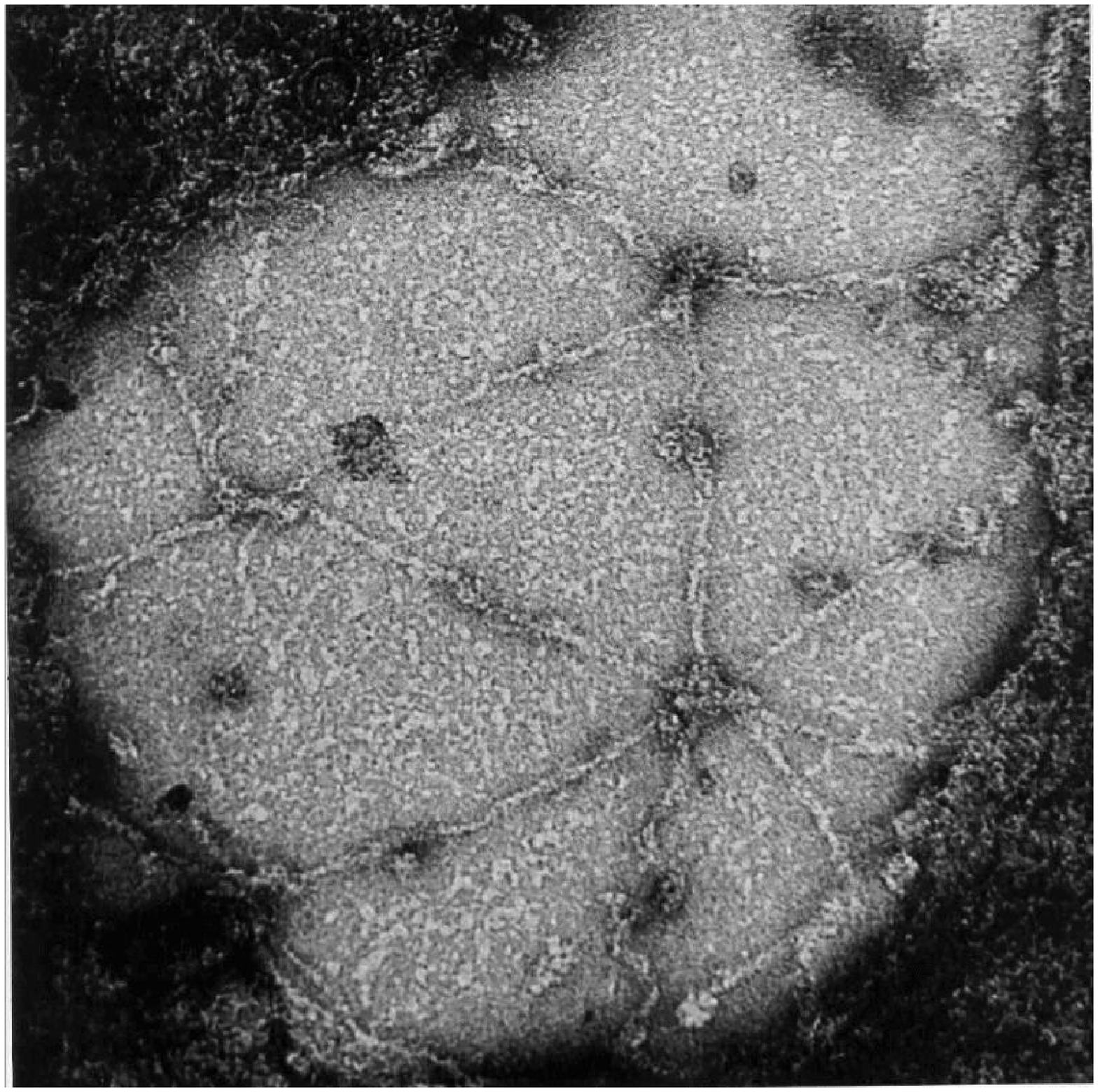}}
\caption{An electron micrograph of a region of the erythrocyte
cytoskeleton. The skeleton is negatively stained (magnification
365,000) and has been artificially spread to a surface area nine to
ten times as great as in the native membrane \cite{Byers}. }
\label{fig__spectrin1}
\epsfxsize=3in
\centerline{\epsfbox{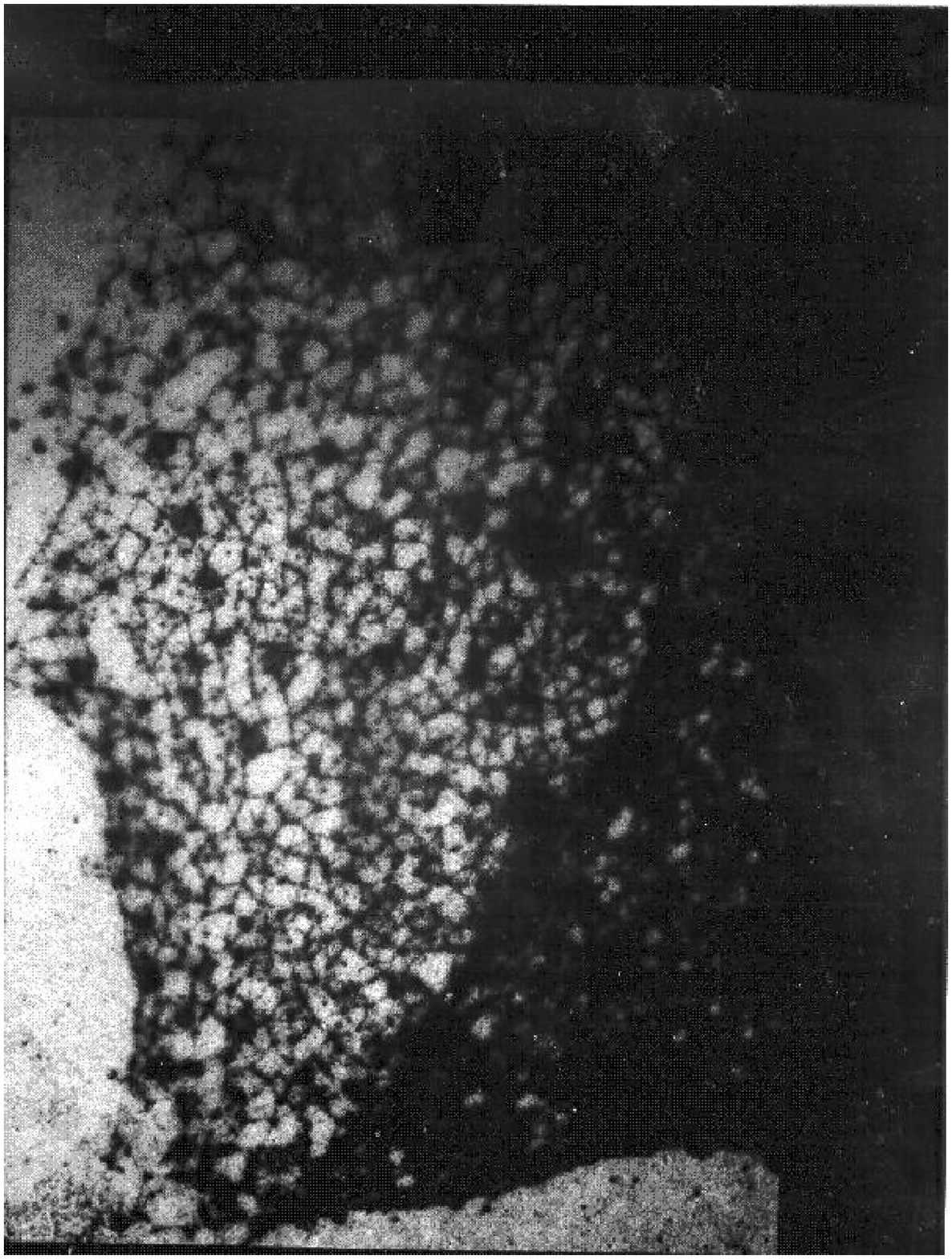}}
\caption{An extended view of the crystalline spectrin/actin
network which forms the cytoskeleton of the red blood cell membrane 
\cite{Branton2}. }
\label{fig__spectrin2}
\end{figure}

The cytoskeletons of cell membranes are beautiful and naturally occurring 
crystalline membranes that are essential to cell membrane stability
and functionality. The simplest and most thoroughly studied example 
is the cytoskeleton of mammalian erythrocytes (red blood cells). 
The human body has roughly $5 \times 10^{13}$ red blood cells. 
The red blood cell cytoskeleton is a fishnet-like
network of triangular plaquettes formed primarily by the proteins {\em
spectrin} and {\em actin}. The links of the mesh are spectrin
tetramers (of length approximately 200 nm) and the nodes are short
actin filaments (of length 37 nm and typically 13 actin monomers long)
\cite{Skel:93,Branton}, as seen in Fig.\ref{fig__spectrin1} and 
Fig.\ref{fig__spectrin2}. There are
roughly 70,000 triangular plaquettes in the mesh altogether and the
cytoskeleton as a whole is bound by ankyrin and other proteins 
to the cytoplasmic side of the fluid
phospholipid bilayer which constitutes the other key component of the
red blood cell membrane.   

\begin{figure}[htb]
\epsfxsize=3in
\centerline{\epsfbox{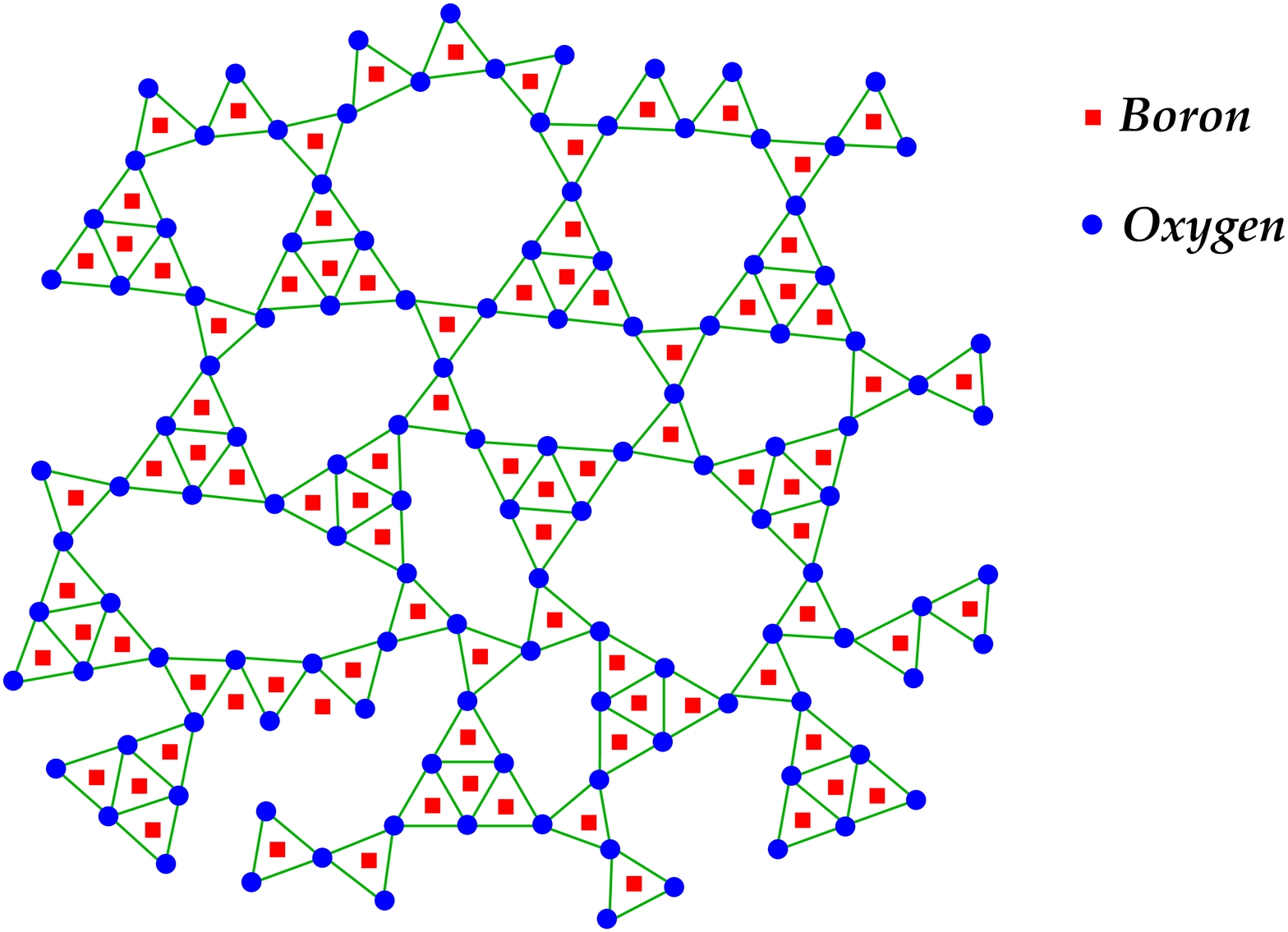}}
\caption{The sheet molecule $B_2O_3$}
\label{fig__sheet_mol}
\end{figure}

There are also inorganic realizations of crystalline membranes.
Graphitic oxide (GO) membranes are micron size sheets of solid carbon
with thicknesses on the order of 10\AA, formed by exfoliating carbon
with a strong oxidizing agent. 
Their structure in an aqueous suspension has been examined by several
groups \cite{Hwa,Wen,Zasa:94}.
Metal dichalcogenides such as MoS$_2$ have also been observed to form
rag-like sheets \cite{Chianelli}.
Finally similar structures occur in the large sheet molecules, shown 
in Fig.\ref{fig__sheet_mol}, believed to be an ingredient in glassy
$B_2O_3$.  

In contrast to crystalline membranes, fluid membranes are
characterized by vanishing shear modulus and dynamical connectivity.
They exhibit significant shape fluctuations controlled by an effective bending
rigidity parameter. 
   
\begin{figure}[htb]
\epsfxsize=3in
\centerline{\epsfbox{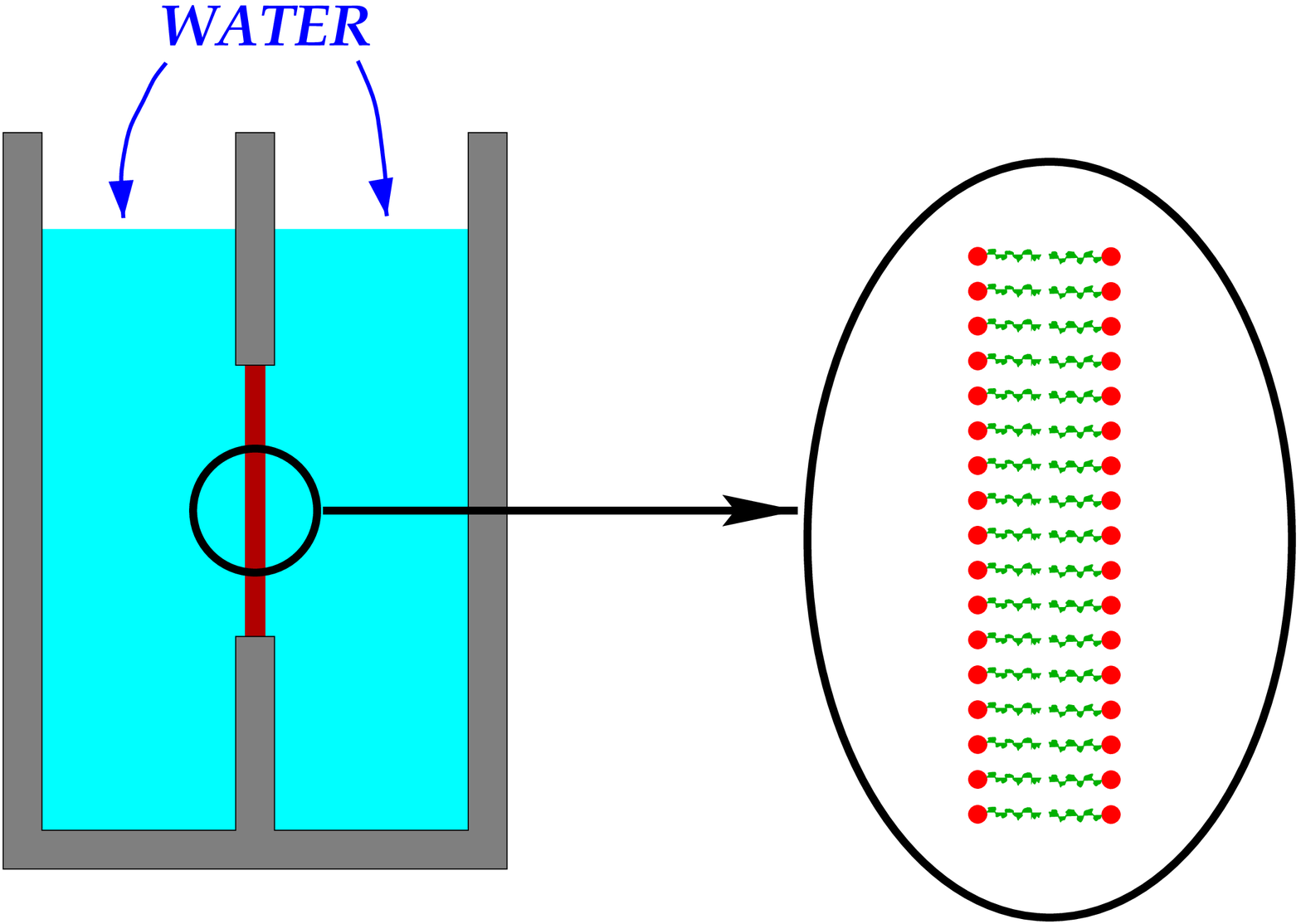}}
\caption{Schematic of experimental procedure to make a {\em black}
membrane.}
\label{fig__black}
\end{figure}

\begin{figure}[htb]
\epsfxsize=2in
\centerline{\epsfbox{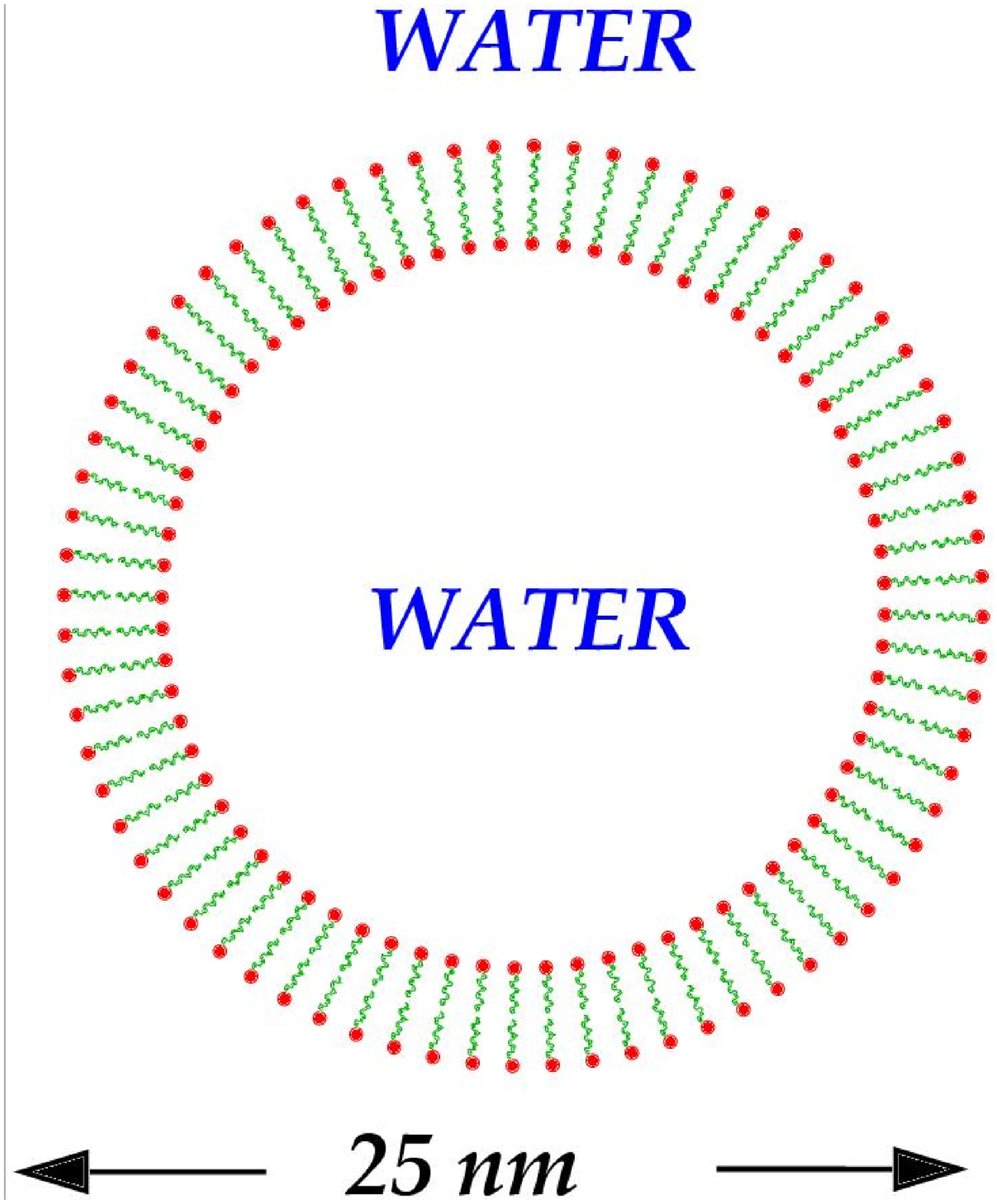}}
\caption{The structure of a liposome with its pure lipid spherical bilayer}
\label{fig__liposome}
\end{figure}

A rich source of physical realizations of fluid membranes is found in 
{\em amphiphilic} systems \cite{Lipowsky,GS:94,Peliti}. Amphiphiles are
molecules with a two-fold character {--} one part is hydrophobic and
another part hydrophilic. The classic examples are lipid molecules,
such as phospholipids, which have polar or ionic head groups
(the hydrophilic component) and hydrocarbon tails (the hydrophobic
component). Such systems are observed to self-assemble into a
bewildering array of ordered structures, such as monolayers, planar
(see Fig.\ref{fig__black}) and spherical bilayers (vesicles or
liposomes) (see Fig.~\ref{fig__liposome}) as well as lamellar, hexagonal and
bicontinuous phases \cite{Gruner}. In each case the basic ingredients are thin
and highly flexible surfaces of amphiphiles. The lipid bilayer of cell
membranes may itself be viewed as a fluid membrane with considerable
disorder in the form of membrane proteins (both peripheral and
integral) and with, generally, an attached crystalline cytoskeleton,
such as the spectrin/actin mesh discussed above. 

A complete understanding of these biological membranes will require a
thorough understanding of each of its components (fluid and
crystalline) followed by the challenging problem of the coupled system
with thermal fluctuations, self-avoidance, potential anisotropy and
disorder. The full system is currently beyond the
scope of analytic and numerical methods but there has been
considerable progress in the last fifteen years.

\begin{figure}[hp]
\epsfxsize=3in
\centerline{\epsfbox{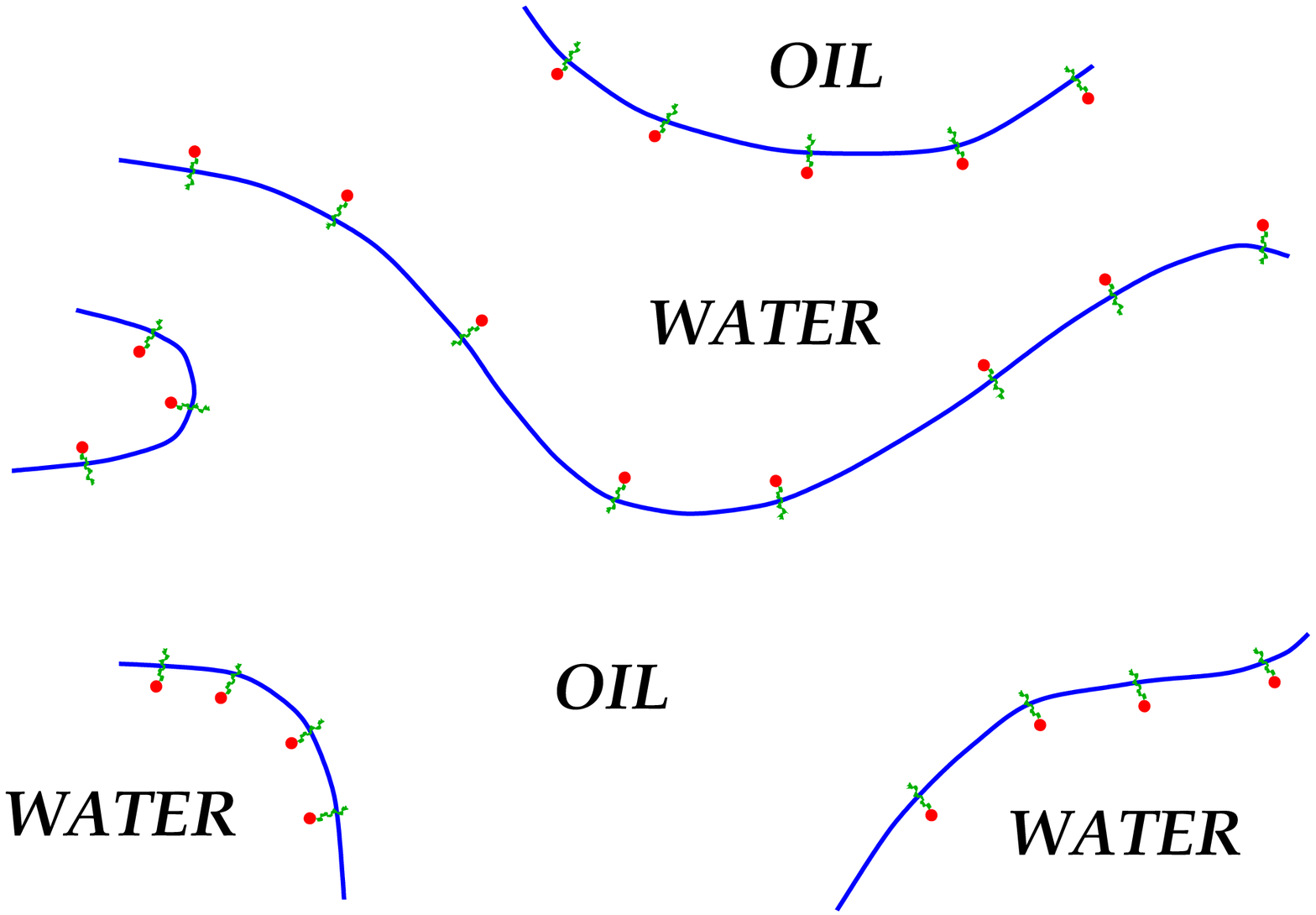}}
\caption{The structure of a microemulsion formed by the addition of
surfactant to an oil-water mixture.}
\label{fig__water_oil}
\epsfxsize=3in
\centerline{\epsfbox{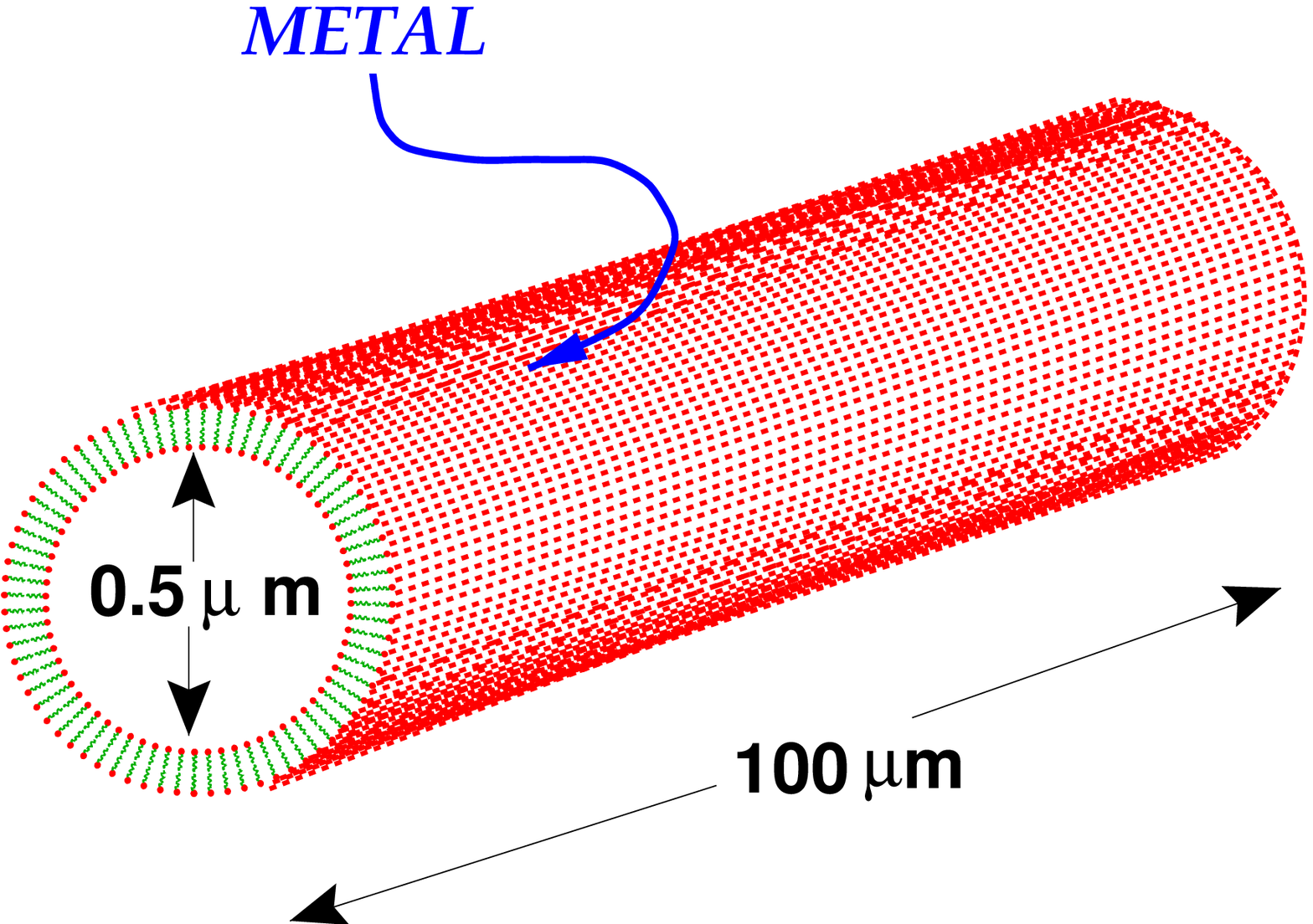}}
\caption{Metal-coated fluid microcylinders (tubules) formed by chiral lipids.}
\label{fig__tubule}
\end{figure}

Related examples of fluid membranes arise when the surface tension 
between two normally immiscible substances, such as oil and water, 
is significantly lowered by the surface action of amphiphiles
(surfactants), which preferentially orient with their polar heads 
in water and their hydrocarbon tails in oil.        
For some range of amphiphile concentration both phases can span the
system, leading to a bicontinuous complex fluid known as a {\em
microemulsion}. The oil-water interface of a microemulsion is a rather 
unruly fluid surface with strong thermal fluctuations \cite{Safran} 
(see Fig.\ref{fig__water_oil}).

\begin{figure}[hp]
\epsfxsize=3in
\centerline{\epsfbox{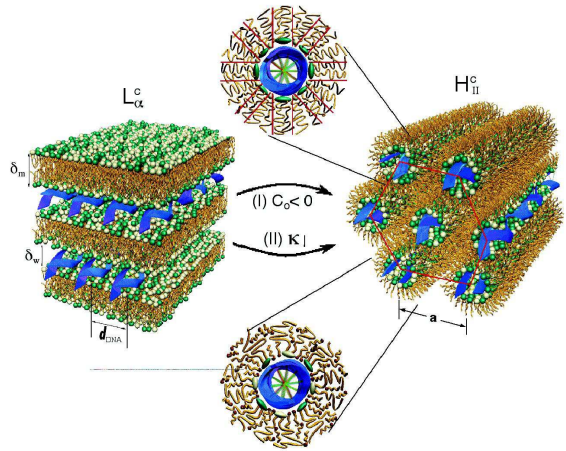}}
\caption{The lamellar and inverted hexagonal DNA-membrane complexes from the work of \cite{Safinya2}}
\label{fig__Cyrusall}
\end{figure}

The structures formed by membrane/polymer complexes are 
of considerable current theoretical, experimental and medical
interest. To be specific it has recently been found that mixtures of 
cationic liposomes (positively charged vesicles) and linear DNA chains
spontaneously self-assemble into a coupled two-dimensional smectic
phase of DNA chains embedded between lamellar lipid 
bilayers \cite{Safinya1,Salditt}.
For the appropriate regime of lipid
concentration the same system can also form an inverted
hexagonal phase with the DNA encapsulated by cylindrical columns of
liposomes\cite{Safinya2} (see Fig.\ref{fig__Cyrusall}). 
In both these structures the liposomes may
act as non-viral carriers (vectors) for DNA with many 
potentially important applications in gene therapy \cite{Crystal:95}. 
Liposomes themselves have long been studied and utilized 
in the pharmaceutical industry as drug carriers \cite{Needham}.
On the materials science side the self-assembling ability of membranes
is being exploited to fabricate microstructures for advanced material
development. One beautiful example is the use of chiral-lipid based
fluid microcylinders (tubules) as a template for metallization.
The resultant hollow metal {\em needles} may be half a micron in
diameter and as much as a millimeter in length
\cite{Schnur:93,SEJSSS:96}, as illustrated in Fig.\ref{fig__tubule}. 
They have potential applications as, for example, cathodes for vacuum
field emission and microvials for controlled release \cite{Schnur:93}.    

\section{The Renormalization Group}\label{SECT__RG}

The Renormalization Group (RG) has provided an extremely general 
framework that has unified whole areas of physics and chemistry \cite{WiKo:74}.
It is beyond the scope of this review to discuss the RG formalism in 
detail but there is an ample literature to which we refer the reader
(see the articles in this issue).
It is the goal of this review to apply the RG framework to the
statistical mechanics of membranes, and for this reason we briefly emphasize and review
some well known aspects of the RG and its related $\vap$-expansion.

The RG formalism elegantly shows that the large distance properties
(or equivalently low $p$-limit) of different models are actually 
governed by the properties of the corresponding Fixed Point (FP). 
In this way one can compute observables in a variety of models,
such as a molecular dynamics simulation or a continuum Landau 
phenomenological approach, and obtain the same
long wavelength result. The main idea is to encode the effects 
of the short-distance degrees of freedom in redefined couplings. 
A practical way to implement such a program 
is the Renormalization Group Transformation (RGT), which provides an explicit 
prescription for integrating out all the 
high $p$-modes of the theory. One obtains the large-distance universal term
of any model by applying a very large ($\infty$ to be rigorous)
number of RGTs.

The previous approach is very general and simple but presents
the technical problem of the proliferation in the number of operators 
generated along the RG flow. There are established techniques to 
control this expansion, one of the most successful ones being the
$\vap$-expansion. The $\vap$-expansion may also performed via 
a field theoretical approach using Feynman diagrams and dimensional 
regularization within a minimal subtraction scheme, which we briefly discuss
below. Whereas it is true that this technique is rather abstract 
and intuitively not very close to the physics of the model, 
we find it computationally much simpler. 

Generally we describe a particular model by several fields 
$\{ \phi,\chi,\cdots \}$ and we construct the Landau free energy by
including all terms compatible with the symmetries and introducing 
new couplings $(u,v,\cdots)$ for each term. The Landau free energy
may be considered in arbitrary dimension $d$, and then, one usually
finds a Gaussian FP (quadratic in the fields) which is infrared stable  
above a critical dimension ($d_U$). Below $d_U$ there are one or
several couplings that define relevant directions. 
One then computes all physical quantities as
a function of $\vap\equiv d_{U}-d$, that is, as perturbations of
the Gaussian theory.

In the field theory approach, we introduce a renormalization constant
for each field $(Z_{\phi},Z_{\chi},\cdots)$ and a renormalization 
constant $(Z_{u},Z_{v},\cdots)$ for each relevant direction below $d_U$.
If the model has symmetries, there are some relations among observables
(Ward identities) and some of these renormalization constants 
may be related. This not only reduces their number but also has the
added bonus of providing cross-checks in practical calculations. 
Within dimensional regularization,
the infinities of the Feynman diagrams appear as poles in $\vap$,
which encode the short-distance details of the model. If we use
these new constants ($Z$'s) to absorb the poles in $\vap$, thereby producing
a complete set of finite Green's functions, we have 
succeeded in carrying out the RG program of including the appropriate short-distance 
information in redefined couplings and fields. This particular
prescription of absorbing only the poles in $\vap$ in the $Z$'s is called the
Minimal Subtraction Scheme (MS), and it considerably
simplifies practical calculations. 

As a concrete example, we consider the theory of a single scalar field
$\phi$ with two independent coupling constants. The one-particle 
irreducible Green's function has the form
\be\label{RG__Green}
\Gamma^{N}_R({\bf k}_i;u_R,v_R,M)=Z^{N/2}_{\phi} 
\Gamma^N({\bf k}_i;u,v; \frac{1}{\vap}) \ ,
\ee
where the function on the left depends on a new parameter $M$, which
is unavoidably introduced in eliminating  the poles in $\vap$. The associated
correlator also depends on redefined couplings $u_R$ and $v_R$. 
The rhs depends on the poles in $\vap$, but its only dependence on $M$ 
arises through $Z_{\phi}$. This observation allows one to write  
\bea\label{RG_CA}
&&M\frac{d}{d M}\left( Z_{\phi}^{-N/2} \Gamma_R^{(N)}\right)=
\\\nonumber
&=&
\left( M \frac{\parp}{\parp M}+\beta(u_R)\frac{\parp}{\parp u_R}
+\beta(v_R)\frac{\parp}{\parp v_R}-\frac{N}{2} \Gamma_{\phi} \right)
\Gamma_R^{(N)}=0 \ ,
\eea
where
\bea\label{define_set_RG}
u_R=M^{-\vap}F(Z_{\phi},Z_{\chi},\cdots|Z_{u}) u & \ , &
v_R=M^{-\vap}F(Z_{\phi},Z_{\chi},\cdots|Z_{v}) v
\\\nonumber
\beta_u(u_R,v_R)=\left.\left(M\frac{\parp u_R}{\parp M}\right)
\right|_{u,v} 
& \ , &
\beta_v(u_R,v_R)=\left.\left(M\frac{\parp v_R}{\parp M}\right)
\right|_{u,v} 
\nonumber\\\nonumber
\gamma_{\phi}&=&\left.\left(M\frac{\parp \ln Z_{\phi} }{\parp M}\right)
\right|_{u,v} \ .
\eea
The $\beta$-functions control the running of the coupling by
\be\label{run__FP}
M\frac{d u_R}{d M}=\beta_u(u_R,v_R) \ , \
M\frac{d v_R}{d M}=\beta_v(u_R,v_R)
\ee
The existence of a FP, at which couplings cease to flow, requires 
$\beta(u_R^{\ast},v_R^{\ast})=0$ for all $\beta$-functions of the model.
Those are the most important aspects of the RG we wanted to review.
In Appendix~\ref{APP_RGstuff} we derive more appropriate expressions
of the RG-functions for practical convenience. For a detailed
exposition of the $\vap$-expansion within the field theory framework
we refer to the excellent book by Amit\cite{Amit:84}.

\section{Crystalline Membranes}\label{SECT__POLYMEM}

\begin{figure}[htb]
  \epsfxsize=4 in \centerline{\epsfbox{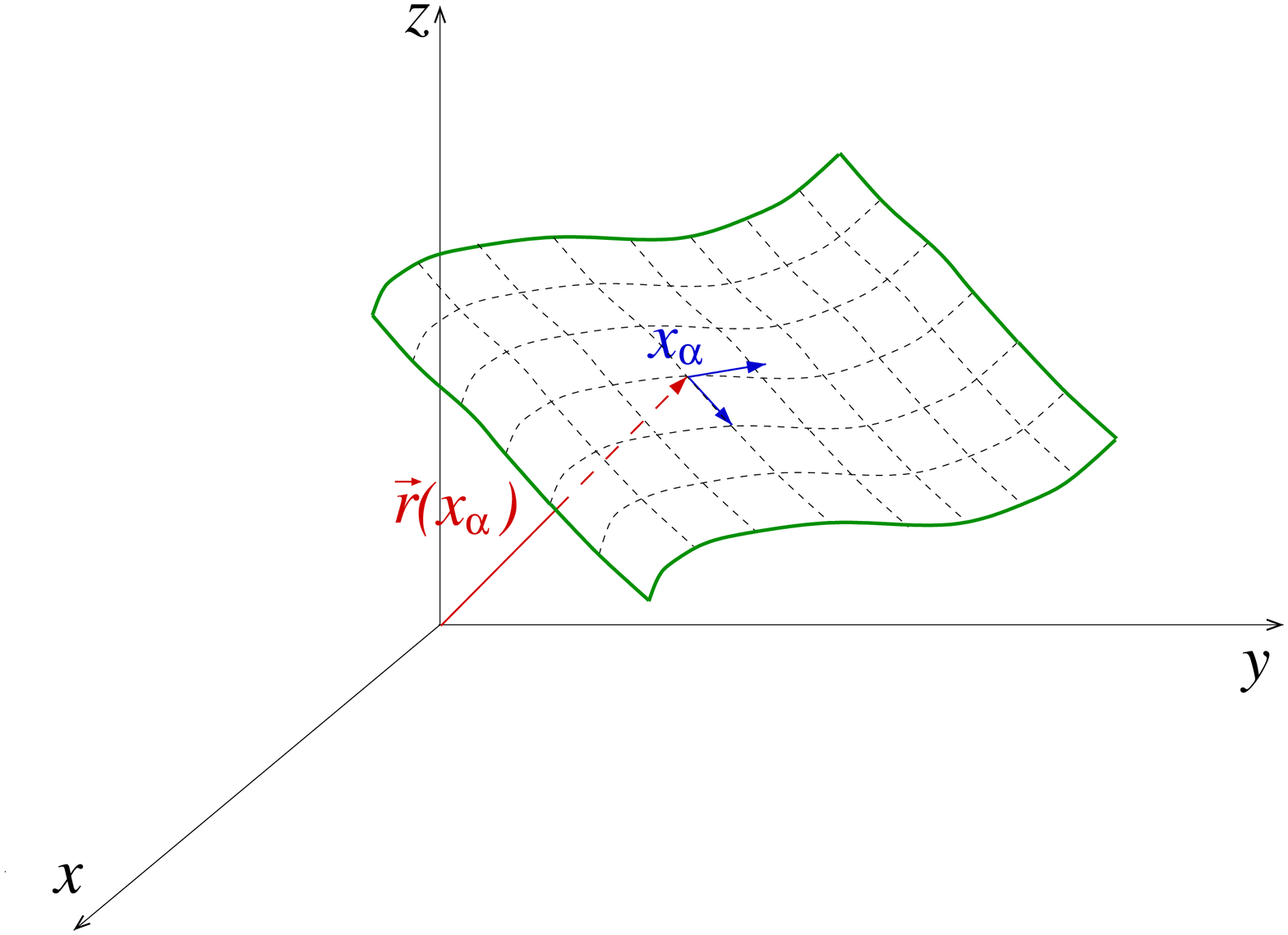}}
  \caption{Representation of a membrane.}
  \label{fig__quadr}
  \end{figure}

A crystalline membrane is a two dimensional fish-net structure with
bonds (links) that never break - the connectivity of the monomers
(nodes) is fixed. It is useful to keep the discussion general and
consider $D$-dimensional objects embedded in $d$-dimensional space.
These are described by a $d$-dimensional vector 
${\vec r}({\bf x})$, with ${\bf x}$ the  $D$-dimensional internal
coordinates, as illustrated in Fig.\ref{fig__quadr}. 
The case $(d=3,D=2)$ corresponds to the physical crystalline membrane. 

To construct the Landau free energy of the model, one must recall that the 
free energy must be invariant under global translations, so the 
order parameter is given by derivatives of the embedding ${\vec r}$, that is
${\vec t}_{\alpha}=\frac{\parp{\vec r}}{\parp u_{\alpha}}$,
with $\alpha=1,\cdots,D$. This latter condition, together with the
invariance under rotations (both in internal and bulk space), 
give a Landau free energy \cite{NP:87,Jer1,PKN:88} 
\bea\label{LAN_CR_VER}
F({\vec r})&=&\int d^D{\bf x} \left[ 
\frac{1}{2}\kappa (\parp_{\alpha}^2 {\vec r})^2+
\frac{t}{2}(\parp_{\alpha} {\vec r})^2+u(\parp_{\alpha} {\vec r} 
\parp_{\beta} {\vec r} )^2+v(\parp_{\alpha} {\vec r} 
\parp^{\alpha} {\vec r})^2 \right]
\nonumber\\
&+&\frac{b}{2}
\int d^D{\bf x}\, d^D{\bf y} \delta^d({\vec r}({\bf x})-
{\vec r}({\bf y})) \ ,
\eea
where higher order terms may be shown to be irrelevant at
long wavelength, as discussed later. The physics in Eq.(\ref{LAN_CR_VER}) 
depends on five parameters,

\begin{itemize}
\item{$\kappa$, \underline{bending rigidity} :} This is the coupling
to the extrinsic curvature (the square of the Gaussian mean
curvature). Since reparametrization invariance is broken for
crystalline membranes, this term may be replaced by its
long-wavelength limit. For large and positive bending 
rigidities flatter surfaces are favored.

\item{$t,u,v$, \underline{elastic constants} :} These coefficients 
encode the microscopic elastic properties of the membrane. In a flat
phase, they may be related to the Lam\'e coefficients of Landau elastic
theory (see sect.~\ref{SUB__subflat}).

\item{$b$, \underline{Excluded volume or self-avoiding coupling} : } 
This is the coupling that imposes an energy penalty for the membrane to
self-intersect. The case $b=0$, i.\ e.\ no self-avoidance, 
corresponds to a {\em phantom} model.

\end{itemize}

\begin{figure}[htb]
  \epsfxsize=5 in \centerline{\epsfbox{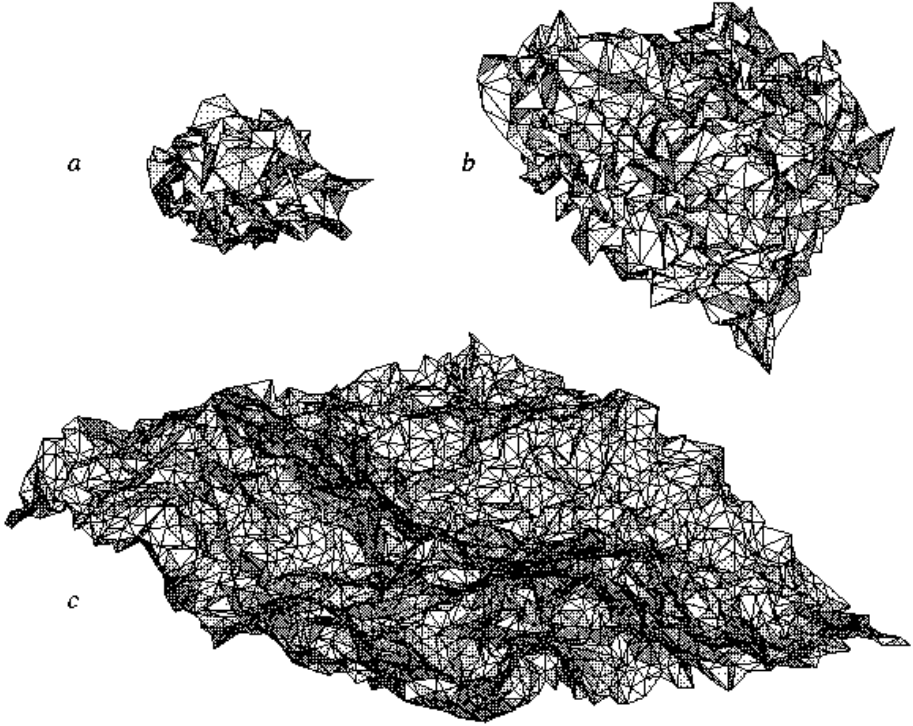}}
  \caption{Examples of a) crumpled phase, b) crumpling transition (crumpled 
   phase) and c) a flat phase. Results correspond to a numerical simulation
   of the phantom case \cite{BCFTA:96} and gives a very intuitive physical picture 
   of the different phases.}
  \label{fig__PHASES}
  \end{figure}

\noindent We generally expand ${\vec r}({\bf x})$ as
\be\label{mf_variable}
{\vec r}({\bf x})=(\zeta {\bf x}+{\bf u}({\bf x}),h({\bf x})) \ ,
\ee
with ${\bf u}$ the $D$-dimensional phonon in-plane modes, and
$h$ the $d-D$ out-of-plane fluctuations. If $\zeta=0$ the model 
is in a rotationally invariant crumpled phase, where the typical 
surfaces have fractal dimension, and there is no real 
distinction between the in-plane phonons and out-of plane modes. 
For a pictorial view, see cases a) and b) in Fig.\ref{fig__PHASES}.

If $\zeta \neq 0$ the membrane is flat up to small fluctuations
and the full rotational symmetry is spontaneously broken. The fields 
$h$ are the analog of the Goldstone bosons and they have different naive 
scaling properties than ${\bf u}$. See Fig.\ref{fig__PHASES} for a
visualization of a typical configuration in the flat phase.

We will begin by studying the phantom case first.
This simplified model may even be relevant to physical systems since
one can envision membranes that self-intersect (at least over some
time scale). 
One can also view the model as a fascinating toy model 
for understanding the more physical self-avoiding case 
to be discussed later. Combined analytical and numerical studies have
yielded a thorough understanding of the phase diagram of phantom
crystalline membranes. 

\subsection{Phantom}\label{SUBSECT__PHAN}

The Phantom case corresponds to setting $b=0$ in the free energy 
Eq.(\ref{LAN_CR_VER}):
\be\label{LAN_CR_PH}
F({\vec r})=\int d^D{\bf x} \left[ 
\frac{1}{2}\kappa (\parp_{\alpha}^2 {\vec r})^2+
\frac{t}{2}(\parp_{\alpha} {\vec r})^2+u(\parp_{\alpha} {\vec r} 
\parp_{\beta} {\vec r} )^2+v(\parp_{\alpha} {\vec r} 
\parp^{\alpha} {\vec r})^2 \right] \ .
\ee

\begin{figure}[htb]
  \epsfxsize=4 in \centerline{\epsfbox{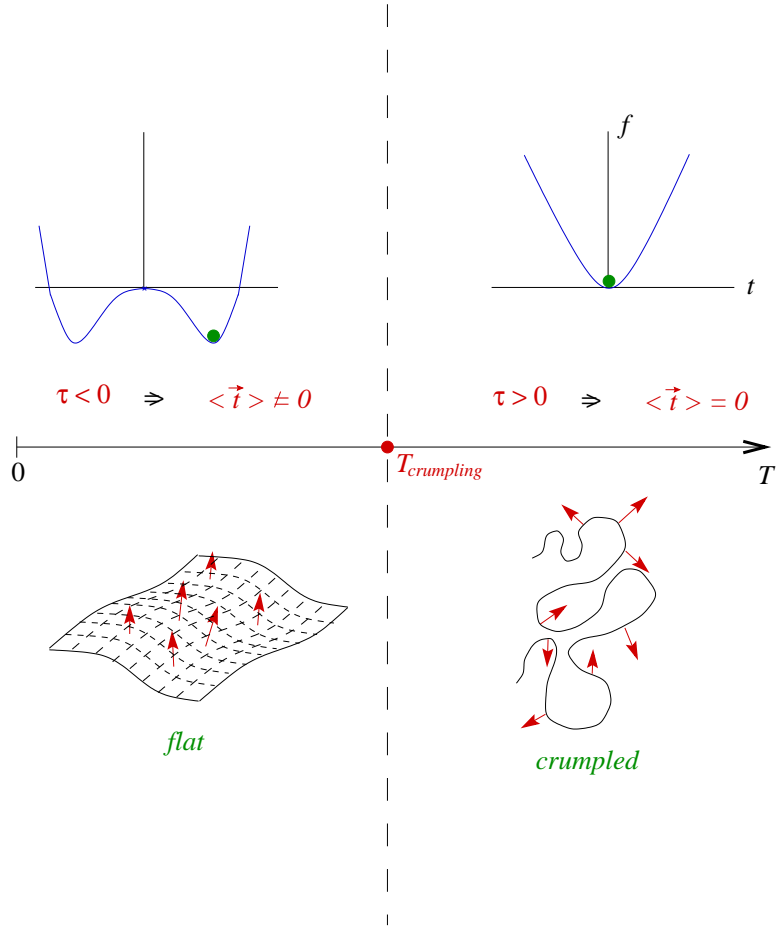}}
  \caption{Mean field solution for crystalline membranes.}
  \label{fig__mfsoln}
  \end{figure}

The mean field effective potential, using the decomposition of 
Eq.(\ref{mf_variable}),  becomes
\be\label{mf_eq_ph}
V(\zeta)=D \zeta^2(\frac{t}{2}+(u+vD)\zeta^2) \ ,
\ee
with solutions
\be\label{mf_zeta}
\zeta^2 = \left\{
             \begin{array}{r@{\quad:\quad}l}
	      0 & t \ge 0 \\
	      -\frac{t}{4(u+vD)} & t < 0 \ .
	      \end{array}
	      \right. 
\ee
There is, consequently, a flat phase for $t  < 0$ and a crumpled phase for 
$t >0$, separated by a crumpling transition at $t=0$ (see Fig.\ref{fig__mfsoln}).

\begin{figure}[htb]
  \epsfxsize=4 in \centerline{\epsfbox{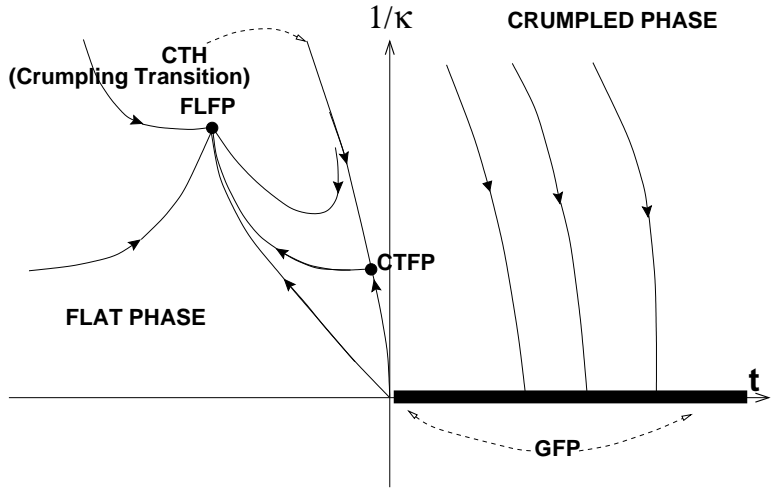}}
  \caption{Schematic plot of the phase diagram for phantom
    membranes. GFP are the equivalent FPs describing the crumpled phase.
    The crumpling transition is described by the Crumpling transition 
    critical line (CTH), which contains the Crumpling Transition FP (CTFP). 
    The Flat phase is described by the (FLFP).}
  \label{fig__PHAN}
  \end{figure}

The actual phase diagram agrees qualitatively with the phase 
diagram of the model shown schematically in Fig.\ref{fig__PHAN}.
The crumpled phase is described by a line of equivalent FPs(GFP). There is 
a general hyper-surface, whose projection onto the $\kappa-t$ plane 
corresponds to a one-dimensional curve (CTH), which corresponds to 
the crumpling transition. Within the CTH there is an infrared stable 
FP (CTFP) which describes the large distance properties 
of the crumpling transition. Finally,
for large enough values of $\kappa$ and negative values of $t$, the system
is in a flat phase described by the corresponding infra-red stable
FP (FLFP) \footnote{The FLFP is actually a line of equivalent fixed
points.}. Although the precise phase diagram turns out to be slightly 
more complicated than the one depicted in Fig.\ref{fig__PHAN}, the 
additional subtleties do not modify the general picture.
 
The evidence for the phase diagram depicted in Fig.\ref{fig__PHAN} 
comes from combining the results of a variety of analytical and
numerical calculations. We present in detail the results obtained
from the $\vap$-expansion since they have wide applicability and allow
a systematic calculation of the $\beta$-function and the critical
exponents. We also describe briefly results obtained with other
approaches.

\subsubsection{The crumpled phase}\label{SUBSUBSECT__isocrphase}

In the crumpled phase, the free energy Eq.(\ref{LAN_CR_PH}) for 
$D \ge 2$ simplifies to
\be\label{LAN_CR_PH_IRR}
F({\vec r})=\frac{t}{2} \int d^D {\bf x} (\parp_{\alpha} {\vec r})^2
+\mbox{Irrelevant terms} \ ,
\ee
since the model is completely equivalent to a linear sigma model in
$D \le 2$ dimensions having $O(d)$ symmetry, and therefore all derivative 
operators in ${\vec r}$ are irrelevant by power counting. The parameter 
$t$ labels equivalent Gaussian FPs, as depicted in 
Fig.\ref{fig__PHAN}. In RG language, it defines a  
completely marginal direction. This is true provided the condition 
$t>0$ is satisfied. The large distance properties of this phase are
described by simple Gaussian FPs and therefore the connected Green's
function may be calculated exactly with result
\be\label{CR_gg}
G({\bf x}) \sim \left\{
             \begin{array}{c  l}
	      |{\bf x}|^{2-D} & D\neq 2 \\
	       \log |{\bf x}| & D=2
	      \end{array}
	      \right. 
\ee
The associated critical exponents may also be computed exactly. 
The Hausdorff dimension $d_H$, or equivalently the size exponent $\nu=D/d_H$, 
is given (for the membrane case $D=2$) by 
\be\label{CR_Haus}
d_H=\infty \  (\nu = 0)  \rightarrow R_G^2 \sim \log L \ .
\ee
The square of the radius of gyration $R_G^2$ scales logarithmically 
with the membrane size $L$. This result is in complete 
agreement with numerical simulations of tethered membranes in 
the crumpled phase where the logarithmic behavior of the radius 
of gyration 
is accurately checked
\cite{KKN:86,KKN:87,BEW:89,ADJ:89,RK:90,HW:91,WS:93,BET:94,W:96,BCFTA:96}.
Reviews may be found in \cite{GK1:97,GK2:97}.

\subsubsection{The Crumpling Transition}

The Free energy is now given by
\be\label{LAN_CRTR_PH}
F({\vec r})=\int d^D{\bf x} \left[ 
\frac{1}{2}(\parp_{\alpha}^2 {\vec r})^2+
u(\parp_{\alpha} {\vec r} \parp_{\beta} {\vec r} )^2+
\hat{v}(\parp_{\alpha} {\vec r} \parp^{\alpha} {\vec r})^2 \right] \ ,
\ee
where the dependence on $\kappa$ may be included in the couplings 
$u$ and $\hat{v}$. With the leading term having two derivatives, the directions 
defined by the couplings $u$ and $\hat{v}$ are relevant by naive power counting 
for $D \le 4$. This shows that the model is 
amenable to an $\vap$-expansion with $\vap=4-D$. For practical purposes,
it is more convenient to consider the coupling $v=\hat{v}+\frac{u}{D}$.
We provide the detailed derivation of the corresponding $\beta$ functions in  
Appendix~\ref{APP__CT__FP}. The result is 
\be\label{beta_FLAT_TR}
\begin{array}{c c r}
\beta_u(u_R,v_R) &=&  -\vap u_R+\frac{1}{8 \pi^2}\left\{
(d/3+65/12)u^2_R+6u_R v_R+4/3 v^2_R \right\} \\
\beta_v(u_R,v_R) &=&  -\vap v_R+\frac{1}{8 \pi^2}\left\{
21/16u^2_R+21/2 u_R v_R+(4d+5) v^2_R \right\} 
\end{array}
\ee
Rather surprisingly, this set of $\beta$ functions does not possess a FP, 
except for $d > 219$. This result would suggest that the crumpling 
transition is first order for $d=3$. Other estimates, however, give 
results which are consistent with the crumpling transition being 
continuous. These are 
\begin{itemize}
\item{Limit of large elastic constants}\cite{DG:88}: 
The Crumpling transition is approached from the flat phase, 
in the limit of infinite elastic constants.
The model is
\be\label{CR_LGD}
H_{NL}=\int d^D\sigma \frac{\kappa}{2} (\Delta {\vec r})^2 \ ,
\ee
with the further constraint 
${\parp_{\alpha} \vec r} {\parp_{\beta} \vec r}= \delta_{\alpha \beta}$.
Remarkably, the $\beta$-function may be computed within a large $d$ expansion,
yielding a continuous crumpling transition with size exponent at 
the transition (for $D=2$)
\be\label{nu_exp}
d_H=\frac{2d}{d-1} \rightarrow \nu=1-\frac{1}{d} \ .
\ee
\item{SCSA Approximation}\cite{LDR:92}: The Schwinger-Dyson equations for the model 
given by Eq.(\ref{LAN_CRTR_PH}) are truncated to include up to four 
point vertices. The result for the Hausdorff dimension and size exponent is
\be\label{CR_SCSA}
  d_H=2.73 \rightarrow \nu=0.732 \ .
\ee

\item{MCRG Calculation} \cite{ET1:96}: The crumpling transition is studied using 
MCRG (Monte Carlo Renormalization Group) techniques. Again, the 
transition is found to be continuous with exponents
\be\label{CR_MCRG}
  d_H=2.64(5) \rightarrow \nu=0.85(9) \ .
\ee
\end{itemize}

Each of these three independent estimates give a continuous crumpling 
transition with a size exponent in the range $\nu \sim 0.7 \pm .15$.
It would be interesting to understand how the $\vap$-expansion must be
performed in order to reconcile it with these results.

  \begin{figure}[htb]
  \epsfxsize=4 in \centerline{\epsfbox{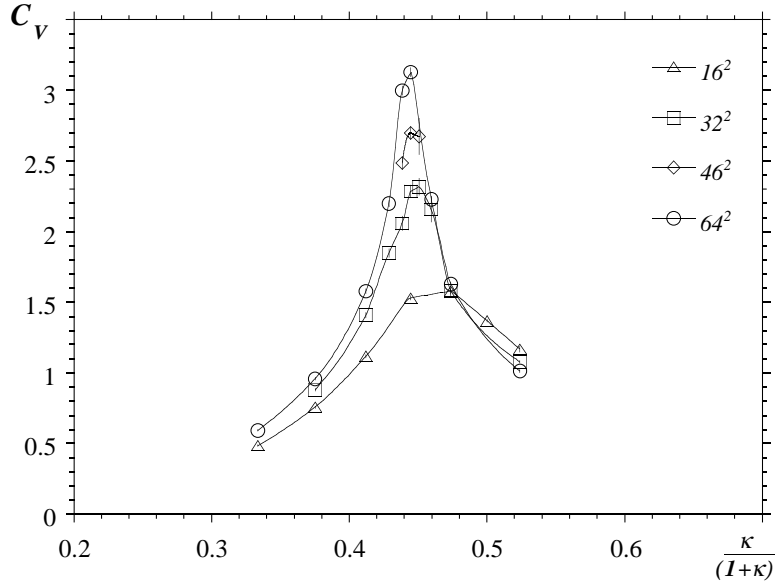}}
  \caption{Plot of the specific heat observable \cite{BCFTA:96}. 
           The growth of the specific heat peak with system size 
	   indicates a continuous transition.}
  \label{fig__CR__cont}
  \end{figure}

Further evidence for the crumpling transition being continuous is 
provided by numerical simulations 
\cite{KKN:86,KKN:87,BEW:89,ADJ:89,RK:90,HW:91,WS:93,BET:94,W:96,BCFTA:96} 
where the analysis of observables
like the specific heat (see Fig.\ref{fig__CR__cont}) or the radius of gyration 
radius give textbook continuous phase transitions, although the precise 
value of the exponents at the transition are difficult to pin down.  
Since this model has also been explored numerically with different 
discretizations on several lattices, there is clear evidence 
for universality of the crumpling transition \cite{BET:94}, again consistent with a 
continuous transition. In Appendix~\ref{APP__Dis} we present more
details of suitable discretizations of the energy for numerical
simulations of membranes. 

\subsubsection{The Flat Phase}\label{SUB__subflat}

\begin{figure}[htb]
  \epsfxsize=4 in \centerline{\epsfbox{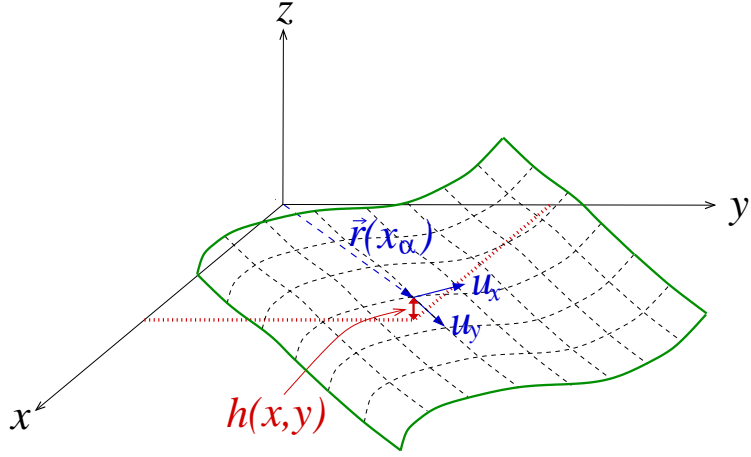}}
  \caption{Coordinates for fluctuations in the flat phase}
  \label{fig__quadr_u}
  \end{figure}

In a flat membrane (see Fig.\ref{fig__quadr_u}), we consider the strain tensor
\be\label{def_strain}
u_{\alpha \beta}= \partial_{\alpha} u_{\beta}+\partial_{\beta} u_{\alpha}
+\partial_{\alpha} h \partial_{\beta} h \ .
\ee
The free energy Eq.(\ref{LAN_CR_PH}) becomes
\be\label{LAN_FL_PH}
F({\bf u},h)=\int d^D {\bf x} \left[
\frac{\hat{\kappa}}{2} (\parp_{\alpha} \parp_{\beta} h )^2+
\mu u_{\alpha \beta} u^{\alpha \beta} + \frac{\lambda}{2} (u^{\alpha}_{\alpha})^2
\right] \ ,
\ee
where we have dropped irrelevant terms. 
One recognizes the standard Landau Free energy of elasticity theory,
with Lam\'e coefficients $\mu$ and $\lambda$, plus an extrinsic
curvature term, with bending rigidity $\hat \kappa$. These 
couplings are related to the original ones
in Eq.(\ref{LAN_CR_PH}) by $\mu=u\zeta^{4-D}$, $\lambda=2 v \zeta^{4-D}$, 
$\hat{\kappa}=\kappa \zeta^{4-D}$ and 
$t=-4(\mu+\frac{D}{2}\lambda)\zeta^{D-2}$.  

The large distance properties of the flat phase for crystalline membranes are
completely described by the Free energy Eq.(\ref{LAN_FL_PH}).
Since the bending rigidity may be scaled out at the crumpling
transition, the free energy becomes a function of $\frac{\mu}{\kappa^2}$ and
$\frac{\lambda}{\kappa^2}$. The $\beta$-function for the couplings $u,v$
at $\kappa=1$ may be calculated within an $\vap$-expansion, which we 
describe in detail in Appendix~\ref{APP__FP__FP}. Let us recall that 
the dependence on $\kappa$ may be trivially restored at any stage. The 
result is
\bea\label{beta_CR_TR}
\beta_{\mu}(\mu_R,\lambda_R)&=&-\vap \mu_R
      +\frac{\mu_R^2}{8 \pi^2}(\frac{d_c}{3}+20A)  
      \\ \nonumber
\beta_{\lambda}(\mu_R,\lambda_R) &=& -\vap  \lambda_R+
      \frac{1}{8 \pi^2}(\frac{d_c}{3} \mu^2_R+2 (d_c+10A) \lambda_R \mu_R
      +2 d_c \lambda^2_R) \ ,
\eea
where $d_c=d-D$, and $A=\frac{\mu_R+\lambda_R}{2 \mu_R+\lambda_R}$.
These $\beta$ functions show four fixed points whose actual 
values are shown in Table~\ref{TAB__FL_EXP}. 

  \begin{figure}[htb]
  \epsfxsize=3 in \centerline{\epsfbox{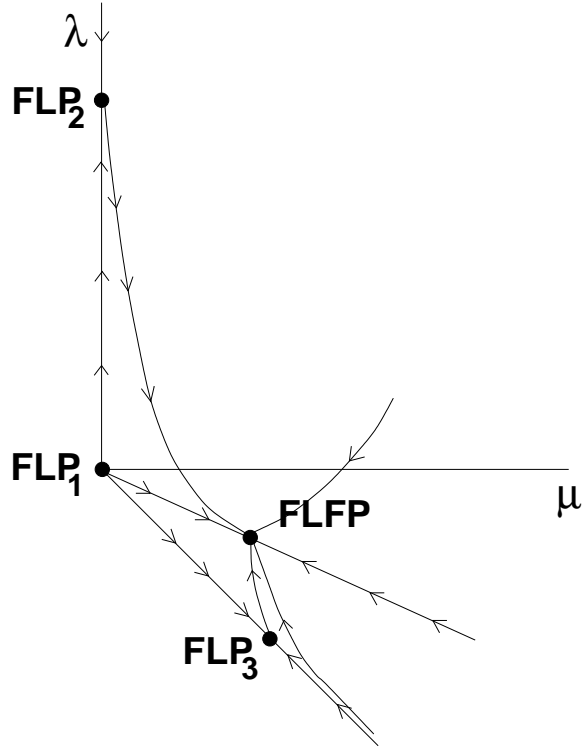}}
  \caption{Phase diagram for the phantom flat phase. There are three
  infra-red unstable FPs, labelled by FLP1, FLP2 and FLP3, but
  the physics of the flat phase is governed by the infra-red stable FP
  (FLFP).}
  \label{fig__PD_FLAT}
  \end{figure}

As apparent from Fig.\ref{fig__PD_FLAT}, the phase diagram of the 
flat phase turns out to be slightly more involved than the one
shown in Fig.\ref{fig__PHAN}, as there are three FPs 
in addition to the FLFP already introduced. These additional FPs
are infra-red unstable, however, and can only be reached for very specific values 
of the Lam\'e coefficients, so for any practical situation we can 
regard the FLFP as the only existing FP in the flat phase.

 \begin{table}[htb]
 \centerline{
 \begin{tabular}{|c|c|c|c|c|}\hline
   FP      &    $\mu^\ast_R$ & $\lambda_R^\ast$ & $\eta$ & $\eta_u$\\\hline
   FP1     &    $0$       &  $0$          & $0$  & $0$             \\\hline 
   FP2     &    $0$       &  $2\vap/d_c$  & $0$  & $0$             \\\hline 
   FP3     &    $\frac{12 \vap}{20+d_c}$  & $\frac{-6 \vap}{20+d_c}$    
           &    $\frac{\vap}{2+d_c/10} $  & $\frac{\vap}{1+20/d_c}$ \\\hline 
   FLFP    &    $\frac{12 \vap}{24+d_c}$  & $\frac{-4 \vap}{24+d_c}$    
           &    $\frac{\vap}{2+d_c/12}$   & $\frac{\vap}{1+24/d_c}$ \\\hline 
  \end{tabular}}
  \caption{The FPs and critical exponents of the flat phase.}
  \label{TAB__FL_EXP}
  \end{table}

\bigskip

\noindent {\bf The properties of the flat phase}
\medskip

The flat phase is a very important phase as will be apparent once we
study the full model, including self-avoidance. For that reason we turn
now to a more detailed study of its most important properties.

Fig.\ref{fig__PHASES} (c) gives an intuitive visualization of
a crystalline membrane in the flat phase. The membrane is essentially a 
flat two dimensional object up to fluctuations in the 
perpendicular direction. The rotational symmetry of the model 
is spontaneously broken, being reduced from $O(d)$ to $O(d-D)\times
O(D)$. The remnant rotational symmetry is realized in 
Eq.(\ref{LAN_FL_PH}) as
\bea\label{sym__trans}
h_i({\bf x}) &\rightarrow & h_i({\bf x})+A^{i \alpha} {\bf x}_{\alpha}
\\\nonumber
u_{\alpha}({\bf x}) &\rightarrow& u_{\alpha} - A^{i \alpha} h_i 
-\frac{1}{2} \delta^{i j} (A^{i \alpha} A^{\beta j} {\bf x}_{\beta})
\ ,
\eea
where $A^{i \alpha}$ is a $D \times (d-D)$ matrix. This relation
is very important as it provides Ward identities which simplify 
enormously the renormalization of the theory.

Let us first study the critical exponents of the model. There are
two key correlators, involving the in-plane and the out-of-plane phonon
modes. Using the RG equations, it is easy to realize that at any
given FP, the low-$p$ limit of the model is given by
\bea\label{low_q_mode}
\Gamma_{u u}({\vec p}) &\sim & |{\vec p}|^{2+\eta_u}
\\\nonumber
\Gamma_{h h}({\vec p}) &\equiv& |{\vec p}|^4 \kappa({\vec p}) \sim
|{\vec p}|^{4-\eta} \ ,
\eea
where the last equation defines the anomalous elasticity 
$\kappa({\vec p})$ as a function of momenta ${\vec p}$. These two
exponents are not independent, since they satisfy the scaling 
relation \cite{AL:88} 
\be\label{FLAT__scaling}
\eta_u=4-D-2\eta \ ,
\ee
which follows from the Ward identities associated with the remnant
rotational symmetry (Eqn.(\ref{sym__trans}). Another important exponent is the 
roughness exponent $\zeta$, which measures the fluctuations transverse to
the flat directions. It can be expressed as $\zeta=\frac{4-D-\eta}{2}$.

The long wavelength properties of the flat phase are described by the 
FLFP (see Fig.\ref{fig__PD_FLAT}). Since the FLFP occurs at non-zero renormalized
values of the Lam\'e coefficients, the associated critical exponents
discussed earlier are clearly non-Gaussian. Within an $\vap$-expansion, the values for
the critical exponents are given in Table~\ref{TAB__FL_EXP}. There are
alternative estimates available from different methods. These are
\begin{itemize}
\item{Numerical Simulation:} In \cite{BCFTA:96} a large scale simulation of
the model was performed using very large meshes. The results obtained for the 
critical exponents are very reliable, namely 
\be\label{Sim__exponents}
             \begin{array}{l l  l}
	       \eta_u=0.50(1) & \eta=0.750(5) & \zeta=0.64(2) 
	      \end{array}
\ee
For a review of numerical results see \cite{GK1:97,GK2:97}.
\item{SCSA Approximation:} This consists of suitably truncating the 
Schwinger-Dyson equations to include up to four-point correlation
functions \cite{LDR:92}. The result for general $d$ is
\be\label{SCSA_flat}
\eta(d)=\frac{4}{d_c+(16-2d_c+d^2_c)^{1/2}} \ ,
\ee
which for $d=3$ gives
\be\label{SCSA_flat_d3}
  \begin{array}{l l  l}
   \eta_u=0.358 & \eta=0.821 & \zeta=0.59 
  \end{array}
\ee
\item{Large d expansion:} The result is \cite{DG:88}
\be\label{Larged_flat}
\eta=\frac{2}{d} \rightarrow \eta(3)=2/3
\ee
\end{itemize}
We regard the results of the numerical simulation as our most accurate 
estimates, since we can estimate the errors. The results obtained from the
SCSA, which are the best analytical estimate, are in acceptable agreement
with simulations.

Finally there are two experimental measurements 
of critical exponents for the flat phase of crystalline membranes.
The static structure factor of the red blood cell cytoskeleton 
(see Sect.\ref{SECT__Intro}) has been measured by small-angle x-ray
and light scattering, yielding a roughness exponent of
$\zeta=0.65(10)$ \cite{Skel:93}. Freeze-fracture electron microscopy
and static light scattering of the conformations of graphitic oxide
sheets (Sect.\ref{SECT__Intro}) revealed flat sheets with a fractal
dimension $d_H=2.15(6)$. Both these values are in good agreement with
the best analytic and numerical predictions, but the errors are still too 
large to discriminate between different analytic calculations.  

The Poisson ratio of a crystalline membrane (measuring the transverse 
elongation due to a longitudinal stress \cite{Landau7}) is universal and within the
SCSA approximation, which we regard as the more accurate analytical estimate,
is given by
\be\label{FL_PR}
\sigma(D)=-\frac{1}{D+1} \rightarrow \sigma(2)=-1/3 \ ,
\ee
This result has been accurately checked in numerical simulations 
\cite{FBGT:97}. Rather remarkably, it turns out to be negative. 
Materials having a negative Poisson ratio are called {\em auxetic}. 
This highlights potential applications of crystalline membranes to 
materials science since auxetic materials have a wide variety 
of potential applications as gaskets, seals etc.

Finally another critical regime of a flat membrane is achieved by
subjecting the membrane to external tension \cite{GDLP:88,GDLP:89}.
This allows a low temperature phase in which the membrane has a domain
structure, with distinct domains corresponding to flat phases with 
different bulk orientations. This describes, physically, a {\em
buckled} membrane whose equilibrium shape is no longer planar.

\subsection{Self-avoiding}\label{Sub_SECT__SA} 

Self-avoidance is a necessary interaction in any realistic description
of a crystalline membrane. It is introduced in the form of a delta-function
repulsion in the full model Eq.(\ref{LAN_CR_VER}). We have already analyzed the 
phantom case and explored in detail the distinct phases. 
The question before us now is the effect of self-avoidance on each
of these phases.

The first phase we analyze is the flat phase. Since self-intersections
are unlikely in this phase, it is intuitively clear that self-avoidance should
be irrelevant. This may also be seen if one neglects 
the effects of the in-plane phonons. In the self-avoiding term for the flat phase
we have
\bea\label{SA__term__sim}
&&\frac{b}{2}
\int d^D{\bf x}\, d^D{\bf y} \delta^d({\vec r}({\bf x})-
{\vec r}({\bf y}))
\\\nonumber
&=&\frac{b}{2}\int d^D{\bf x}\, d^D{\bf y} 
\delta^{D}(\zeta({\bf x}-{\bf y})+{\bf u}({\bf x})-{\bf u}({\bf y}) )
\delta^{d-D}(h({\bf x})-h({\bf y})) 
\\\nonumber
&\sim&\frac{b}{2}\int d^D{\bf x}\, d^D{\bf y} 
\delta^{D}(\zeta({\bf x}-{\bf y}))
\delta^{d-D}(h({\bf x})-h({\bf y}))=0 \ ,
\eea
as the trivial contribution where the membrane equals itself is 
eliminated by regularization. The previous argument receives 
additional support from numerical simulations in the
flat phase, where it is found that self-intersections are extremely rare
in the typical configurations appearing in those simulations.  It seems 
clear that self-avoidance is most likely an irrelevant operator, in the 
RG sense, of the FLFP. Nevertheless, it would be very interesting if one
could provide a more rigorous analytical proof for this statement.
A rough argument can be made as follows.  
Shortly we will see that the Flory approximation for self-avoiding
membranes predicts a fractal dimension $d_H=2.5$. For bulk dimension
$d$ exceeding 2.5 therefore we expect self-avoidance to be irrelevant.
A rigorous proof of this sort remains rather elusive, as it involves
the incorporation of both self-avoidance and non-linear elasticity, and this remains 
a difficult open problem.

The addition of self-avoidance in the crumpled phase consists of 
adding the self-avoiding interaction to the free energy 
of Eq.(\ref{LAN_CR_PH_IRR})  
\be\label{CRU__SA}
F({\vec r})=\frac{1}{2}\int d^D {\bf x} (\parp_{\alpha} {\vec r}({\bf x}))^2
+\frac{b}{2}\int d^D{\bf x}\, d^D{\bf y} \delta^d({\vec r}({\bf x})-
{\vec r}({\bf y})) \ ,
\ee
which becomes the natural generalization of the Edwards' model for polymers to 
$D$-dimensional objects. Standard power counting shows that the 
GFP of the crumpled phase is infra-red unstable 
to the self-avoiding perturbation for
\be\label{SA__eps}
\vap(D,d)\equiv 2D-d\frac{2-D}{2} > 0 \ ,
\ee
which implies that self-avoidance is a relevant perturbation for $D=2$-objects
at any embedding dimension $d$. The previous remarks make it apparent that it
is possible to perform an $\vap$-expansion of the model 
\cite{Dup:87,AL:87,KN3:87,KN1:88}.
In Appendix~\ref{APP__SA__CP}, we present the calculation of 
the $\beta$-function at lowest order in $\vap$ using 
the MOPE (Multi-local-operator-product-expansion) formalism \cite{DDG:94,DDG:97}.
The MOPE formalism has the advantage that it is more easily 
generalizable to higher orders in $\vap$, and enables concrete proofs showing
that the expansion may be carried out to all orders. At lowest order,
the result for the $\beta$-function is
\bea\label{SA__CR__B}
\beta_b(b_R)&=&-\vap b_R+\frac{(2-D)^{-1+\frac{d}{2}}}{(4\pi)^{\frac{d}{2}}}
\left(\frac{2 \pi^{\frac{D}{2}}}{\Gamma(D/2)}\right)^{2+\frac{d}{2}}\left[
\frac{\Gamma(\frac{D}{2-D})^2}{\Gamma(\frac{2D}{2-D})}+
\frac{d}{2}\frac{(2-D)^2}{2D}\right]\frac{b^2_R}{2}
\nonumber\\
&\equiv&-\vap b_R+a_1 b_R^2 \ .
\eea
The infra-red stable FP is given at lowest order in $\vap$ by
$b^{\ast}_R=\frac{\vap}{a_1}$, which clearly shows that the GFP of the 
crumpled phase is infra-red unstable in the presence of self-avoidance.

\begin{figure}[htb]
  \epsfxsize=4 in \centerline{\epsfbox{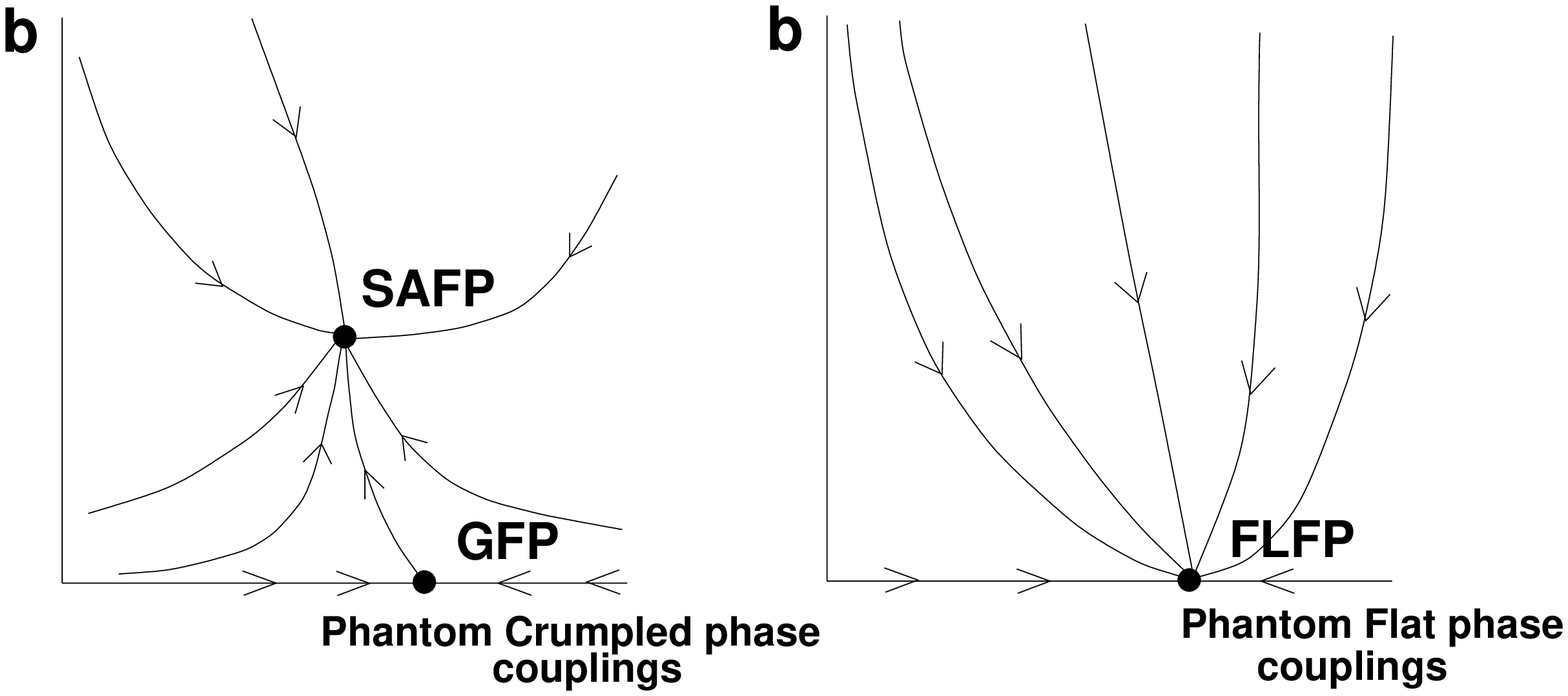}}
  \caption{The addition of Self-avoidance at the Crumpled and Flat 
   phases.}
  \label{fig__CR_SA}
  \end{figure}

The preceding results are shown in Fig.\ref{fig__CR_SA} and may be
summarized as
\begin{itemize}
\item{The flat phase of self-avoiding crystalline membranes is exactly the 
same as the flat phase of phantom crystalline tethered membranes.}
\item{The crumpled phase of crystalline membranes is destabilized by the 
presence of any amount of self-avoidance.}
\end{itemize}

The next issue to elucidate is whether this new SAFP describes a 
crumpled self-avoiding phase or a flat one and to give a more quantitative
description of the critical exponents describing the universality class.
Supposing that the SAFP is, in fact, flat we must understand its
relation to the FLFP describing the physics of the flat phase 
and the putative phase transitions between these two.

\subsubsection{The nature of the SAFP}

\begin{figure}[htb]
  \epsfxsize=3 in \centerline{\epsfbox{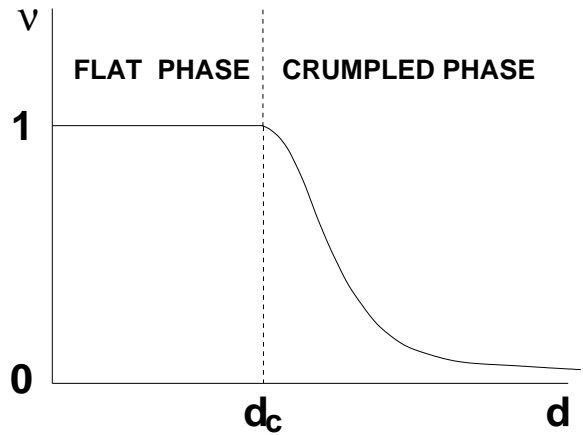}}
  \caption{The size exponent as a function of $d$. 
	   There is a critical dimension $d_c$ below which the
           crumpled phase disappears.}
  \label{fig__SA_scenario}
  \end{figure}

Let us study in more detail the model described in Eq.(\ref{CRU__SA}). The
key issue is whether this model still admits a crumpled phase, and if
so to determine the associated size exponent. 
On general grounds we expect that there is a critical dimension
$d_c$, below which there is no crumpled phase (see Fig.\ref{fig__SA_scenario}).

An estimate for the critical dimension may be obtained from a Flory
approximation in which minimizes the free energy obtained by replacing 
both the elastic and self-avoiding terms with the radius of gyration 
raised to the power of the appropriate scaling dimensions. Within 
the Flory treatment a $D$-dimensional membrane is in a crumpled phase, 
with a size exponent given by
\be\label{Flory_est}
\nu=(D+2)/(d+2) \ .
\ee
From this it follows that $d_c=D$ (see Fig.\ref{fig__SA_scenario}).
The Flory approximation, though very accurate for polymers ($D=1$), remains 
an uncontrolled approximation. 

In contrast the $\vap$-expansion provides a systematic determination
of the critical exponents. For the case of membranes, however, some extrapolation
is required, as the upper critical dimension is infinite.
This was done in \cite{Hwa:90}, where it is shown that reconsidering 
the $\vap$-expansion as a double expansion in $\vap$ and $D$, critical 
quantities may be extrapolated for $D=2$-dimensional objects. At lowest
order in $\vap$, the membrane is in a crumpled phase.  The enormous task 
of calculating the next correction ($\vap^2$) was successfully carried 
out in \cite{DaWi:96}, employing more elaborate extrapolation methods than 
those in \cite{Hwa:90}. Within this calculation, the $d=3$ membrane is 
still in a crumpled phase, but with a size exponent now closer to 1. It 
cannot be ruled out that the $\vap$-expansion, successfully carried out to 
all orders could give a flat phase $\nu=1$. In fact, 
the authors in \cite{DaWi:96,WiDa:97} present some arguments in favor of a 
scenario of this type, with a critical dimension $d_c\sim 4$.

Other approaches have been developed with different results. A Gaussian 
approximation was developed in \cite{Gou:91,LeD:92}. The size exponent of 
a self-avoiding membrane within this approach is
\be\label{Gauss_est}
\nu=4/d ,
\ee
and since one has $\nu > 1$ for $d \leq 4$, one may conclude that the
membrane is flat for $d \leq d_c=4$. Since we cannot determine the accuracy of 
the Gaussian approximation this estimate must be viewed largely as interesting 
speculation. Slightly more elaborate arguments of this type \cite{GuPa:92} 
yield an estimated critical dimension $d_c=3$.

\vskip 0.5cm
\noindent {\bf Numerical simulations}
\vskip 0.5cm

We have seen that numerical simulations provide good support for
analytic results in the case of phantom membranes. 
When self-avoidance is included, numerical simulations become
invaluable, since analytic results are harder to come by.
It is for this reason that we discuss them in greater detail than in 
previous sections.

A possible discretization of membranes with excluded volume
effects consists of a network of $N$ particles arranged in a triangular
array. Nearest neighbors interact with a potential
\be\label{tether_pot}
V_{NN}({\vec r})=\left\{ \begin{array}{c c} 
          0 & \mbox{for $|{\vec r}|< b$ } \\
	  \infty & \mbox{for $|{\vec r}| > b $}
	                 \end{array} \right.
			 \ ,
\ee
although some authors prefer a smoothened version, with the same 
general features. The quantity $b$ is of the order of a few lattice 
spacings. This is a lattice version of the elastic term in
Eq.(\ref{CRU__SA}). The discretization of the self-avoidance 
is introduced as a repulsive hard sphere potential, now acting 
between any two atoms in the membrane, instead 
of only nearest neighbors. A hard sphere repulsive potential is, for example,
\be\label{exc_potential}
V_{Exc}({\vec r})=\left\{ \begin{array}{c c} 
          \infty & \mbox{for $|{\vec r}|< \sigma $} \\
	  0 & \mbox{ for $|{\vec r}| > \sigma $ }
	                 \end{array} \right.
			 \ ,
\ee
where $\sigma$ is the range of the potential, and $\sigma < b$. Again, some
smoothened versions, continuous at $|{\vec r}|=\sigma$, have also been
considered. This model may be pictured as springs, defined by the 
nearest-neighbor potential Eq.(\ref{tether_pot}), with excluded volume 
effects enforced by balls of radius $\sigma$
(Eq.(\ref{exc_potential})). This model represents a lattice discretization of
Eq.(\ref{CRU__SA}).
 
Early simulations of this type of model \cite{KKN:86,KKN:87} provided a first estimate 
of the size exponent at $d=3$ fully compatible with the Flory estimate
Eq.(\ref{Flory_est}). The lattices examined were not very large,
however, and subsequent simulations with larger volumes \cite{PB:88,ARP:89} found that 
the $d=3$ membrane is actually flat. This result is even more
remarkable if one recalls that there is no explicit bending 
rigidity.

  \begin{figure}[htb]
  \epsfxsize=3 in \centerline{\epsfbox{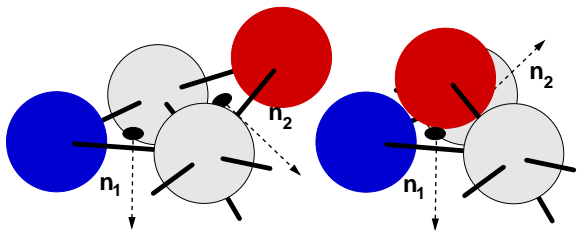}}
  \caption{Visualization of bending rigidity generated by a hard sphere potential.
  Normals ${\vec n}_1$ and ${\vec n}_2$ cannot be anti-parallel.}
  \label{fig__SA__ext}
  \end{figure}

The flat phase was a very surprising result, in some conflict with 
the insight provided from the analytical estimates discussed in the previous
subsection. An explanation for it came from the observation \cite{AN1:90}
that excluded volume effects induce bending rigidity, as depicted in
Fig.\ref{fig__SA__ext}. The reason is that the excluded volume effects
generate a non-zero expectation value for the bending rigidity, since the
normals can be parallel, but not anti-parallel
(see~\ref{fig__SA__ext}). 
This induced bending rigidity was estimated and found to be big enough 
to drive the self-avoiding membrane well within the flat phase of 
the phantom one. This means that this particular discretization of the
model renders any potential SAFP inaccessible and the physics is
described by the FLFP. In \cite{AN2:90} the structure function of the
self-avoiding model is numerically computed and found to compare well 
with the analytical structure function for the flat phase of phantom 
crystalline membranes, including comparable roughness exponents.

The natural question then to ask is whether it is possible to reduce
the bending rigidity sufficiently to produce a crumpled self-avoiding
phase. Subsequent studies addressed this issue in various ways.
The most natural way is obviously to reduce the range of the potential 
sufficiently that the induced bending rigidity is within the crumpled
phase. This is the approach followed in \cite{BLLP:89}. 
The flat phase was found to persist to very small 
values of $\sigma$, with eventual signs of a crumpled phase. 
This crumpled phase may essentially be due to the elimination at
self-avoidance at sufficiently small $\sigma$. A more comprehensive
study, in which the same limit is performed this time with an excluded volume potential 
which is a function of the internal distance along the lattice \cite{KK:93},
concluded that for large membranes, inclusion of excluded volume effects, no
matter how small, leads to flatness. A different approach to weakening the
flat phase, bond dilution \cite{GM:90}, found that the flat phase persists 
until the percolation critical point. In conclusion the bulk of
accumulated evidence indicates that flatness is an intrinsic
consequence of self-avoidance. If this is indeed correct the
SAFP coincides with FLFP and this feature is an inherent consequence
of self-avoidance, rather than an artifact of discretization.   

Given the difficulties of finding a crumpled phase with a repulsive potential, 
simulations for larger values of the embedding space dimension $d$ have also
been performed \cite{Grest:91,BP:94}. These simulations show clear evidence that
the membrane remains flat for $d=3$ and $4$ and undergoes a crumpling transition
for $d \geq 5$, implying $d_c \geq 4$.

An alternative approach to incorporating excluded volume effects corresponds
to discretize a surface with a triangular lattice and imposing the self-avoidance 
constraint by preventing the triangular plaquettes from inter-penetrating. 
This model has the advantage that is extremely flexible, since there 
is no restriction on the bending angle of adjacent plaquettes
(triangles) and therefore no induced bending rigidity (see \ref{APP__Dis}).

The first simulations of the plaquette model \cite{B:91} found a size exponent
in agreement with the Flory estimate Eq.~\ref{Flory_est}. A subsequent 
simulation \cite{KG:93}  disproved this result, and found a size 
exponent $\nu=0.87$, higher than the Flory estimate, but below one. 
More recent results using larger lattices and more sophisticated
algorithms seem to agree completely with the results obtained from the
ball and spring models \cite{BCTT:2000}.

Further insight into the lack of a crumpled phase for self-avoiding
crystalline membranes is found in the study of folding 
\cite{DiFGu1:94,DiFGu2:94,BDGG:95,BDGG:96,BGGM:97,CGP1:96,CGP2:96}.
This corresponds to the limit of infinite elastic constants studied
by David and Guitter with the further approximation that the space of
bending angles is discretized. One quickly discovers that the
reflection symmetries of the allowed
folding vertices forbid local folding (crumpling) of surfaces.
There is therefore essentially no entropy for crumpling. There is,
however, local unfolding and the resulting statistical mechanical
models are non-trivial. The lack of local folding is the discrete 
equivalent of the long-range curvature-curvature interactions that
stabilize the flat phase. The dual effect of the integrity of the
surface (time-independent connectivity) and self-avoidance is so
powerful that crumpling seems to be impossible in low embedding dimensions.  
   
\subsubsection{Attractive potentials}

Self-avoidance, as introduced in Eq.(\ref{exc_potential}) is a totally 
repulsive force among monomers. There is the interesting possibility of
allowing for attractive potentials also. This was pioneered in  
\cite{AN1:90} as a way to escape to the induced bending rigidity 
argument (see Fig.\ref{fig__SA__ext}), since an attractive potential 
would correspond to a negligible (or rather a negative) bending 
rigidity. Remarkably, in \cite{AN1:90}, a compact (more crumpled) 
self-avoiding phase was found, with fractal dimension close to 3.

This was further studied in \cite{AK:91}, where it was found that with 
an attractive Van der Waals potential, the 
crystalline membrane underwent a sequence of folding transitions leading to a
crumpled phase. In \cite{LP:92} similar results were found, but instead of
a sequence of folding transitions a crumpled phase was found with an
additional compact (more crumpled phase) at even lower temperatures. 
Subsequent work gave some support to this scenario \cite{GP:94}.

On the analytical side, the nature of the $\Theta$-point for membranes
and its relevance to the issue of attractive interactions 
has been addressed in \cite{DaWi:95}.

We think that the study of a tether with an attractive potential remains
an open question begging for new insights.
A thorough understanding of the nature of the compact phases produced by
attractive interactions would be of great value.

\subsubsection{The properties of the SAFP}

\begin{figure}[htb]
\epsfxsize=3in
\centerline{\epsfbox{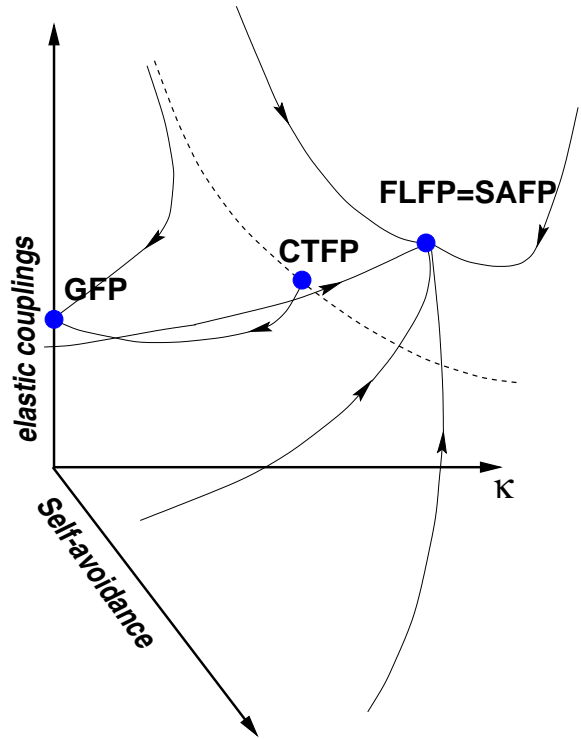}}
\caption{The conjectured phase diagram for self-avoiding crystalline
membranes in $d=3$. With any degree of self-avoidance the flows are to
the flat phase fixed point of the phantom model (FLFP).}
\label{fig__SApd}
\end{figure}

The enormous efforts dedicated to study the SAFP have not resulted in a
complete clarification of the overall scenario since the existing 
analytical tools do not provide a clear picture. Numerical results
clearly provide the best insight. For the physically relevant case 
$d=3$, the most plausible situation is that there is no crumpled phase 
and that the flat phase is identical to the flat phase of the phantom
model. For example, the roughness exponents
$\zeta_{SA}$ from numerical simulations of self-avoidance at $d=3$ using
ball-and-spring models \cite{Grest:91} and the roughness exponent at the FLFP, 
Eq.(\ref{Sim__exponents}), compare extremely well
\be\label{COMP_SA_FL}
\zeta_{SA}=0.64(4) \ , \ \zeta=0.64(2) \ ,
\ee
So the numerical evidence allows us to conjecture that the SAFP is exactly
the same as the FLFP, and that the crumpled self-avoiding phase is
absent in the presence of purely repulsive potentials (see Fig.\ref{fig__SApd}). 
This identification of fixed points enhances the significance of the
FLFP treated earlier. 
It would be very helpful if analytical tools were developed to 
further substantiate this statement.

\section{Anisotropic Membranes}\label{SECT__POLYMEM_ANI}

An anisotropic membrane is a crystalline membrane having the property 
that the elastic or the bending rigidity properties in one 
distinguished direction are different from those in the $D-1$
remaining directions. As for the isotropic case we keep the 
discussion general and describe the membrane
by a $d$-dimensional ${\vec r}({\bf x}_{\perp},y)$, where now the
$D$ dimensional coordinates are split into $D-1 \ \ {\bf x}_{\perp}$
coordinates and the orthogonal distinguished direction $y$.

The construction of the Landau free energy follows the same steps
as in the isotropic case. Imposing translational invariance,
$O(d)$ rotations in the embedding space and $O(D-1)$
rotations in internal space, the equivalent of Eq.(\ref{LAN_CR_VER})
is now
\bea\label{LGW}
F(\vec r({\bf x}))&=& \frac{1}{2} \int d^{D-1}{\bf x}_{\perp} dy \left[
\kappa_{\perp}(\partial_{\perp}^2 \vec r)^2 + \kappa_y (\pary^2 \vec
r)^2  \right.
\nonumber\\
&& + \kappa_{\perp y} \pary^2 \vec r \cdot \parp_{\perp}^2 \vec r +
t_{\perp}(\parp_{\alpha}^{\perp} \vec r)^2 + t_y(\pary \vec r)^2
\nonumber\\
&& + \frac{u_{\perp \perp}}{2}(\parp_\alpha^{\perp} \vec r \cdot
\parp_{\beta}^{\perp} \vec r)^2 + \frac{u_{yy}}{2}(\pary \vec r \cdot
\pary \vec r)^2
\nonumber\\
&& + u_{\perp y} (\parp_{\alpha}^{\perp} \vec r \cdot \pary \vec r)^2
+ \frac{v_{\perp \perp}}{2}(\parp_{\alpha}^{\perp} \vec r \cdot
\parp_{\alpha}^{\perp} \vec r)^2
\nonumber\\
&& \left.  + v_{\perp y}(\parp_{\alpha}^{\perp} \vec r)^2
(\pary \vec r)^2 \right]
\nonumber\\
&&+ \frac{b}{2} \int d^D {\bf x} \int d^D {\bf x}^\prime \delta^d
(\vec r({\bf x}) - \vec r({\bf x}^\prime)).
\eea
This model has eleven parameters, representing distinct physical 
interactions:
\begin{itemize}
\item{ $ \kappa_{\perp},\kappa_y,\kappa_{\perp y} $ 
\underline{bending rigidity}:} the anisotropic versions of 
the isotropic bending rigidity splits into three distinct terms.

\item{ $t_{\perp},t_y, u_{\perp \perp}, u_{yy}, v_{\perp \perp}, v_{\perp y}$
\underline{elastic constants}:} there are six quantities describing the 
microscopic elastic properties of the anisotropic membrane.

\item{ $b$, \underline{self-avoidance coupling}:} This particular
term is identical to its isotropic counterpart. 
\end{itemize}

\begin{figure}[htb]
  \epsfxsize=4 in \centerline{\epsfbox{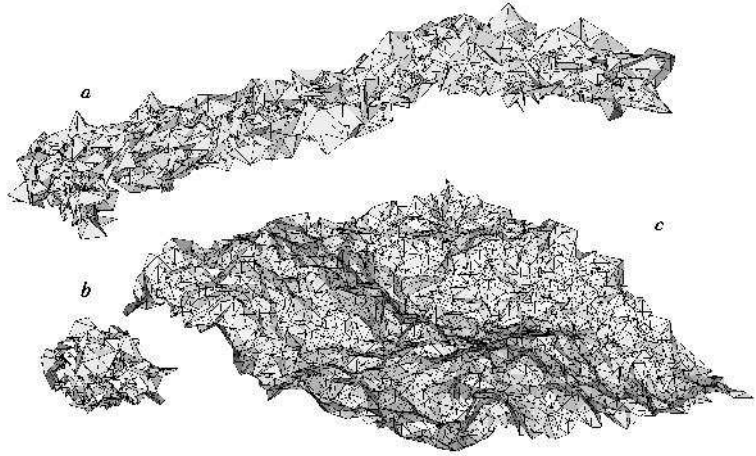}}
  \caption{Examples of a) the tubular phase b) the crumpled phase and
  c) the flat phase of an anisotropic phantom crystalline membrane
  taken from the simulations of \cite{BFT:97}}
  \label{fig__CRTUFL}
  \end{figure}

Following the same steps as in the isotropic case, we split
\be\label{mf_variable_any}
{\vec r}({\bf x})=(\zeta_{\perp} {\bf x}_{\perp}+{\bf u}_{\perp}({\bf x}),
\zeta_y y +u_y({\bf x}), h({\bf x})) \ ,
\ee
with ${\bf u}_{\perp}$ being the $D-1$-dimensional phonon in-plane modes, 
$u_y$ the in-plane phonon mode in the distinguished direction $y$ and
$h$ the $d-D$ out-of-plane fluctuations. If $\zeta_{\perp}=\zeta_y=0$,
the membrane is in a crumpled phase and if both
$\zeta_{\perp} \ne 0$ and $\zeta_y \neq 0$ the membrane is in a flat 
phase very similar to the isotropic case (how similar will be
discussed shortly). There is, however, the possibility that 
$\zeta_{\perp}=0$ and $\zeta_y \ne 0$ or $\zeta_{\perp} \ne 0$ and
$\zeta_y = 0$. This describes a completely new phase, in which the membrane
is crumpled in some internal directions but flat in the remaining ones.
A phase of this type is called a {\em tubular phase} and does not
appear when studying isotropic membranes. 
In Fig.\ref{fig__CRTUFL} we show an intuitive visualization of a
tubular phase along with the corresponding flat and crumpled phases 
of anisotropic membranes.

We will start by studying the phantom case. We show, using both
analytical and numerical arguments, that the phase diagram contains
a crumpled, tubular and flat phase. The crumpled and flat phases are 
equivalent to the isotropic ones, so anisotropy turns out to be an
irrelevant interaction in those phases. The new physics is contained
in the tubular phase, which we describe in detail, both 
with and without self-avoidance.

\subsection{Phantom}\label{SECT__phase_ani}

\subsubsection{The Phase diagram}\label{SubSECT__tub}

We first describe the mean field theory phase diagram and then 
the effect of fluctuations. 
There are two situations depending on the particular 
values of the function $\Delta$, which depends on the 
elastic constants $u_{\perp \perp},v_{\perp y},u_{yy}$ and $v_{\perp \perp}$.
Since the derivation is rather technical, we refer to Appendix~\ref{MF__Appen} 
for the details.

\begin{figure}[htb]
  \epsfxsize=5 in \centerline{\epsfbox{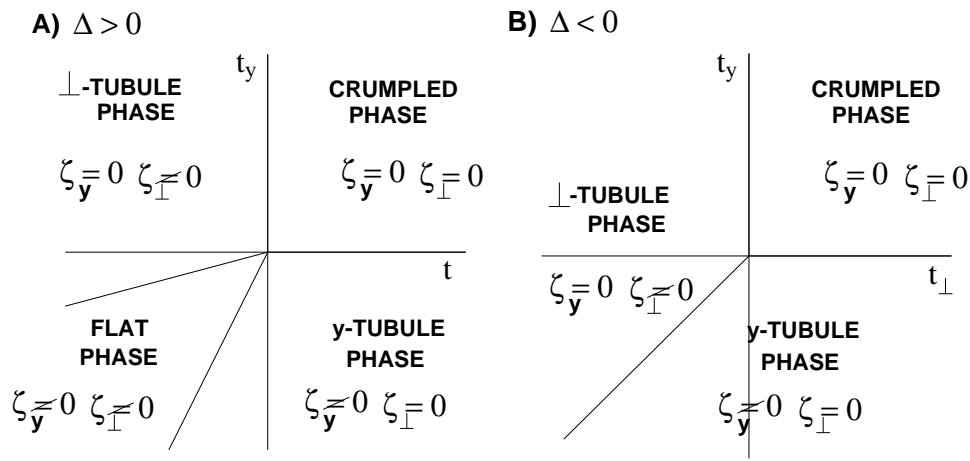}}
  \caption{The phase diagram for anisotropic phantom membranes} 
  \label{fig__MF_an}
  \end{figure}

\begin{itemize}
{\item Case A} ($ \Delta > 0$): the mean field solution displays all possible phases.
When $t_y > 0$ and $t_{\perp}>0$ the model is in a crumpled phase. 
Lowering the temperature, one of the $t$ couplings becomes 
negative, and we reach a tubular phase (either $\perp$ or $y$-tubule).
A further reduction of the temperature eventually leads to a flat phase.

{\item Case B} ($ \Delta < 0$): in this case the flat phase disappears from the mean
field solution. Lowering further the temperature leads to a continuous 
transition from the crumpled phase to a tubular phase. Tubular phases
are the low temperature stable phases in this regime.
\end{itemize}
This mean field result is summarized in Fig.\ref{fig__MF_an}. 

Beyond mean field theory, the Ginsburg criterion applied to this model
tells us that the phase diagram should be stable for physical membranes
$D=2$ at any embedding dimension $d$, so the mean field scenario should 
give the right qualitative picture for the full model.

Numerical simulations have spectacularly confirmed this result. We have 
already shown in Fig.\ref{fig__CRTUFL} the results from the numerical 
simulation in \cite{BFT:97}, where it was shown that changing the
temperature generates a sequence of transitions crumpled-to-tubular 
and tubular-to-flat, in total agreement with case A) in the mean field result
illustrated in Fig.\ref{fig__MF_an}. 

We now turn to a more detailed study of both the crumpled and
flat anisotropic phases. Since we have already studied crumpled and 
flat phases we just outline how those are modified when anisotropy
is introduced.

\subsubsection{The Crumpled Anisotropic Phase}\label{SubSECT__cr_ani}

In this phase $t_y>0$ and $t_{\perp} >0$, and 
the free energy Eq.(\ref{LGW}) reduces for $D \ge 2$ to 
\be\label{crum_ani_fe}
F(\vec r({\bf x}))=\frac{1}{2} \int d^{D-1}{\bf x}_{\perp} dy \left[
t_{\perp}(\parp_{\alpha}^{\perp} \vec r)^2 + t_y(\pary \vec r)^2
\right]+\mbox{Irrelevant} \ .
\ee
By redefining the $y$ direction as $y^{\prime}=\frac{t_{\perp}}{t_y} y$
this reduces to Eq.(\ref{LAN_CR_PH_IRR}), with $t\equiv t_{\perp}$. We have
proved that anisotropy is totally irrelevant in this particular phase.

\subsubsection{The Flat Phase}\label{SubSECT__Fl_ani}

This phase becomes equivalent to the isotropic case as well. Intuitively,
this may be obtained from the fact that if the membrane is flat, the
intrinsic anisotropies are only apparent at short-distances, and therefore
by analyzing the RG flow at larger and larger distances the membrane
should become isotropic. This argument may be made slightly more 
precise \cite{Toner:88}.

\subsection{The Tubular Phase}\label{SECT__tubular}

We now turn to the study of the novel tubular phase, both in the
phantom case and with self-avoidance. Since the physically relevant
case for membranes is $D=2$ the $y$-tubular and $\perp$-tubular 
phase are the same. So we concentrate on the properties of the
$y$-tubular phase.

\begin{figure}[htb]
\epsfxsize=3in \centerline{\epsfbox{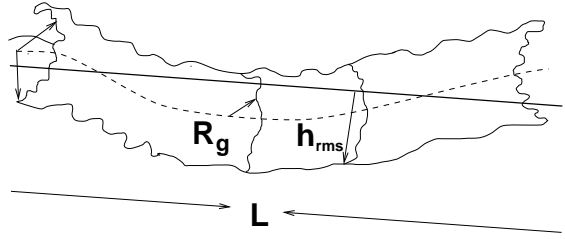}}
\caption{A schematic illustration of a tubular configuration
indicating the radius of gyration $R_g$ and the height 
fluctuations $h_{rms}$.}
\label{fig__tubdef}
\end{figure}

The key critical exponents characterizing the tubular phase are
the size (or Flory) exponent $\nu$, giving the scaling of the tubular
diameter $R_g$ with the extended ($L_y$) and transverse ($L_{\perp}$)
sizes of the membrane, and the roughness exponent $\zeta$ associated with 
the growth of height fluctuations $h_{rms}$ (see Fig.\ref{fig__tubdef}):
\bea
\label{nuzeta}
R_g(L_{\perp},L_y) & \propto & L_{\perp}^{\nu} S_R(L_y/L_{\perp}^z)\\
\nonumber
h_{rms}(L_{\perp},L_y) & \propto & L_y^{\zeta} S_h(L_y/L_{\perp}^z)
\eea
Here $S_R$ and $S_h$ are scaling functions \cite{RT:95,RT:98}
and $z$ is the anisotropy exponent.
 
\begin{figure}[htb]
   \epsfxsize=4in \centerline{\epsfbox{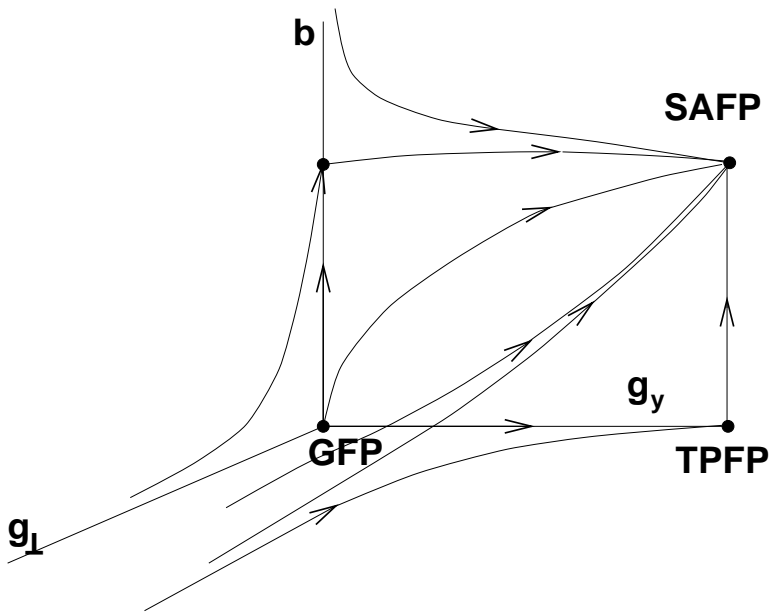}}
   \caption{The phase diagram for self-avoiding anisotropic membranes
   with the Gaussian fixed point (GFP), the tubular phase fixed point
   (TPFP) and the self-avoidance fixed point (SAFP).}
   \label{fig__BG}
   \end{figure}

The general free energy described in Eq.(\ref{LGW}) may be simplified
considerably in a $y$-tubular phase. The analysis required is 
involved and we refer the interested reader to \cite{BG:97,BT:99}.
We just quote the final result. It is
\bea\label{free_EG}
F(u,\vec h)&=&\frac{1}{2}\int d^{D-1}{\bf x}_{\perp} dy \left[
\kappa (\pary^2 \vec h)^2+t(\parp_{\alpha} \vec h)^2 \right.
\nonumber\\
&&+
g_{\perp}(\parp_{\alpha} u+\partial_{\alpha} \vec h \pary \vec h )^2
\nonumber\\
&&+\left.
g_y(\pary u+\frac{1}{2}(\pary \vec h)^2)^2
\right] 
\nonumber\\
&&+
\frac{b}{2}\int dy d^{D-1}{\bf x}_{\perp}d^{D-1}{\bf x}_{\perp}^\prime
\delta^{d-1}(\vec{h}({\bf x}_{\perp},y)-\vec{h}({\bf x}_{\perp}^\prime,y)) 
\ .
\eea
Comparing with Eq.(\ref{LGW}), this free energy
does represent a simplification as the number of couplings has been
reduced from eleven to five. Furthermore, the coupling $g_{\perp}$ is irrelevant 
by standard power counting. The most natural assumption is to 
set it to zero. In that case the phase diagram one obtains is 
shown in Fig.\ref{fig__BG}. Without self-avoidance $b=0$,
the Gaussian Fixed Point (GFP) is unstable and the infra-red stable
FP is the tubular phase FP (TPFP). Any amount of self-avoidance,
however, leads to a new FP, the Self-avoiding Tubular FP (SAFP),
which describes the large distance properties of self-avoiding tubules.
   
\begin{figure}[htb]
\epsfxsize=4in \centerline{\epsfbox{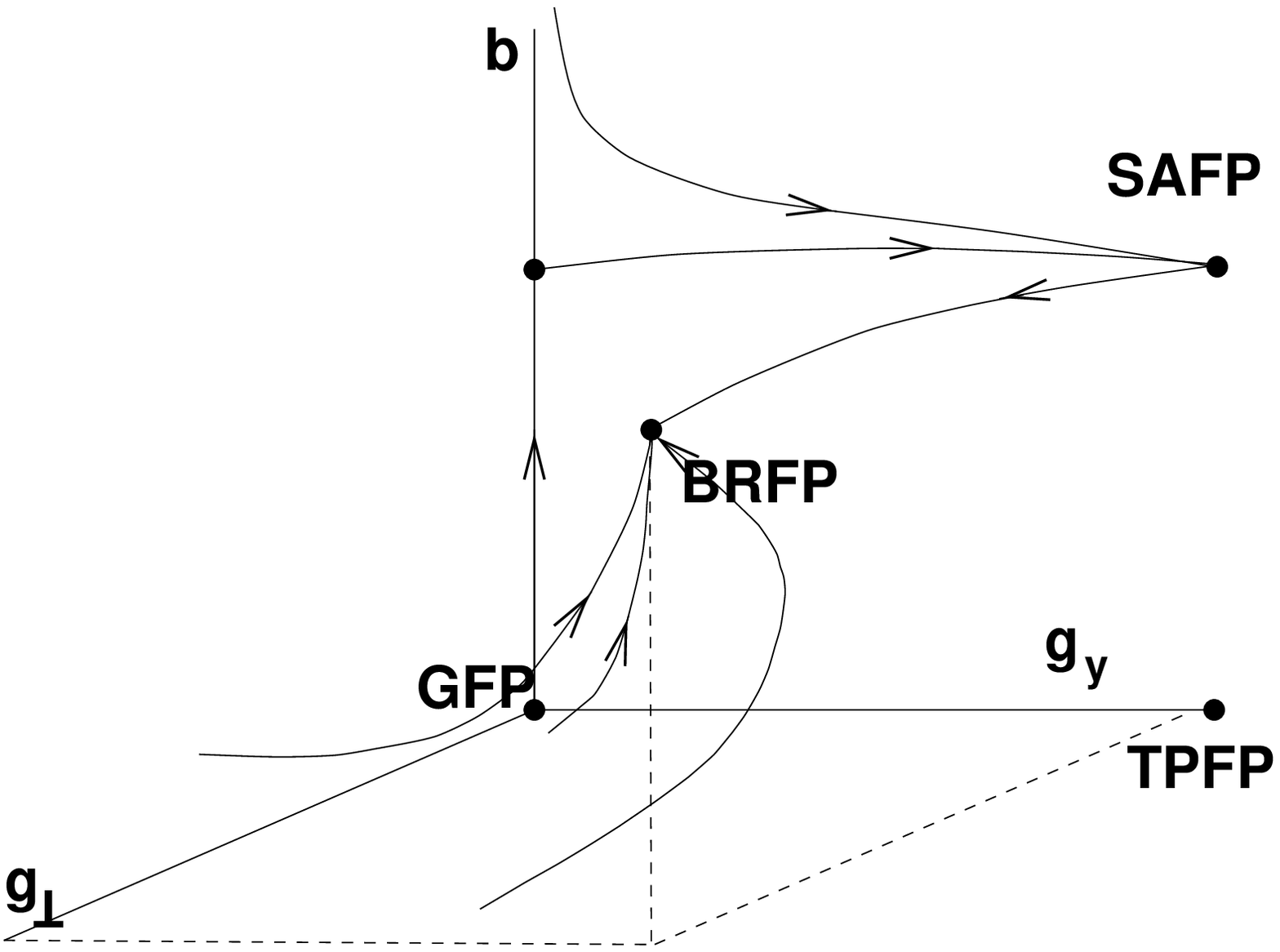}}
\caption{The phase diagram for self-avoiding anisotropic membranes
with the Gaussian fixed point (GFP), the tubular phase fixed point
(TPFP), the self-avoidance fixed point (SAFP) and the bending rigidity
fixed point (BRFP).}
\label{fig__RT}
\end{figure}

We just mention, though, that other authors advocate a different
scenario \cite{RT:98}. 
For sufficiently small embedding  dimensions $d$, including the
physical $d=3$ case, these authors suggest the existence of a new bending rigidity renormalized
FP (BRFP), which is the infra-red FP describing the actual properties 
of self-avoiding tubules (see Fig.~\ref{fig__RT}). 

Here we follow the arguments presented in \cite{BT:99} and consider 
the model defined by Eq.(\ref{free_EG}) with the $g_{\perp}$-term as the model describing the large
distance properties of tubules. One can prove then than there
are some general scaling relations among the critical exponents.
All three exponents may be expressed in terms of a single exponent 
\bea\label{Ani_scaling}
\zeta&=&\frac{3}{2}+\frac{1-D}{2z} \nonumber\\
&\nu&=\zeta z \ .
\eea

Remarkably, the phantom case as described by Eq.(\ref{free_EG}) can be  
solved exactly. The result for the size exponent is 
\be\label{Phantom_ani_size}
\nu_{Phantom}(D)=\frac{5-2D}{4} \ , \nu_{Phantom}(2)= \frac{1}{4}
\ee
with the remaining exponents following from the scaling
relations Eq.(\ref{Ani_scaling}).

 \begin{table}[htb]
 \centerline{
 \begin{tabular}{|c||l|l|l|l|l|}
 \multicolumn{1}{c}{$d$}     &
 \multicolumn{1}{c}{$\nu$}   & \multicolumn{1}{c}{$\nu_F$ } &
 \multicolumn{1}{c}{$\nu_V$} & \multicolumn{1}{c}{$\nu_{Flory}$} \\\hline
 8  & $0.333(5)  $ & $0.34(1)$  & $0.34(1)$  & $0.333$ \\\hline
 7  & $0.374(8)  $ & $0.39(2)$  & $0.39(2)$  & $0.375$ \\\hline
 6  & $0.42(1)   $ & $0.44(2)$  & $0.44(4)$  & $0.429$ \\\hline
 5  & $0.47(1)   $ & $0.51(3)$  & $0.51(5)$  & $0.500$ \\\hline
 4  & $0.54(2)   $ & $0.60(4)$  & $0.60(6)$  & $0.600$ \\\hline
 3  & $0.62(2)   $ & $0.71(6)$  & $0.70(9)$  & $0.750$ \\\hline
 \end{tabular}}
 \caption{Different extrapolations for the size exponent at different 
  embedding dimensions $d$ \cite{BT:99}. The first column gives the corrections
  to the mean field result, the second corrections to the Flory estimate and
  the third corresponds to corrections to the Gaussian approximation. The 
  last column quotes the Flory estimate for comparison.}
 \label{tab__EXP_comp}
 \end{table}

The self-avoiding case may be treated with techniques similar to those
in isotropic case. The size exponent may be estimated within a 
Flory approach. The result is
\be\label{Flory_ani}
\nu_{Flory}=\frac{D+1}{d+1} \ .
\ee
The Flory estimate is an uncontrolled approximation. Fortunately,
a $\vap$-expansion, adapting the MOPE technique described for the
self-avoiding isotropic case to the case of tubules, is also
possible \cite{BG:97,BT:99}. The $\beta$-functions are computed and provide evidence for
the phase diagram shown in Fig.\ref{fig__BG}. Using rather involved
extrapolation techniques, it is possible to obtain estimates
for the size exponent, which are shown 
in Table~\ref{tab__EXP_comp}.  The rest of the exponents may be 
computed from the scaling relations.

\section{Defects in membranes: The Crystalline-Fluid transition and
Fluid membranes}\label{SECT__Defects}

\begin{figure}[htb]
  \epsfxsize= 4 in  \centerline{\epsfbox{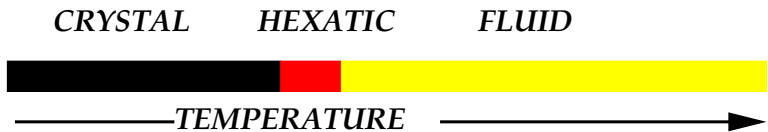}}
  \caption{Two stage melting according to KTNHY theory.}
  \label{fig__KTNHY}
  \end{figure}

A flat crystal melts into a liquid when the temperature is increased.
This transition may be driven by the sequential liberation
of defects, as predicted by the KTNHY theory. The KTNHY theory is 
schematically shown in Fig.\ref{fig__KTNHY}. With increasing
temperature, a crystal melts first to an intermediate hexatic 
phase via a continuous transition, and finally goes to a conventional 
isotropic fluid phase via another continuous transition. 

We will not review here either the KTNHY theory or the 
experimental evidence in its favor {--} \cite{Nelson:83}. We just
want to emphasize here that the KTNHY theory is in general agreement
with existing experiments, although there are two main points 
worth keeping in mind when studying the more difficult case of fluctuating
geometries. 1) The experimental evidence for the existence of the 
hexatic phase is not completely settled in those transitions which 
are continuous. 2) Some 2D crystals (like Xenon absorbed on graphite)
melt to a fluid phase via a first order transition without any 
intermediate hexatic phase. 

The straight-forward translation of the previous results to the
tethered membrane would suggest a similar scenario. There would then be
a crystalline to hexatic transition and a hexatic to fluid transition, 
as schematically depicted in Fig.\ref{fig__KTNHY}. Although the 
previous scenario is plausible, there are no solid experimental
or theoretical results that establish it. From the theoretical 
point of view, for example, an important open problem is how to 
generalize the RG equations of the KTNHY theory to the case 
of fluctuating geometry.
The situation looks even more uncertain experimentally, especially 
considering the elusive nature of the hexatic phase even in the case 
of flat monolayers.

In this review we will assume the general validity 
of the KTNHY scenario and we describe models of hexatic membranes, as well
as fluid membranes. The study of the KTNHY theory in fluctuating
geometries is a fascinating and challenging problem that deserves
considerable effort. In this context,
let us mention recent calculations of defects on
frozen topographies \cite{BNT:99}, which show that even in the more simplified
case when the geometry is frozen, defects proliferate in an attempt to
screen out Gaussian curvature, even at zero temperature, and organize 
themselves in rather surprising and unexpected structures. These results
hint at a rich set of possibilities for the more general case of
fluctuating geometries.

\subsection{Topological Defects}

A crystal may have different distortions from its ground state. 
Thermal fluctuations are the simplest. Thermal fluctuations are
small displacements from the ground state, and therefore one 
may bring back the system to its original positions by local
moves without affecting the rest of the lattice. There are more 
subtle lattice distortions though, where the lattice cannot be
taken to its ground state by local moves. These are the topological
defects. There are different possible topological defects that
may occur on a lattice, but we just need to consider dislocations
and disclinations. Let us review the most salient features.

  \begin{figure}[htb]
  \epsfxsize= 2 in  \centerline{\epsfbox{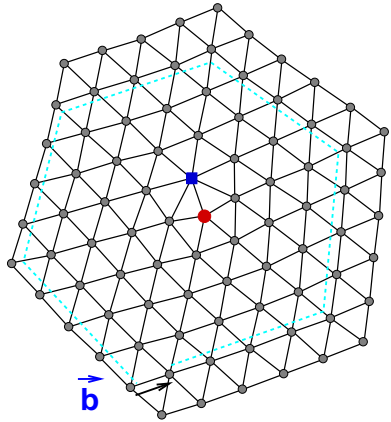}}
  \caption{Example of a dislocation showing the breaking of the 
  translational holonomy measured by the Burgers vector $\vec{b}$.} 
  \label{fig__disl}
  \end{figure}
  
\begin{figure}[htb]
  \centerline{\epsfxsize= 2 in  \epsfbox{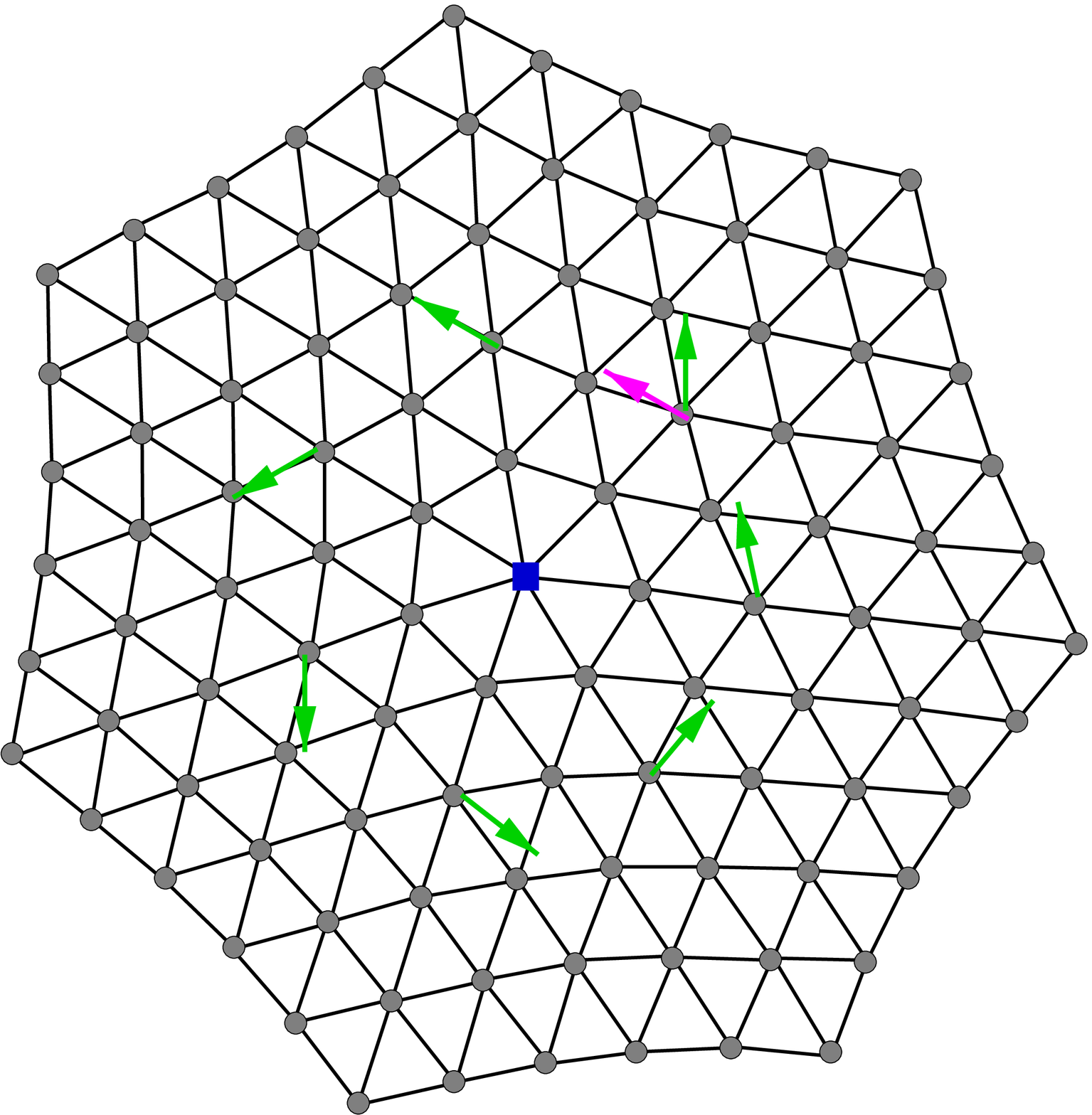}
  \epsfxsize= 2 in  \epsfbox{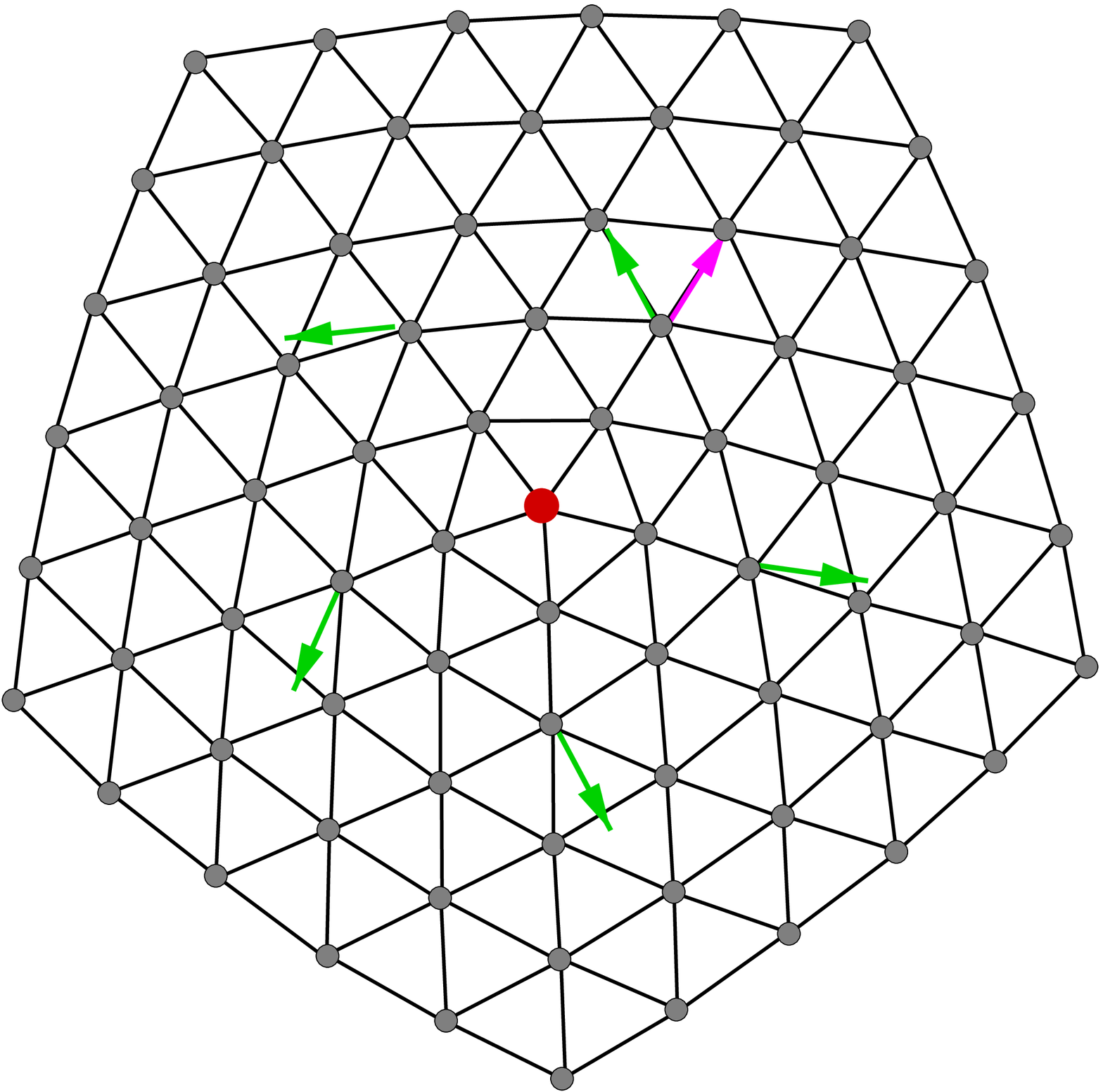}}
  \caption{Example of a minus disclination (left figure) and a plus 
  disclination (right figure). The orientational holonomy is broken by 
  $\pm\frac{\pi}{3}$ respectively.}
  \label{fig__discl}
  \end{figure}

\begin{itemize}
\item{\underline{Dislocation}}: represents the breaking of the 
translational holonomy. A path that would naturally close in a 
perfect lattice fails to close by a vector ${\vec b}$, the Burgers
vector, as illustrated in Fig.\ref{fig__disl}. In a flat monolayer, the energy is 
\be\label{DISl__EN}
E=\frac{K_0 {\vec b}^2}{8 \pi} \ln(\frac{R}{a}) \ ,
\ee
where $K_0$ is the Young Modulus. It diverges {\em logarithmically}
with system size.
\item{\underline{Disclination}}: represents the breaking of the
rotational holonomy. The bond angle around the point defect is a multiple
of the natural bond angle in the ground state ($\frac{\pi}{3}$ in
a triangular lattice), as illustrated in Fig.\ref{fig__discl} for a $+$ and $-$
disclination. The energy for a disclination in a flat monolayer is
given by
\be\label{Discl__EN}
E = \frac{K_0 s^2}{32 \pi} R^2 \ .
\ee
Note the {\em quadratic} divergence of the energy with system size $R$.
\end{itemize}

Inspection of Fig.\ref{fig__disl} shows that a dislocation
may be regarded as a tightly bound +,- disclination pair. 

\subsubsection{Topological Defects in fluctuating geometries}

The problem of understanding topological defects when the geometry 
is allowed to fluctuate was addressed in \cite{SN:88} (see \cite{DRNles} 
for a review). The important new feature is that the energy of a
disclination defect may be lowered considerably if the membrane buckles
out-of-the-plane. That is, the membrane trades elastic energy for 
bending rigidity. The energy for a buckled free disclination is given by
\be\label{new__discl_ENER}
E = f(\kappa,K_0,q_i) \ln(\frac{R}{a}) \ ,
\ee
where $f$ is some complicated function that may be evaluated numerically
for given values of the parameters. It depends explicitly on $q_i$, which
implies that the energies for positive and negative disclinations may be
different, unlike the situation in flat space. The extraordinary reduction in
energy from $R^2$ to $\ln R$ is possible because the buckled membrane
creates positive Gaussian curvature for the
plus-disclination and negative curvature for the negative-disclination. 
This is a very important physical feature of defects on curved surfaces.
The defects attempt to screen out like-sign curvature, and analogously,
like-sign defects may force the surface to create like-sign curvature
in order to minimize the energy.

The reduction in energy for a dislocation defect is even more remarkable, 
since the energy of a dislocation becomes a constant, independent of 
the system size, provided the system is larger than a critical radius 
$R_c$. Again, by allowing the possibility of out-of-plane buckling, a 
spectacular reduction in energy is achieved (from $R^2$ to $\ln R$).

The study of other topological defects, e.g. vacancies, interstitials,
and grain boundaries, may be carried out along the same lines. Since
we are not going to make use of it, we refer the reader to the excellent review
in \cite{DRNles}. 

\subsubsection{Melting and the hexatic phase}

The celebrated Kosterlitz-Thouless argument shows that defects will
necessarily drive a 2D crystal to melt. The entropy of a 
dislocation grows logarithmically with the system size, so for 
sufficiently high temperature, entropy will dominate over the dislocation
energy (Eq.(\ref{DISl__EN})) and the crystal will necessarily melt.
If the same Kosterlitz-Thouless is applied now to 
a tethered membrane, the entropy is still growing logarithmically
with the system size, while the energy becomes independent of the system
size, as explained in the previous subsection, so any finite temperature 
drive the crystal to melt, and the low temperature phase of a tethered 
membrane will necessarily be a fluid phase, either hexatic if the KTNHY melting
can be applied, or a conventional fluid if a first order transition takes
place, or even some other more perverse possibility. This problem has
also been investigated in numerical simulations \cite{GK:97}, which provide some
concrete evidence in favor of the KTNHY scenario, although the issue 
is far from being settled.

It is apparent from these arguments that a hexatic membrane 
is a very interesting and possibly experimentally relevant 
membrane to understand. 

\subsection{The Hexatic membrane}

The hexatic membrane is a fluid membrane that, in contrast to a
conventional fluid, preserves the
orientational order of the original lattice (six-fold (hexatic) for
a triangular lattice). The mathematical
description of a fluid membrane is very different from those with 
crystalline order. Since the description cannot depend on internal
degrees of freedom, the free energy must be invariant under  
reparametrizations of the internal coordinates (that is, should depend
only on geometrical quantities, or in more mathematical terminology,
must be diffeomorphism invariant). The corresponding free energy was proposed
by Helfrich \cite{Helf:73} and it is given by
\be\label{extr_curv}
\frac{{\cal H}_{H}}{T}=\mu \int \sqrt{g} +
\frac{\kappa}{2}\int d {\bf x} \sqrt{g} {\vec H}^2 \ ,
\ee
where $\mu$ is the bare string tension, $\kappa$ the bending rigidity,
$g$ is the determinant of the metric of the surface 
\be\label{metric__srf}
g_{\mu \nu}({\bf x})=\partial_{\mu} {\vec r}({\bf x}) 
\partial_{\nu} {\vec r}({\bf x})
\ee
and ${\vec H}$ is the mean Gaussian curvature of the surface.
For a good description of the differential geometry relevant to the
study of membranes we refer to \cite{D:92}
A hexatic membrane has an additional degree of 
freedom, the bond angle, which is introduced as a field on the 
surface $\theta$.
The hexatic free energy \cite{NP:90} is
obtained from adjoining to the fluid case of Eq.(\ref{extr_curv}), the
additional energy of the bond angle
\be\label{hexatic_energy}
{\cal H}_{hex}/T=\frac{K_A}{2}\int d {\bf x} \sqrt{g}
g^{\mu \nu}(\partial_{\mu} \theta+\Omega_{sing}-\Omega^L_{\mu})
(\partial_{\nu} \theta+\Omega_{sing}-\Omega^L_{\nu})
\ee
where $K_A$ is called the hexatic stiffness, and
$\Omega_{\mu}$ is the connection two form of the metric, which
may be related to the Gaussian curvature of the surface by
\be\label{metric__con}
K({\bf x})=\frac{1}{\sqrt{g}}\epsilon^{\mu \nu}\partial_{\mu} 
\Omega_{\nu}
\ee
and $\Omega_{sing}$ is similarly related to the topological defect
density 
\bea\label{hexatic_energy_2}
s({\bf x})&=&\frac{1}{\sqrt{g}}\epsilon^{\mu \nu}\partial_{\mu} 
\Omega_{sing \nu}
\nonumber\\
s({\bf x})&=&\frac{\pi}{3}\frac{1}{\sqrt{g}}\sum_{i=1}^N q_i 
\delta({\bf x},{\bf x}_i) \ .
\eea
From general theorems on differential geometry one has the relation
\be\label{Diff_geom}
\int \sqrt{g} s({\bf x})=4 \pi \rightarrow \sum_{i=1}^i q_i = 2 \chi \ ,
\ee
where $\chi$ is the Euler characteristic of the surface.

Therefore the total free energy for the hexatic membrane is given by
\bea\label{Hex__Mem}
{\cal H}/T&=&\mu \int \sqrt{g}+
\frac{\kappa}{2}\int d {\bf x} \sqrt{g} {\vec H}^2+ 
\\\nonumber
&+& \frac{K_A}{2}\int d {\bf x} \sqrt{g}
g^{\mu \nu}(\partial_{\mu} \theta+\Omega_{sing}-\Omega^L_{\mu})
(\partial_{\nu} \theta+\Omega_{sing}-\Omega^L_{\nu}) \ .
\eea
The partition function is therefore
\bea\label{new_part_func}
{\cal Z}(\beta)&=&\sum_{N_{+},N_{-}} \frac{\delta_{N_+-N_{-},2 \chi}}
{N_{+}!N_{-}!} y^{N_++N_-} \times
\\\nonumber
&& \int D[{\vec r}] D[{\theta}]
\int \prod_{\mu=1}^{N_+} d {\bf
x}^+_{\mu}\sqrt{g} \prod_{\nu=1}^{N_-} d {\bf x}_{\nu}^{-} \sqrt{g}
e^{-{\cal H}({\vec r}({\bf x}),\theta({\bf x}))/T} \ ,
\eea
where $y$ is the fugacity of the disclination density.
The partition function includes a discrete sum over allowed topological
defects, those satisfying the topological constraint Eq.(\ref{Diff_geom}), and
a path integral over embeddings ${\vec r}$ and bond angles $\theta$.
The previous model remains quite intractable since the sum over defects
interaction is very difficult to deal with.

  \begin{figure}[htb]
  \epsfxsize= 4 in  \centerline{\epsfbox{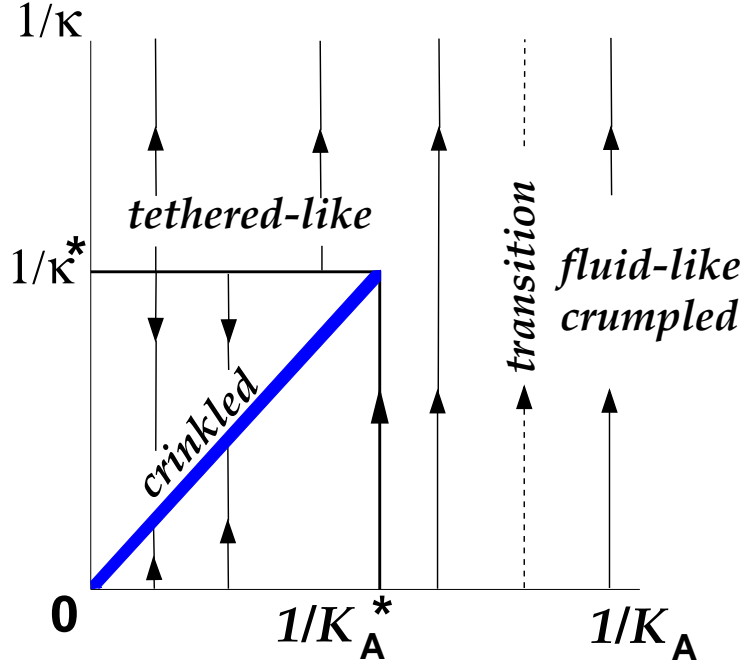}}
  \caption{Phase diagram according to \cite{GuKa:90}. The hexatic 
  membrane interpolates between a crystalline membrane and a fluid one.
  For small rigidity and large hexatic stiffness the RG flows towards 
  a fluid crumpled phase. For small hexatic stiffness and large 
  $\kappa$ it flows towards a tethered like phase, whereas for 
  both large $\kappa$ and $K_A$ the flow is to the crinkled phases
  described by the non-trivial FP of Eq.(\ref{beta__function}).}
  \label{fig__hexatic}
  \end{figure}

Fortunately, the limit of very low fugacity $y \rightarrow 0$ is 
analytically tractable as was shown in the beautiful paper \cite{DGP:87}. 
The RG functions can be computed within a combined large $d$ and large
bending rigidity expansion. The $\beta$ functions in that limit is given 
by
\bea\label{beta__function}
\beta(\alpha)&=&\frac{1}{4 \pi K_A}\left( -D \alpha^2+\frac{3}{4}\alpha^3+
{\cal O}(1/K_A^2)\right) \ ,
\nonumber\\
\beta(K_A)&=&0
\eea
where $\alpha=1/\kappa$. The physics of hexatic membranes in the limit 
of very low fugacity is very rich and show a line of fixed points 
parametrized by the hexatic stiffness $K_A$. The normal-normal correlation 
function, for example, reads \cite{DGP:87}
\be\label{Hex__norm_norm}
\langle {\vec n}({\bf r}) {\vec n}({\bf 0}) \rangle \sim 
|{\bf x}|^{-\eta} \ ,
\ee
with $\eta=\frac{2}{3 \pi} d (d-2)\frac{k_B T}{K_A}$. The FPs of 
Eq.(\ref{beta__function}) describe a new {\em crinkled} phase, 
more rigid than a crumpled phase but more crumpled than a flat one. 
The Hausdorff dimension at the crinkled phase is given by \cite{DGP:87} 
\be\label{crink__phase}
d_H=2+\frac{d(d-2)}{3 \pi} \frac{k_B T}{K_A} \ .
\ee
From the RG point of view, the properties of these crinkled phases
are really interesting, since they involve a line of Fixed Points which
are inequivalent in the sense that the associated critical exponents depend continuously 
on $K_A$, a situation reminiscent of the $XY$-model. In \cite{GuKa:90}
the phase diagram is discussed, and the authors propose the scenario
depicted in Fig.\ref{fig__hexatic}. How these scenarios are modified when
the fugacity is considered is not well established and we refer the reader to the 
original papers \cite{PaLu2:96,PaLu3:96,GuKa:90}. 

The shape fluctuations of hexatic vesicles, for large defect core
energies, have also been investigated \cite{ML:91,Evans:95}. 

\section{The Fluid Phase}\label{SECT__Fluid}

The study of fluid membranes is a broad subject, currently under intense 
experimental and theoretical work. The Hamiltonian is given by
\be\label{Fluid_phase}
{\cal H}/T=\mu \int \sqrt{g}+
\frac{\kappa}{2}\int d {\bf x} \sqrt{g} {\vec H}^2+ 
\frac{\hat{\kappa}}{2}\int \sqrt{g} K \ ,
\ee
This corresponds to the Helfrich hamiltonian together with a term that
allows for topology changing interactions.
For fixed topology it is well-known
\cite{PL:85,Poly:86,Forster:86,Klein1:86} 
that the one loop beta function
for the inverse-bending rigidity has a fixed point only at $\kappa=0$,
which corresponds to the bending rigidity being irrelevant at large
length scales. The RG flow of the bending rigidity is given by
\be\label{FluidRG}
\kappa(l) = \kappa_o - \frac{3T}{4\pi} {\rm ln} (l/a) \ , 
\ee
where a is a microscopic cutoff length.
The fluid membrane is therefore crumpled, for
arbitrary microscopic bending rigidity $\kappa_0$, at length scales beyond
a persistence length which grows exponentially with $\kappa_0$. 
For a fluid membrane out-of-plane fluctuations cost no elastic energy
(the membrane flows internally to accommodate the deformation)
and the bending rigidity is therefore softened by thermal undulations
at all length scales, rather than stiffening at long length scales as
in the crystalline membrane.
 
So far we have assumed an infinite membrane,
which is not always a realistic assumption. A thickness may be 
taken into account via a spontaneous
extrinsic curvature ${\vec H}_0$. The model described by 
Eq.~\ref{Fluid_phase} gets replaced then by
\be\label{Fluid_phase_2}
{\cal H}/T=\mu \int \sqrt{g}+
\frac{\kappa}{2}\int d {\bf x} \sqrt{g} ({\vec H- \vec H_0})^2+ 
\frac{\hat{\kappa}}{2}\int \sqrt{g} K \ .
\ee
Further effects of a finite membrane size for spherical topology have
been discussed in \cite{MM:94,Morse:94,MM:95}.
The phase diagram of fluid membranes when topology change is allowed  
is fascinating and not completely understood. A complete description 
of these phases goes beyond the scope of this review {--} we refer the 
reader to \cite{GK1:97,GK2:97,GS:94,Safran} and references therein.

\section{Conclusions}\label{SECT__CONC}

In this review we have described the distinct universality classes of membranes 
with particular emphasis on crystalline membranes. In each case we 
discussed and summarized the key models describing the interactions of
the relevant large distance degrees of freedom (at the micron
scale). The body of the review emphasizes qualitative and descriptive 
aspects of the physics with technical details presented in extensive
appendices. We hope that the concreteness of these calculations
gives a complete picture of how to extract relevant physical information 
from these membrane models.

We have also shown that the phase diagram of the phantom crystalline 
membrane class is theoretically very well understood both by analytical and 
numerical treatments. To complete the picture it would be 
extremely valuable to find experimental realizations for this particular 
system. An exciting possibility is a system of cross-linked DNA
chains together with restriction enzymes that catalyze cutting and
rejoining \cite{BEN}. 
The difficult chemistry involved in these experiments is not yet 
under control, but we hope that these technical problems will be 
overcome in the near future.

There are several experimental realizations of self-avoiding polymerized
membranes discussed in the text. 
The experimental results compare very well with the 
theoretical estimates from numerical simulations. As a future theoretical
challenge, analytical tools need to be sharpened since they fail to provide a clear
and unified picture of the phase diagram. On the experimental side, there
are promising experimental realizations of tethered membranes which will 
allow more precise results than those presently available. Among them 
there is the possibility of very well controlled synthesis of DNA networks  
to form physical realizations of tethered membranes.

The case of anisotropic polymerized membranes has also been described in
some detail. The phase diagram contains a new tubular phase which 
may be realized in nature. There is some controversy about the precise
phase diagram of the model, but definite predictions for the critical
exponents and other quantities exist. Anisotropic membranes are also
experimentally relevant. They may be created in the laboratory 
by polymerizing a fluid membrane in the presence of 
an external electric field. 

Probably the most challenging problem, both theoretically and experimental, 
is a complete study of the role of defects in polymerized membranes. There
are a large number of unanswered questions, which include 
the existence of hexatic phases, the properties  
of defects on curved surfaces and its relevance to the possible
existence of more complex phases. 
This problem is now under intense experimental investigation.
In this context, let us mention very recent experiments on Langmuir
films in a presumed hexatic phase \cite{Fischer}. The coalescence of 
air bubbles with the film exhibit several puzzling features which are
strongly related to the curvature of the bubble.

Crystalline membranes also provide important insight into the 
fluid case, since any crystalline membrane eventually becomes fluid at
high temperature. The physics of fluid membranes is a complex and
fascinating subject in itself which goes beyond the scope of this
review. We highlighted some relevant experimental realizations and gave a quick
overview of the existing theoretical models. Due to its relevance in
many physical and biological systems and its potential
applications in material science, the experimental and theoretical 
understanding of fluid membranes is, and will continue to be, one of the 
most active areas in soft condensed matter physics.

We have not been able in this review, simply for lack of time, to
address the important topic of the role of disorder.
We hope to cover this in a separate article.   

We hope that this review will be useful for physicists trying to get a 
thorough understanding of the fascinating field of membranes. 
We think it is a subject with significant prospects for new and exciting
developments. 

\bigskip
\noindent{Note}: The interested reader may also find additional
material in a forthcoming review by Wiese \cite{Wiese:00}, 
of which we have seen only the table of contents. 

\bigskip
\centerline{\bf Acknowledgements}
\medskip

This work was supported by the U.S. Department of Energy under 
contract No. DE-FG02-85ER40237.
We would like to thank Dan Branton and Cyrus Safinya for providing us
images from their laboratories and Paula Herrera-Sikl\'ody for 
assistance with the figures. MJB would like to thank Riccardo
Capovilla, Chris Stephens, Denjoe O'Connor and the other
organizers of RG2000 for the opportunity to attend a wonderful meeting
in Taxco, Mexico. 

\newpage

\appendix

\section{Useful integrals in dimensional regularization}\label{APP__SA__Int}

In performing the $\vap$-expansion, we will be considering integrals
of the form
\be\label{int__form}
I_{\alpha_1,\cdots,\alpha_n}(a,b)({\vec p})=\int d^D \hat{q}
\frac{q_{\alpha_1} \cdots q_{\alpha_n}}{(\vec{p}+\vec{q})^{2a}\vec{q}^{2b}}
\ ,
\ee
These integrals may be computed exactly for general 
$D,a,b$ and $\alpha=1,\cdots,N$. The result will be published elsewhere.
We will content ourselves by quoting what we need, the poles in $\vap$,
for the integrals that appear in the diagrammatic calculations. We just
quote the results
\be\label{dim_reg_int}
I_{\alpha_1 \alpha_2}(2,2)=-\frac{1}{8 \pi^2 p^4}p_{\alpha_1}p_{\alpha_2} 
\frac{1}{\vap}
\ee
\be
I_{\alpha_1 \alpha_2 \alpha_3}(2,2)=\frac{1}{8 \pi^2 p^4}
p_{\alpha_1}p_{\alpha_2}p_{\alpha_3} \frac{1}{\vap}
\ee
\bea
&& I_{\alpha_1 \alpha_2 \alpha_3 \alpha_4}(2,2)=-\frac{1}{8 \pi^2 p^4}
\left(p_{\alpha_1}p_{\alpha_2}p_{\alpha_3}p_{\alpha_4}-\right.
\nonumber\\ &&\left. \frac{p^4}{24}(
\delta_{\alpha_1 \alpha_2}\delta_{\alpha_3 \alpha_4}+
\delta_{\alpha_1 \alpha_3}\delta_{\alpha_2 \alpha_4}+  
\delta_{\alpha_1 \alpha_4}\delta_{\alpha_2 \alpha_3})\right)\frac{1}{\vap}
\eea
\bea
I_{\alpha_1 \alpha_2 \alpha_3}(2,1)&=&-\frac{1}{8 \pi^2 p^2}
(\frac{p^2}{6}(p_{\alpha_1}\delta_{\alpha_2 \alpha_3}
+p_{\alpha_2}\delta_{\alpha_1 \alpha_3}+
p_{\alpha_3}\delta_{\alpha_2 \alpha_1})
\nonumber\\ &&
-p_{\alpha_1}p_{\alpha_2}p_{\alpha_3}) \frac{1}{\vap}
\eea

\section{Some practical identities for RG quantities}\label{APP_RGstuff}

The beta functions defined in Eq.~\ref{define_set_RG} may be 
re-expressed as
\be\label{APP_RG_bet}
\left(\begin{array}{cc} 
\beta_u(u_R,v_R) \\ \beta_v(u_R,v_R) 
\end{array}\right)=-\vap \left(\begin{array}{cc} 
\frac{\parp \ln u}{\parp u_R} & \frac{\parp \ln u}{\parp v_R} \\
\frac{\parp \ln v}{\parp u_R} & \frac{\parp \ln v}{\parp v_R} 
\end{array}\right)^{-1} 
\left(\begin{array}{cc} 
1 \\ 1 \end{array}\right)
\ee
The previous expression may be further simplified noticing
\be\label{APP_REP}
A=\left(\begin{array}{cc} 
\frac{\parp \ln u}{\parp u_R} & \frac{\parp \ln u}{\parp v_R} \\
\frac{\parp \ln v}{\parp u_R} & \frac{\parp \ln v}{\parp v_R} 
\end{array}\right)=\left(\begin{array}{cc} 
\frac{1}{u_R} & 0 \\ 0 & \frac{1}{v_R} 
\end{array}\right)+D  \ ,
\ee
so that 
\bea\label{APP_RG_invA}
A^{-1}&=&\left(1+\left(\begin{array}{cc} 
u_R & 0 \\ 0 & v_R 
\end{array} \right) D \right)^{-1}
\left(\begin{array}{cc} 
u_R & 0 \\ 0 & v_R 
\end{array}\right)
\\\nonumber
&=&
\left(\begin{array}{cc} u_R & 0 \\ 0 & v_R \end{array}\right)-
\left(\begin{array}{cc} u_R & 0 \\ 0 & v_R \end{array}\right)D
\left(\begin{array}{cc} u_R & 0 \\ 0 & v_R \end{array}\right)+
\cdots
\eea
where the last result follows from Taylor-expanding. These formulas
easily allow to compute the corresponding $\beta$-functions. If 
\bea\label{APP_RG_COU}
u&=&M^{\vap}\left[ u_R+\frac{1}{\vap}(a_{11} u^2_R+a_{12} u_R v_R+
a_{13} v^2_R) \right] \\\nonumber
v&=&M^{\vap}\left[ v_R+\frac{1}{\vap}(a_{21} u^2_R+a_{22} u_R v_R+
a_{23} v^2_R) \right] \ ,
\eea
from Eq.~\ref{APP_RG_invA} and Eq.~\ref{APP_RG_bet} we easily derive
the leading two orders in the couplings
\bea\label{APP_RG_beta}
\beta_u(u_R,v_R)&=&-\vap u_R+a_{11} u^2_R+a_{12} u_R v_R+
a_{13} v^2_R \\\nonumber
\beta_v(u_R,v_R)&=&-\vap v_R+a_{21} u^2_R+a_{22} u_R v_R+
a_{23} v^2_R \ .
\eea

The formula for $\gamma$ in Eq.~\ref{define_set_RG} may also be
given a more practical expression. It is given by 
\be\label{APP_prac_exp}
\gamma=(\beta_u \frac{\parp}{\parp u_R}+
\beta_v \frac{\parp}{\parp v_R}) \ln Z_{\phi} \ , 
\ee

Those are the formulas we need in the calculations we present in this
review.

\section{Discretized Model for tethered membranes}\label{APP__Dis}

In this appendix we present appropriate discretized models for 
numerical simulation of tethered membranes. The surface is discretized by
a triangular lattice defined by its vertices $\{{\vec r}\}_{a=1,\cdots}$,
with a corresponding discretized version of the Landau elastic term
Eq.~\ref{LAN_FL_PH} given by \cite{SN:88}
\be\label{discr__ver__flat}
F_{s}=\frac{\beta}{2}\sum_{\langle a b \rangle} ( |{\vec r}_a-{\vec r}_b|-1)^2
\ ,
\ee
where $\langle a,b \rangle$ are nearest-neighbor vertices.
If we write ${\vec r}_a={\bf x}_a+{\bf u}_a$ with ${\bf x}_a$ defining the 
vertices of a perfectly regular triangular lattice and ${\bf u}$ the 
small perturbations around it, one gets
\be\label{discr__tensor}
|{\vec r}_a-{\vec r}_b|=1+u_{\alpha\beta} x^{\alpha} x^{\beta} +\cdots 
\ ,
\ee
with $u_{\alpha \beta}$ being a discretized strain tensor and we 
we have neglected higher order terms.
Plugging the previous expression into Eq.~\ref{discr__ver__flat} and
passing from the discrete to the continuum language we obtain
\be\label{fin__dis_ver}
F_{s}=\frac{\sqrt{3}}{8}\beta \int d^2 {\bf x} (2 u^2_{\alpha \beta}+
u^2_{\alpha \alpha})
\ee
which is the elastic part of the free energy Eq.~\ref{LAN_FL_PH} with
$\lambda=\mu=\frac{\sqrt{3}}{4} \beta$. 

The bending rigidity term is written in the continuum as
\be\label{bending__rig}
S_{ext}=\int d^2 {\bf x} \sqrt{g} K^{\mu}_{\alpha \beta}
K_{\mu}^{\alpha \beta}=\int d^2 {\bf u} \sqrt{g} g^{\alpha \beta}
\nabla_{\alpha} {\vec n} \nabla_{\beta} {\vec n}
\ee
where ${\vec n}$ is the normal to the surface and $\nabla$ is the 
covariant derivative (see \cite{D:89} for a detailed description of
these geometrical quantities). We discretize the
normals form the the previous equation by
\be\label{bend__rig_dis}
\int d^2 {\bf u} \sqrt{g} g^{\alpha \beta}
\nabla_{\alpha} {\vec n} \nabla_{\beta} {\vec n}
\rightarrow
\sum_{\langle a b \rangle} ({\vec n}_a - {\vec n}_b)^2=
2 \sum_{\langle a b \rangle} (1-{\vec n}_a {\vec n}_b)
\ee

The two terms Eq.~\ref{discr__ver__flat} and Eq.~\ref{bend__rig_dis}
provide a suitable discretized model for a tethered membranes. 
However, in actual simulations, the even more simplified discretization
\be\label{sim__disc}
F=\sum_{\langle a, b \rangle} ({\vec r}_a-{\vec r}_b)^2+
\kappa \sum_{\langle i, j \rangle}(1-{\vec n}_i {\vec n}_j)
\ee
is preferred since it is simpler and describes the same universality 
class (see \cite{BCFTA:96} for a discussion).
Anisotropy may be introduced in this model by ascribing distinct 
bending rigidities to bending across links in different intrinsic
directions \cite{BFT:97}.

Self-avoidance can be introduced in this model by imposing that the 
triangles that define the discretized surface cannot
self-intersect. There are other possible discretizations of self-avoidance
that we discus in sect.~\ref{Sub_SECT__SA}.

In order to numerically simulate the model Eq.~\ref{sim__disc} different
algorithms have been used. A detailed comparison of the 
performance of each algorithm may be found in \cite{TF:98}.

\section{The crumpling Transition}\label{APP__CT__FP}

\begin{figure}[htb]
  \epsfxsize=3 in \centerline{\epsfbox{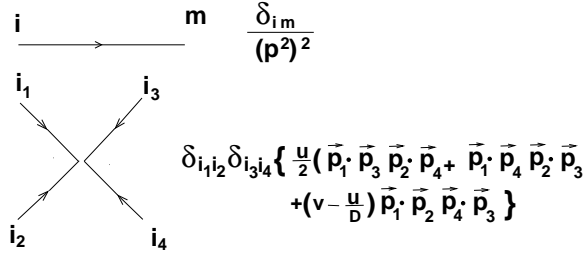}}
  \caption{Feynman rules for the model at the crumpling transition}
  \label{fig__Feyn_CT}
  \end{figure}

The Free energy is given by Eq.~\ref{LAN_CRTR_PH}
\be\label{APP_CRTR_PH}
F({\vec r})=\int d^D{\bf x} \left[ 
\frac{1}{2}(\parp_{\alpha}^2 {\vec r})^2+
u(\parp_{\alpha} {\vec r} \parp_{\beta} {\vec r} -
\frac{\delta_{\alpha \beta}}{D}(\parp_{\alpha \vec r})^2 )^2+
v(\parp_{\alpha} {\vec r} \parp^{\alpha} {\vec r})^2 \right] \ ,
\ee
where the dependence on $\kappa$ is trivially scaled out. 
The Feynman rules for the model are given in Fig.\ref{fig__Feyn_CT}.

We need three renormalized constants, namely $Z$, $Z_u$ and $Z_v$ in order
to renormalize the theory. We define the renormalized quantities by
\bea\label{App_CR__RG}
{\vec r}&=&Z^{-1/2} {\vec r}  
\\\nonumber
u_R=M^{-\vap}Z^2 Z_{u}^{-1} u & \ , &
v_R=M^{-\vap}Z^2 Z_{v}^{-1} v \ .
\eea
Then Eq.~\ref{APP_CRTR_PH} becomes
\bea\label{REN_CRTR_APP}
F({\vec r})&=&\int d^D{\bf x} \left[ 
\frac{Z}{2}(\parp_{\alpha}^2 {\vec r_R})^2+
M^{\vap} Z_{u} u(\parp_{\alpha} {\vec r_R} \parp_{\beta} {\vec r_R} -
\frac{\delta_{\alpha \beta}}{D}(\parp_{\alpha} \vec r_R)^2 )^2
\right. +
\nonumber \\ &+& \left. M^{\vap} Z_{v} v(\parp_{\alpha} {\vec r_R} \parp^{\alpha} 
{\vec r_R})^2 \right] \ ,
\eea

\begin{figure}[htb]
  \epsfxsize=4 in \centerline{\epsfbox{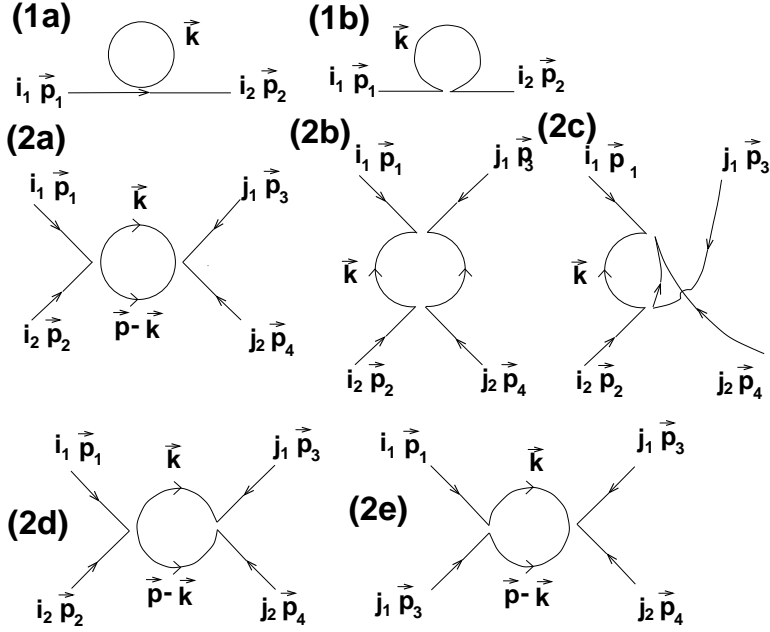}}
  \caption{Diagrams to consider at one loop}
  \label{fig__one_loop_CT}
  \end{figure}

In order to compute the renormalized couplings, one must compute all
relevant diagrams at one loop. Those are depicted in 
fig.~\ref{fig__one_loop_CT}. Within dimensional regularization, 
diagrams (1a) and (1b) are zero, which in turns imply that the renormalized
constant is $Z=1$ at leading order in $\vap$, similarly as in linear
$\sigma$ models.

Using the integrals in dimensional regularization (see
Sect.~\ref{APP__SA__Int}) Diagram (2a) 
gives the result
\be\label{CR__TR__RES2A}
\frac{d}{8 \pi^2} \frac{1}{\vap} \delta^{i_1 i_2} \delta^{j_1 j_2}
\left\{ \frac{u^2}{24}( \vec p_1\cdot\vec p_3 \vec p_2\cdot\vec p_4+
\vec p_1\cdot\vec p_4 \vec p_2\cdot\vec p_3)+
(v^2-\frac{u^2}{48})\vec p_1\cdot\vec p_2 \vec p_3\cdot\vec p_4
\right\}
\ee

And diagram (2b) and (2c) may be computed at once, since the 
result of (2c) is just (2b) after interchanging 
$\vec p_3 \leftrightarrow \vec p_4$, so the total result 
(2a)+(2b) is

\bea\label{CR__TR__RES2BC}
&& \frac{1}{16 \pi^2} \frac{1}{\vap} \delta^{i_1 i_2} \delta^{j_1 j_2}
\left\{ (\frac{61}{96}u^2+\frac{7}{12} uv+\frac{v^2}{6})
( \vec p_1\cdot\vec p_3 \vec p_2\cdot\vec p_4\right.+
\vec p_1\cdot\vec p_4 \vec p_2\cdot\vec p_3)\nonumber\\
&&\left. +(\frac{v^2}{6}+\frac{u v}{12}+\frac{u^2}{96})
\vec p_1\cdot\vec p_2 \vec p_3\cdot\vec p_4 \right\}
\eea

And the result for (2d) and (2e) is just identical, so the total 
result (2d)+(2e) is 

\bea\label{CR__TR__RES2DE}
&& \frac{1}{16 \pi^2} \frac{1}{\vap} \delta^{i_1 i_2} \delta^{j_1 j_2}
\left\{ (\frac{u^2}{24}+\frac{1}{6} uv)
( \vec p_1\cdot\vec p_3 \vec p_2\cdot\vec p_4\right.+
\vec p_1\cdot\vec p_4 \vec p_2\cdot\vec p_3)\nonumber\\
&&\left. +(v^2+\frac{13}{6}u v-\frac{u^2}{48})
\vec p_1\cdot\vec p_2 \vec p_3\cdot\vec p_4 \right\}
\eea
Adding up all these contributions taking into account 
the different combinatorial factors (4 the first contribution, 8 the
last two ones) and recalling $Z=1$, we get
\bea\label{CR__rencoup}
u&=& M^{\vap}\left[ u_R+\frac{1}{\vap}\frac{1}{8 \pi^2}\left( 
(\frac{d}{3}+\frac{65}{12})u^2_R+6u_R v_R+\frac{4}{3} v^2_r\right)\right]
\nonumber\\
v&=&M^{\vap}\left[ v_R+\frac{1}{\vap}\frac{1}{8 \pi^2}\left( 
\frac{21}{16} u_R +\frac{21}{2} u_R v_R+(4d+5)v^2_R\right)\right] \ .
\eea
The resultant $\beta$-functions are then readily obtained by applying
Eq.(\ref{APP_RG_beta}).  

\section{The Flat Phase}\label{APP__FP__FP}

\begin{figure}[htb]
  \epsfxsize=4 in \centerline{\epsfbox{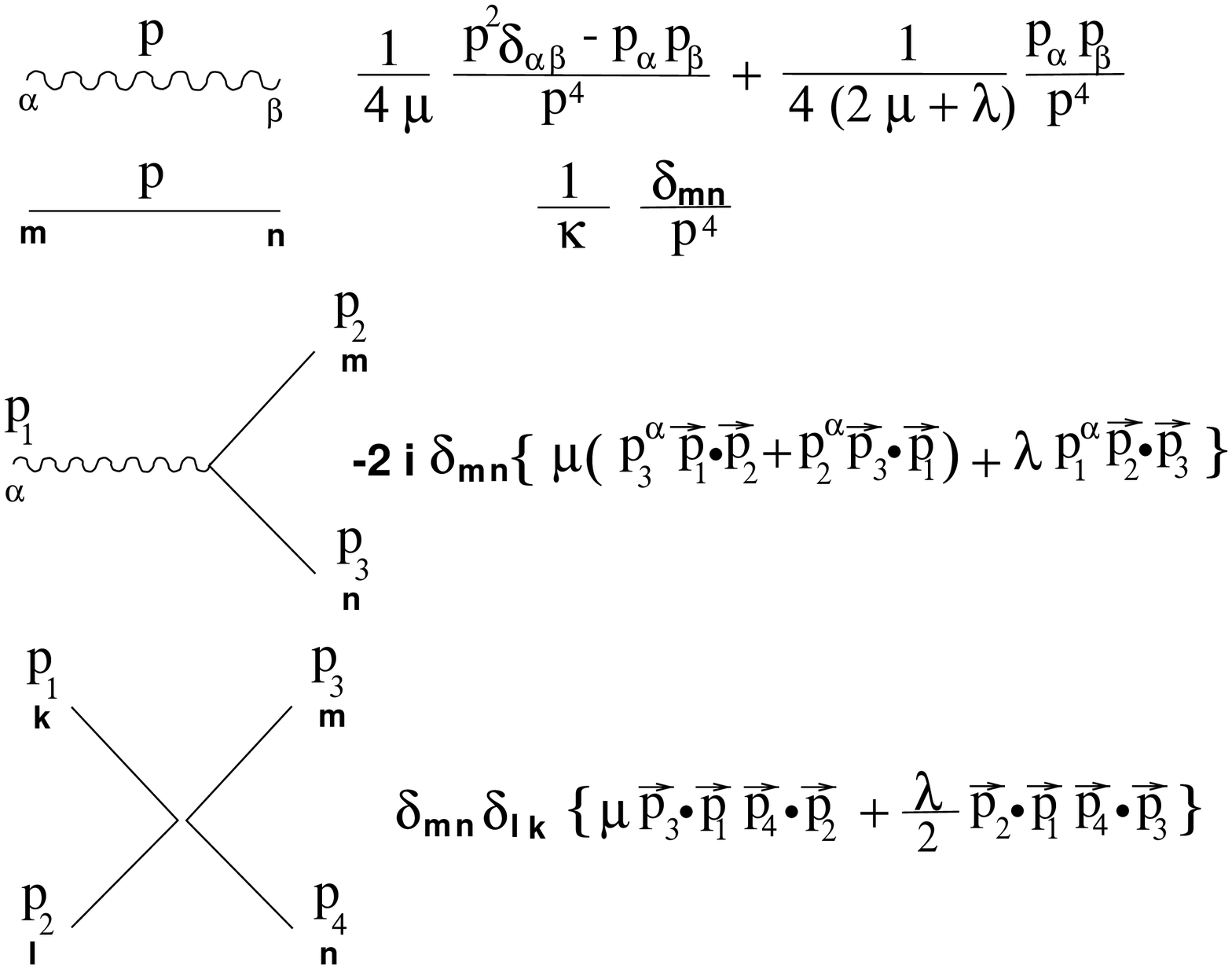}}
  \caption{Feynman rules for the model in the flat phase}
  \label{fig__Feyn_FLA}
  \end{figure}

The free energy is given in Eq.~\ref{LAN_FL_PH},and it is given 
by
\be\label{LAN_FL_PH_APP}
F({\bf u},h)=\int d^D {\bf x} \left[
\frac{\hat{\kappa}}{2} (\parp_{\alpha} \parp_{\beta} h )^2+
\mu u_{\alpha \beta} u^{\alpha \beta} + \frac{\lambda}{2} (u^{\alpha}_{\alpha})^2
\right] \, .
\ee
The Feynman rules are shown in fig.~\ref{fig__Feyn_FLA}, it is apparent that 
the in-plane phonons couple different from the out-of-plane, which play the
role of Goldstone bosons.

\begin{figure}[htb]
  \epsfxsize=4 in \centerline{\epsfbox{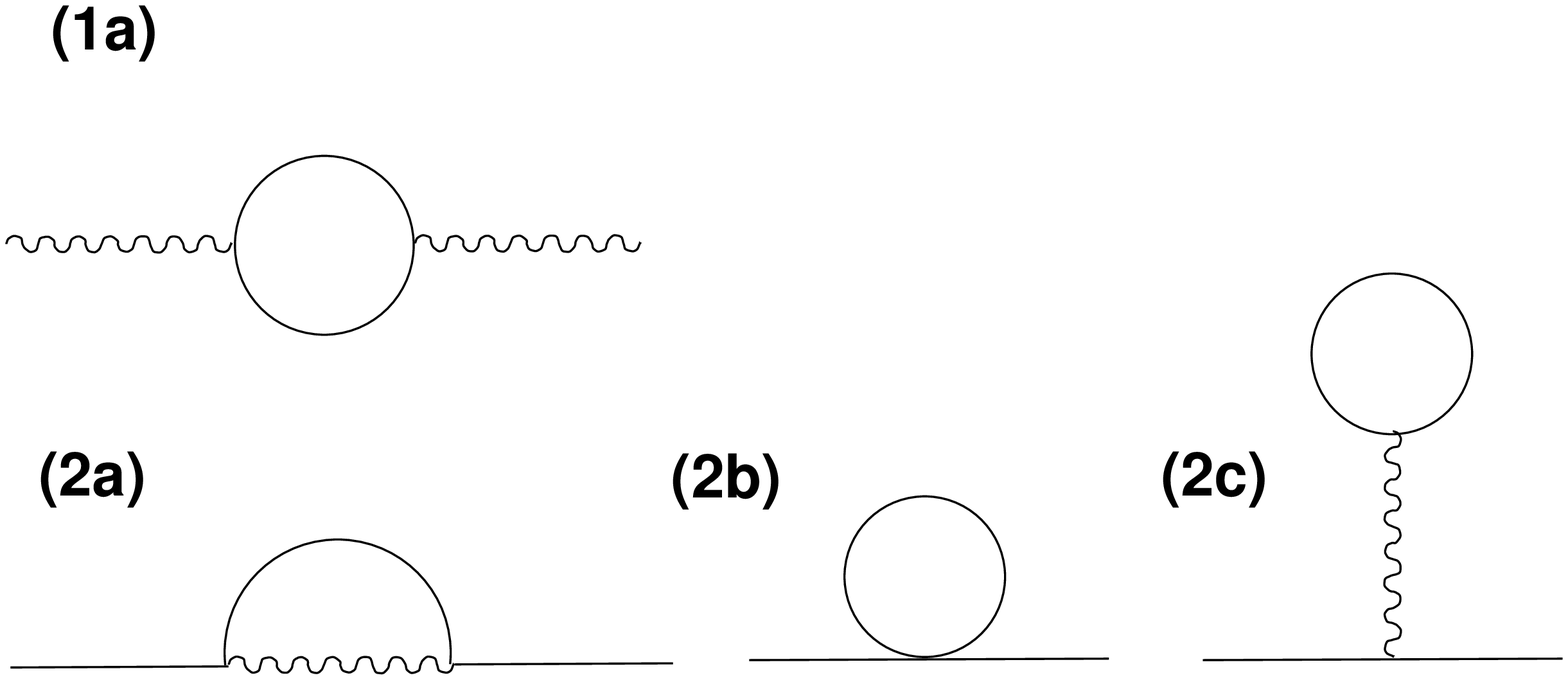}}
  \caption{Diagrams to consider at one loop}
  \label{fig__one_loop_FP}
  \end{figure}

We apply standard field theory techniques to obtain the RG-quantities. 
Using the Ward identities, the theory can be renormalized using 
only three renormalization constants $Z$, $Z_{\mu}$ and $Z_{\lambda}$,
corresponding to the wave function and the two coupling renormalizations. 
Renormalized quantities read
\bea\label{App_def__RG}
h_R=Z^{-1/2} h  & \ , & {\bf u}=Z^{-1} {\bf u}
\\\nonumber
\mu_R=M^{-\vap}Z^2 Z_{\mu}^{-1} \mu & \ , &
\lambda_R=M^{-\vap}Z^2 Z_{\lambda}^{-1} \lambda \ ,
\eea
Then Eq.~\ref{LAN_FL_PH_APP} becomes
\be\label{REN_FL_PH_APP}
F({\bf u},h)=\int d^D {\bf x} \left[
Z (\parp_{\alpha} \parp_{\beta} h_R )^2+
2 M^{\vap} Z_{\mu} \mu_R u_{R \alpha \beta} u^{\alpha \beta}_R + 
M^{\vap} Z_{\lambda} \lambda (u^{\alpha}_{R \alpha})^2
\right] \, .
\ee
We now compute the renormalized quantities from the leading divergences
appearing in the Feynman diagrams. The diagrams to consider 
are given in Fig.\ref{fig__one_loop_FP}. These can 
be computed using the integrals given in Sect.\ref{APP__SA__Int}.

The result of diagram (1a) is given by
\be\label{Dia__1A}
\frac{1}{\vap}\frac{d_c}{6 \pi^2}\left[ 
\mu^2(\delta_{\alpha \beta}-\frac{p_{\alpha} p_{\beta}}{p^2})+
3(\mu^2+2\mu\lambda+2\lambda^2)\frac{p_{\alpha}p_{\beta}}{p^2}\right] p^2
 \ .
\ee
Diagrams (2b) and (2c) are identically zero, so (2a) is the only 
additional diagram to be computed. The result is
\be\label{Dia__2A}
-\frac{1}{\vap}\frac{\delta^{i j}}{8 \pi^2} 
\frac{\mu(\mu+\lambda)}{2 \mu+\lambda} 10 (p^2)^2
\ee
from Eq.~\ref{Dia__2A} and the definitions in Eq.~\ref{App_def__RG}
\be\label{zeta_plain}
Z=1-\frac{1}{\vap}\frac{10}{8 \pi^2}\frac{\mu_R(\mu_R+\lambda_R)}
{2\mu_R+\lambda_R} \ .
\ee
Using the previous result in the diagrams (1a) whose result is in 
Eq.~\ref{Dia__1A} we obtain 
\bea\label{zetamula}
Z_{\mu}&=&1+\frac{1}{\vap}\frac{d_c}{24 \pi^2} \mu_R
\\\nonumber
Z_{\lambda}&=&1+\frac{1}{\vap}\frac{d_c}{24 \pi^2}(\mu^2_R+6\mu_R\lambda_R
+6\lambda_R^2)/\lambda_R
\eea
and we deduce the renormalized couplings
\bea\label{FL__rencoup}
\mu&=& M^{\vap}\left[ \mu_R+\frac{1}{\vap}\left( 
\frac{10}{4 \pi^2}\frac{(\mu_R+\lambda_R)}{2\mu_R+\lambda_R}+
\frac{d_c}{24\pi^2}\right)\mu^2_R \right]
\\\nonumber
\lambda&=&M^{\vap}\left[ \lambda_R+\frac{1}{\vap}\left( 
\frac{10}{4 \pi^2}\frac{(\mu_R+\lambda_R)}{2\mu_R+\lambda_R}\mu_R\lambda_R+
\frac{d_c}{24 \pi^2}(\mu^2_R+6\mu_R\lambda_R+6\lambda^2_R)\right)\right]
\eea
from which the $\beta$-functions trivially follow with the aid of 
Eq.(\ref{APP_RG_beta}).

\section{The Self-avoiding phase}\label{APP__SA__CP}

The model has been introduced in Eq.~\ref{CRU__SA} and is given by
\be\label{APP_CRU__SA}
F({\vec r})=\frac{1}{2}\int d^D {\bf x} (\parp_{\alpha} {\vec r}({\bf x}))^2
+\frac{b}{2}\int d^D{\bf x}\, d^D{\bf y} \delta^d({\vec r}({\bf x})-
{\vec r}({\bf y})) \ ,
\ee
We follow the usual strategy of defining the renormalized quantities by
\bea\label{App_SAdef__RG}
 {\vec r}&=&Z^{1/2} {\vec r}_R
\\\nonumber
 b&=&M^{\vap}Z_b Z^{d/2} b_R  \ ,
\eea
and the renormalized Free energy by 
\be\label{APP_REN__SA}
F({\vec r})=\frac{Z}{2}\int d^D {\bf x} (\parp_{\alpha} {\vec r}_R({\bf x}))^2
+M^{\vap} Z_b 
\frac{b_R}{2}\int d^D{\bf x}\, d^D{\bf y} \delta^d({\vec r}_R({\bf x})-
{\vec r}_R({\bf y})) \ .
\ee
The $\delta$-function being non-local adds some technical
difficulties to the calculation of the renormalized constants $Z$ and $Z_b$. 
There are different approaches available but we will follow the 
MOPE (Multilocal-Operator-Product-Expansion), which we will just explain
in a very simplified version. A rigorous description of the method 
may be found in the literature. 

The idea is to expand the $\delta$-function term in Eq.~\ref{APP_REN__SA}
\be\label{APP_SA_exp}
e^{-F({\vec r})}=e^
{-\frac{Z}{2}\int d^D {\bf x} (\parp_{\alpha} {\vec r}_R({\bf x}))^2}
\times \sum_{n=0}^{\infty} \left(-M^{\vap} Z_b 
\frac{b_R}{2}\int d^D{\bf x}\, d^D{\bf y} \delta^d({\vec r}_R({\bf x})-
{\vec r}_R({\bf y})) \right)^n \ ,
\ee
with this trick, the delta-function term may be treated
as expectation values of a Gaussian free theory. This observation alone
allows to isolate the poles in $\vap$. We write the identity
\be\label{APP__Norm_ord}
e^{i {\vec k}({\vec r}({\bf x}_1)-{\vec r}({\bf x}_2))}=
:e^{i {\vec k}({\vec r}({\bf x}_1)-{\vec r}({\bf x}_2))}:e^{k^2 G(x_1-x_2)}
\ ,
\ee
where $G(x)$ is the two point correlator 
\be\label{APP_two_poin}
-G({\bf x})=-\langle {\vec r}({\bf x}) {\vec r}(0) \rangle =
\frac{|{\bf x}|^{2-D}}{(2-D)S_D} \ ,
\ee
with $S_D$ being the volume of the $D$-dimensional sphere.
The symbol $::$ stands for normal ordering. A normal ordered operator is
non-singular at short-distances, so it may be Taylor-expanded
\bea\label{APP__Norm__Tayl}
e^{i {\vec k}({\vec r}({\bf x}_1)-{\vec r}({\bf x}_2))}&=&
(1+i({\bf x_1-x_2})^{\alpha} ({\vec k} \parp_{\alpha} {\vec r})
\\\nonumber
&& -\frac{1}{2}({\bf x_1-x_2})^{\alpha}({\bf x_1-x_2})^{\beta} 
({\vec k} \parp_{\alpha}{\vec r}) ({\vec k} \parp_{\beta} {\vec r})
+\cdots) e^{k^2 G(x_1-x_2)} \ .
\eea
To isolate the poles in $\vap$ we do not need to consider higher order
terms as it will become clear.
If we now  integrate over ${\vec k}$, we get
\bea\label{APP_MOPE}
\delta^d({\vec r}({\bf x}_1)-\vec r({\bf x}_2))&=&
\frac{1}{(4 \pi)^{d/2}(-G({\bf x}_1-{\bf x}_2))^{d/2}} 1
\\\nonumber
&&-
\frac{1}{4}\frac{({\bf x}_1-{\bf x}_2)^{\alpha}({\bf x}_1-{\bf x}_2)^{\beta}}
{(4 \pi)^{d/2}(-G({\bf x}_1-{\bf x}_2))^{d/2+1}} 
 \parp_{\beta} {\vec r}({\bf x}) \parp_{\alpha} {\vec r}({\bf x})+\cdots
\\\nonumber
&\equiv& C^1({\bf x}_1-{\bf x}_2) 1 + C^{\alpha \beta}({\bf x}_1-{\bf x}_2)   
 \parp_{\beta} {\vec r}({\bf x}) \parp_{\alpha} {\vec r}({\bf x})+\cdots
\eea
where we omit higher dimensional operators in ${\vec r}$, which are 
irrelevant by power counting, so since the theory is renormalizable they
cannot have simple poles in $\vap$. Additionally, we have defined
${\bf x}=\frac{{\bf x}_1+{\bf x}_2}{2}$. 
One recognizes in Eq.~\ref{APP_MOPE} Wilson's Operator product expansion,
applied to the non-local delta-function operator.

\begin{figure}[htb]
  \epsfxsize=4 in \centerline{\epsfbox{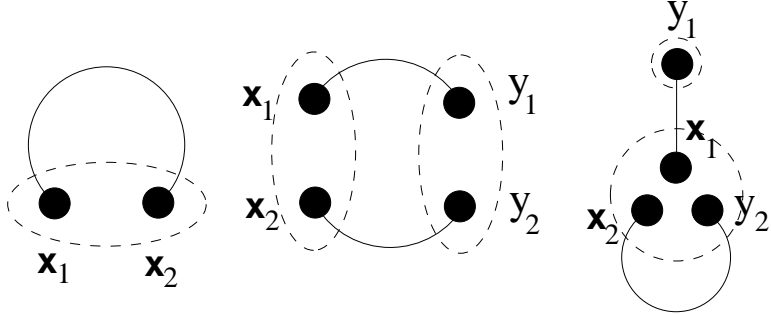}}
  \caption{Diagrammatic expansion to isolate the poles in $\vap$ within 
  the MOPE formalism at lowest non-trivial order. Solid lines represent
  $\delta$-function terms and dashed lines indicate that points inside
  are taken arbitrarily close. Higher orders contributions arise in the
  same way.}
  \label{fig__MOPE}
  \end{figure}

Following the same technique of splitting the operator into a normal
ordered part and a singular part at short-distances, it just takes a
little more effort to derive the OPE for the product of two delta
functions, the result is
\be\label{APP_twodel}
\delta^d({\vec r}({\bf x}_1)-\vec r({\bf y}_1))
\delta^d({\vec r}({\bf x}_2)-\vec r({\bf y}_2))=
C({\bf x}_1-{\bf x}_2,{\bf y}_1-{\bf y}_2) 
\delta^d({\vec r}({\bf x})-\vec r({\bf y}))+\cdots
\ee
with $C({\bf x}_1-{\bf x}_2,{\bf y}_1-{\bf y}_2)= \frac{1}{(4 \pi)^{d/2}
(-G({\bf x}_1-{\bf x}_2)-G({\bf y}_1-{\bf y}_2))^{d/2}}$. The terms
omitted are again higher dimensional by power counting so they.
The OPE Eq.~\ref{APP_MOPE} and Eq.~\ref{APP_twodel} is all we need to 
compute the renormalization constants at lowest order in $\vap$,
but the calculation may be pursued to higher orders in $\vap$. In order
to do that, one must identify where poles in $\vap$ arise. In the 
previous example poles in $\vap$ appear whenever the internal coordinates
(${\bf x}_1$ and ${\bf x}_2$ in Eq.~\ref{APP_MOPE},
${\bf x}_1$,${\bf x}_2$,${\bf y}_3$ and ${\bf y}_2$ in 
Eq.~\ref{APP_twodel}) are pairwise made to coincide. This is diagrammatically
shown in fig.~\ref{fig__MOPE}. It is possible to show, that higher poles
appear in the same way, if more $\delta$-product terms are considered.

Let us consider the first delta-function term corresponding to $n=1$
in the sum Eq.~\ref{APP_SA_exp}. Using Eq.~\ref{APP_MOPE} we have
\bea\label{APP_comZ}
&&-\frac{b_r M^{\vap}}{2} Z_{b} \int d^D{\bf x}\, d^D{\bf y} 
\delta^d({\vec r}_R({\bf x})-{\vec r}_R({\bf y}))
\\\nonumber
&=&-\frac{b_r M^{\vap}}{2} Z_{b} \int d^D{\bf x}\, d^D{\bf y} 
(C^1({\bf x}-{\bf y})+ C^{\alpha \beta}({\bf x}-{\bf y})   
\parp_{\beta} {\vec r}_R({\bf x}) \parp_{\alpha} {\vec r}_R({\bf x})+
\cdots
\\\nonumber
&=&-\frac{b_r M^{\vap}}{2} Z_{b}\int d^D{\bf x} 
\parp_{\alpha} {\vec r}_R({\bf x}) \parp^{\alpha} {\vec r}_R({\bf x})
\int d^D{\bf y} \frac{\delta_{\alpha \beta} C^{\alpha \beta}}{D}+\cdots
\eea
The first term just provides a renormalization of the identity
operator, which we can neglect. From
\be\label{APP_int_in_MOPE}
\int_{|{\bf y}|> 1/M} d^D{\bf y} \frac{\delta_{\alpha \beta} 
C^{\alpha \beta}}{D}({\bf y})=
-\frac{1}{4D}\frac{M^{-\vap}}{\vap} (4 \pi)^{-d/2}(2-D)^{1+d/2}
\left(\frac{2 \pi^{D/2}}{\Gamma(D/2)}\right)^{2+d/2} \ ,
\ee
and we can absorb the pole by $Z$ if we define
\be\label{APP_Z_factor}
Z=1+\frac{b_R}{\vap}\frac{(4 \pi)^{-d/2}}{4D}(2-D)^{1+d/2}
\left(\frac{2 \pi^{D/2}}{\Gamma(D/2)}\right)^{2+d/2} \ ,
\ee

From the short distance behavior in the sum Eq.~\ref{APP_SA_exp}
corresponding to $n=2$ we get
\bea\label{APP_comZb}
&&-\frac{b_r^2 M^{2\vap}}{8} \int d^D{\bf x}_1 \, d^D{\bf y}_1 
d^D{\bf x}_2 \, d^D{\bf y}_2
\delta^d({\vec r}_R({\bf x}_1)-{\vec r}_R({\bf y}_1))
\delta^d({\vec r}_R({\bf x}_1)-{\vec r}_R({\bf y}_2))
\nonumber\\
&&-\frac{b_r^2 M^{2\vap}}{8} \int d^D{\bf x} \, d^D{\bf y} 
\delta^d({\vec r}_R({\bf x})-{\vec r}_R({\bf y}))
\int d^D{\bf z} \, d^D{\bf w} C(z,w)  \ ,
\eea
where, in order to isolate the pole we can perform the following tricks
\bea\label{APP_last__int}
&& \int d^D{\bf z} \, d^D{\bf w} C(z,w)
\\\nonumber
&=& (4 \pi)^{-d/2}S_D^{d/2}(2-D)^{d/2} 
\left(\frac{2 \pi^{D/2}}{\Gamma(D/2)}\right)^2
\int_0^{M^{-1}}dz \int_0^{M^{-1}}d w \frac{z^{D-1}w^{D-1}}
{(z^{2-D}+w^{2-D})^{d/2}}
\\\nonumber
&=& (4 \pi)^{-d/2}S_D^{d/2}(2-D)^{d/2} 
\left(\frac{2 \pi^{D/2}}{\Gamma(D/2)}\right)^2\frac{M^{-\vap}}{(2-D)^2}
\int_0^{1}\int_0^{1}d x d y  \frac{x^{\frac{D}{2-D}} y^{\frac{D}{2-D}}}
{(x+y)^{-d/2}}
\\\nonumber
&=& (4 \pi)^{-d/2}S_D^{d/2}(2-D)^{d/2} 
\left(\frac{2 \pi^{D/2}}{\Gamma(D/2)}\right)^2\frac{M^{-\vap}}{(2-D)^2}
\int_{x^2+y^2 \le 1} d x d y  \frac{x^{\frac{D}{2-D}} y^{\frac{D}{2-D}}}
{(x+y)^{-d/2}}
\\\nonumber
&=& (4 \pi)^{-d/2}S_D^{d/2}(2-D)^{d/2} 
\left(\frac{2 \pi^{D/2}}{\Gamma(D/2)}\right)^2\frac{M^{-\vap}}{(2-D)^3}
\frac{\Gamma(\frac{D}{2-D})^2}{\Gamma(\frac{2D}{2-D})}\frac{1}{\vap}
\eea
since changing the boundary of integration from a square to a circle
does not affect the residue of the pole.
We finally have
\be\label{APP_Zb_final}
Z_b=1+\frac{b_R}{\vap}\frac{1}{2}(2-D)^{-1+d/2} 
\frac{\Gamma(\frac{D}{2-D})^2}{\Gamma(\frac{2D}{2-D})}
\left(\frac{2 \pi^{D/2}}{\Gamma(D/2)}\right)^{2+d/2}
(4 \pi)^{-d/2} \ ,
\ee
and the $\beta$-function follows from the definitions 
Eq.(\ref{App_SAdef__RG}) together with Eq.(\ref{APP_Z_factor}) 
and Eq.(\ref{APP_RG_beta}).  

\section{The mean field solution of the anisotropic case}\label{MF__Appen}

The free energy has been introduced in Eq.~\ref{LGW}. Let us first 
show the constraints on the couplings so that the Free energy
is bounded from below. 
\begin{itemize}
\item{$u_{yy} > 0$:} This follows trivially.
\item{$ u^{\prime}_{\perp \perp} \equiv v_{\perp \perp}+
\frac{u_{\perp \perp}}{D-1} > 0$ :} 
Define $A_{\alpha}^i=\parp_{\alpha} r^i({\bf x})$ then from 
Eq.~\ref{LGW} we
get
\bea\label{ineq_ani_APP}
&&\frac{u_{\perp \perp}}{2} Tr (A A^{T})^2+\frac{v_{\perp \perp}}{2}
(Tr A A^{T})^2\ge(\frac{u_{\perp \perp}}{D-1}+v_{\perp \perp})/2 
(Tr A A^{T})^2
\nonumber \\
&=&\frac{u^{\prime}_{\perp \perp}}{2}(Tr A A^{T})^2 \ ,
\eea
which implies $u_{\perp \perp}^{\prime} > 0$.
\item{$ v_{\perp y} > -(u^{\prime}_{\perp \perp} u_{yy})^{1/2}$ :}
defining ${\vec b}=\partial_y {\vec r}({\bf x})$, It is derived from
\be\label{further_cons_any}
\frac{u^{\prime}_{\perp \perp}}{2}(Tr(A^TA))^2+\frac{u_{yy}}{2} (b^2)^2+
v_{\perp y} b^T b Tr A A^T > 0 \ .
\ee
\end{itemize}

Introducing the variables 
\be\label{def__pot_an}
A=\left(\begin{array}{cc}
         v_{\perp \perp}+\frac{u_{\perp \perp}}{D-1} & v_{\perp y} \\
	 v_{\perp y} & u_{yy}
	 \end{array} \right) \ , \ \
b=(t_{\perp},t_y) 
\ee
and $w=((D-1)\zeta^2_{\perp},\zeta^2_y)$, the mean field effective 
potential may be written as
\be\label{eff_pot_an}
V(w)=\frac{1}{2}L_{\perp}^{D-1} L_y \left[ w \cdot b +
\frac{1}{2} w \cdot A \cdot w \right] \ .
\ee
In this form, it is easy to find the four minima of Eq.~\ref{eff_pot_an},
those are
\begin{enumerate}
\item{Crumpled phase:} 
\be\label{Flat_crumpled_an}
\begin{array}{l}
\zeta^2_{\perp}=0 \\
\zeta^2_{y}=0
\end{array} \  \ \ V_{min}=0
\ee
\item{Flat phase:}
\be\label{Flat_par_an}
\begin{array}{l}
\zeta^2_{\perp}=-\frac{u_{yy} t_{\perp}-v_{\perp} t_y}
{\Delta (D-1)} \\
\zeta^2_{y}=\frac{-v_{\perp y} t_{\perp}+u_{\perp \perp}^{\prime}t_y}{\Delta} 
\end{array} \ \ \ \ V_{min}=-\frac{L_{\perp}^{D-1}L_y}{4 \Delta}
\left[ u_{\perp \perp}^{\prime} t^2_y+u_{yy} t^2_{\perp}-2 v_{\perp y} 
t_{\perp} t_y \right]
\ee
\item{${\perp}$-Tubule:}
\be\label{Perptub_par_an}
\begin{array}{l}
\zeta^2_{\perp}=\frac{-t_{\perp}}{u_{\perp \perp}^{\prime}} 
\\ \zeta^2_{y}=0
\end{array} \ \ \ \  V_{min}=-\frac{L_{\perp}^{D-1}L_y}{4}
\frac{t^2_{\perp}}{u_{\perp \perp}^{\prime}}
\ee
\item{$y$-Tubule:}
\be\label{ytub_par_an}
\begin{array}{l}
\zeta^2_{\perp}=0 
\\ \zeta^2_{y}=-\frac{t_y}{u_{yy}} 
\end{array} \ \ \ \ V_{min}=-\frac{L_{\perp}^{D-1}L_y}{4}\frac{t^2_{y}}{u_yy}
\ee
\end{enumerate}

The regions in which  each of the four minima prevails depend on the
sign of $\Delta$. 
\begin{itemize}
\item{$\Delta > 0$:} 
 Let us see under which conditions the flat phase may occur.  
 We must satisfy the equations 
 \bea\label{App_flat_ani}
 u_{yy}t_{\perp} & < & v_{\perp y} t_{y} 
 \nonumber\\
 u^{\prime}_{\perp \perp} t_y  & < & v_{\perp y} t_{\perp y} 
 \eea
 If $v_{\perp y} > 0$
 this inequalities can only be satisfied if both
 $t_{\perp}$ and $t_y0$ have the same sign. If they
 are positive, Eq.~\ref{App_flat_ani} imply
 $\Delta t_{\perp} < 0$ or  $\Delta t_y < 0$,
 which by the assumption $\Delta > 0$ cannot be satisfied.
 The flat phase exists for $t_y < 0$ and $t_{\perp} < 0$ and 
 satisfying Eq.~\ref{App_flat_ani}. If $t_y > 0$ and $t_{\perp} > 0$
 then the flat phase or the tubular cannot exist (see
 Eq.~\ref{Perptub_par_an} and Eq.~\ref{ytub_par_an}) so those
 are the conditions for the crumpled phase. Any other case is a 
 tubular phase, either ${\perp}$-tubule or $y$-tubule, depending on
 which of the inequalities Eq.~\ref{App_flat_ani} is not satisfied.
 If $v_{\perp y} < 0$ it easily checked from Eq.~\ref{Flat_par_an} that
 the flat phase exists as well and the same analysis apply.

\item{$\Delta < 0$:}
 From inequality Eq.~\ref{further_cons_any} we have $v_{\perp y} > 0$.
 The inequalities are now
 \bea\label{App_flat__ani_2}
  t_{y} & < & \frac{u_{yy}}{v_{\perp}} t_{\perp}  
 \nonumber\\
  t_y  & > & \frac{v_{\perp y}}{u_{\perp \perp}^{\prime} t_{\perp y} }
 \eea
 Now, in order to have a solution for both inequalities we must have 
 $\frac{u_{yy}}{v_{\perp}} > \frac{v_{\perp y}}{u_{\perp \perp}^{\prime}}$
 which requires $\Delta > 0$. This proves that the flat phase cannot
 exist. There is then a crumpled phase for $t_{y}>0$ and $t_{\perp}> 0$
 and tubular phase when either one of this two conditions are not
 satisfied.
\end{itemize}

\newpage 

\bibliography{melt,examples,reviews,thesis}

\end{document}